\documentclass[twocolumn,prl,amsmath,amssymb,floatfix,superscriptaddress]{revtex4-2} 
\usepackage{amsmath}
\usepackage{amssymb}
\usepackage{graphicx}
\usepackage{bm}
\usepackage{color}
\usepackage{hyperref}
\usepackage{graphicx}
\usepackage{upgreek}
\usepackage{scalerel}
\usepackage{multirow}
 
\usepackage{pgffor}
\usepackage{pdfpages}
\usepackage{mathdots}
\usepackage{float}
\usepackage{silence}
\usepackage{physics} 
\usepackage{lscape}   
\usepackage{color, colortbl}
\usepackage[table]{xcolor}
\usepackage{comment}  
\makeatletter
\AtBeginDocument{\let\LS@rot\@undefined}
\makeatother

\begin{document}
\title{Floquet engineering spin triplet states in unconventional magnets}

\author{Pei-Hao Fu}
\email{phy.phfu@gmail.com}
\affiliation{School of Physics and Materials Science, Guangzhou University, Guangzhou 510006, China}

\author{Sayan Mondal}
\affiliation{Department of Physics and Astronomy, Uppsala University, Box 516, S-751 20 Uppsala, Sweden}

\author{Jun-Feng Liu}
\affiliation{School of Physics and Materials Science, Guangzhou University, Guangzhou 510006, China}

 \author{Yukio Tanaka}
 \affiliation{Department of Applied Physics,
 Nagoya University, 464-8603 Nagoya, Japan}
 \affiliation{Research Center for Crystalline Materials Engineering, Nagoya University, 464-8603  Nagoya Japan}

\author{Jorge Cayao}
\email{jorge.cayao@physics.uu.se}
\affiliation{Department of Physics and Astronomy, Uppsala University, Box 516, S-751 20 Uppsala, Sweden}

\date{\today}
\begin{abstract}
We consider unconventional magnets with and without spin-singlet $s$-wave superconductivity and demonstrate the emergence of spin triplet states due to light drives.  In particular, we find that a high-frequency linearly polarized light drive induces a spin-triplet density in $d$-wave altermagnets which does not exist in the static regime and can directly reveal the strength of the altermagnetic field.  In this high-frequency regime, we also show that linearly polarized light enables the formation of odd-frequency spin-triplet superconducting correlations possessing $d$-wave and $s$-wave parities, which can be controlled by the light drive and accessed by measuring the spin density. Moreover, for low-frequency linearly and circularly polarized light drives, we obtain that the types of superconducting correlations are broadened due to the presence of Floquet bands, enabling spin-triplet pairs in $d$- and $p$-wave unconventional magnets, which are absent in the static phase.
\end{abstract}
\maketitle
The control of materials' properties by tailored light is at the forefront of condensed matter physics, not only because it deepens our understanding of matter but also due to its potential for emergent phenomena that are absent in the static regime 
\cite{cayssol2013floquet,bukov2015universal,OkaReview2019,Giovannini_2019,rudner2020band,francesconi2020engineering}. 
Of special interest is the manipulation of properties by time-periodic light drives, where Floquet’s theorem \cite{ASENS_1883_2_12__47_0,PhysRev.138.B979,PhysRevA.7.2203} is used to describe the light effects and hence referred to as Floquet engineering \cite{OkaReview2019}. 
The prospect of Floquet engineering was demonstrated in several systems \cite{bukov2015universal,OkaReview2019,Giovannini_2019,rudner2020band,francesconi2020engineering}, with the most salient example being the prediction of topological phases \cite{PhysRevB.79.081406,PhysRevB.82.235114,lindner2011floquet,cayssol2013floquet,PhysRevLett.116.176401} having Floquet bands due to a non-trivial light-matter interaction \cite{PhysRevA.56.748,PhysRevB.84.235108,rechtsman2013photonic,PhysRevB.88.155129,wang2013observation,mciver2020light,aeschlimann2021survival,mahmood2016selective}.
To a great extent, the intriguing way light interacts with matter is strongly tied to the effect that light has on spin and momentum \cite{PhysRevB.79.081406},  making systems with relativistic spin-orbit coupling (SOC) \cite{galitski2013spin,manchon2015new} promising for novel Floquet phases. 

The recent prediction of unconventional magnets (UMs) has opened a new arena where an unusual SOC appears due to nonrelativistic effects \cite{Bai_review24,jungwirth2024He3Ams,jungwirth2024,FukayaCayaoReviewUMs}.
UMs possess zero net magnetization, as in antiferromagnets, and a nonrelativistic splitting of energy bands, as in ferromagnets, which then induce anisotropic spin-polarized Fermi surfaces \cite{Bai_review24,FukayaCayaoReviewUMs}. UMs can be categorized by the dependence of their magnetic order on momentum \cite{Bai_review24}, which leads to even- and odd-parity UMs. The simplest UMs that lately attracted enormous attention have $d$- and $p$-wave parities, which are commonly referred to as $d$-wave altermagnets (AMs) \cite{landscape22,MazinPRX22} and $p$-wave UMs \cite{brekke24,hellenes2024P}. 
The unique spin-momentum properties of UMs have already enabled exotic phases in normal \cite{NakaNatCommun2019,Hayami19,NakaPRB2020,Hayami20,PhysRevB.101.220403,Bai_review24,jungwirth2024He3Ams,jungwirth2024,Libor011028,GonzalezBetancourt2023,Tschirner2023Saturation,Reichlova2024,Samanta2024Spin,Sun2023Spin,Reja2024,Werner2024High,Farajollahpour2025Light,Ezawa2025Third, Fu2025All,Yang2025Unconventional, Peng2025Ferroelastic, Zhu2025TwoDimensional} and superconducting states \cite{PhysRevB.108.L060508,PhysRevB.108.054511,Maeda_2024,PhysRevLett.131.076003,PhysRevLett.133.226002,PhysRevB.108.224421,PhysRevB.109.245424,PhysRevB.110.L060508,PhysRevB.111.L100507,sukhachov2025,PhysRevLett.133.106601,PhysRevB.111.L121401,PhysRevB.108.205410,PhysRevB.111.184515,PhysRevB.111.064502,PhysRevB.111.165406,PhysRevLett.134.026001,chakraborty2024constr,maeda2025pair,PhysRevB.109.L220505,chatterjee2025inter,PhysRevB.108.075425,sun2025pseudopwave,PhysRevB.110.024503,FukayaCayaoReviewUMs,Nagae2025Flat,Fukaya2025Tunneling,lu2025engineeringsubgapstatessuperconductors,fukaya2025crossedsurfaceflatbands,Fu2025Floquet}. 
Despite the great promise of UMs, most of the focus has been devoted to static systems, leaving the effect of light drives and Floquet engineering still an open question.

 \begin{figure}[!t]
\centering
 \includegraphics[width=0.98\columnwidth]{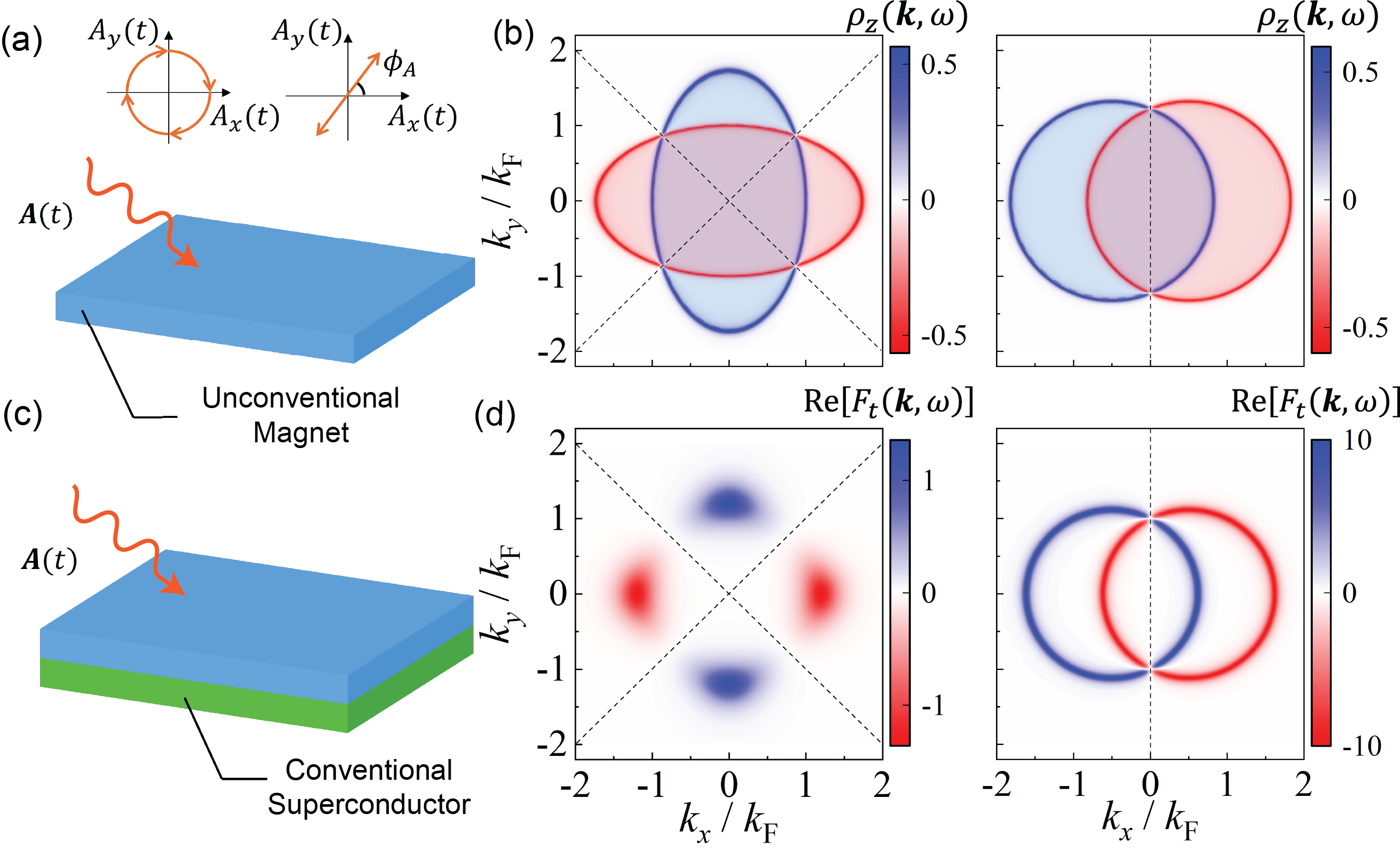}
 \caption{
 (a) Sketch of an UM (blue) under a time-periodic light drive $\bm{A}(t)$ (orange winkle arrow) with circular or linear polarization. 
 (b) Spin densities along $z$ in the static regime of $d$-wave AMs and $p$-wave UMs as a function of $k_{x}$ and $k_{y}$. 
 (c) Sketch of an UM coupled to a conventional spin-singlet $s$-wave superconductor (green) under a time-periodic light drive  $\bm{A}(t)$. 
 (d) Real part of the mixed spin-triplet pair amplitude in the static regime of $d$-wave AMs and $p$-wave UMs with conventional superconductivity.
 Parameters: for (b) $M=0.5$, $\mu=1$, $k_{\rm F}=1$, $\hbar^{2}/2m=1$, $\theta_{d,p}=0$ at frequency $z=0+i 10^{-3}$; for the $d$-wave AM and $p$-wave UM in (d), $\Delta=0.7\mu$ and $\Delta=0.25\mu$, respectively, at $z=0.1\Delta+i10^{-3}$.}
\label{Fig1} 
\end{figure}

In this work, we study the response of $d$-wave AMs and $p$-wave UMs to a time-periodic light drive in the normal and superconducting states. In general, we demonstrate that the interaction between light and UMs is intrinsically unusual due to the nonrelativistic SOC, giving rise to Floquet-engineered spin-triplet states that are absent in equilibrium. In particular, this unique coupling mechanism enables high-frequency linearly polarized light to induce spin-triplet densities and odd-frequency spin-triplet Cooper pairs with $d$- and $s$-wave parities in $d$-wave AMs. 
These light-induced spin-triplet states develop sharp spectroscopic spin-dependent signatures that offer a way to probe the strength and symmetry of the underlying $d$-wave altermagnetism and emergent pairing. We further show that at low drive frequencies, photon emission and absorption processes couple different Floquet bands, thereby generating Floquet spin-triplet states in both $d$-wave and $p$-wave UMs. Our results establish the realization and control of dynamical spin-triplet states in light-driven UMs.

\textit{Models for UMs}.---We are interested in the effect of light on UMs in the normal and superconducting states [Fig.\,\ref{Fig1}(a,c)]. 
We first address  normal state $d$- and $p$-wave UMs under a light drive, modelled by \cite{FukayaCayaoReviewUMs}
\begin{equation}
\label{UMEq1}
H^{j}(\bm{k})=\xi_{\bm{k}}+M^{j}_{\bm{k}} \sigma_{z}\,, 
\end{equation}
where $\xi_{\bm{k}}=\hbar^{2}k^{2}/(2m)-\mu$, $\bm{k}=(k_{x},k_{y})$, $k=|\bm{k}|$, $\mu$ is the chemical potential, and $\sigma_{z}$ the spin Pauli matrix. Here, $M^{j}_{\bm{k}}$ describes the UM with parity $j=(p,d)$ given by \cite{FukayaCayaoReviewUMs}
\begin{equation}
\label{UMs}
\begin{split}
M^{d}_{\bm{k}}&=\bar{M}^{d} \left[2k_{x}k_{y}{\rm sin}(2\theta_{d})+\left(k_{x}^{2}-k_{y}^{2}\right){\rm cos}(2\theta_{d})\right]\,,\\
M^{p}_{\bm{k}}&=\bar{M}^{p}\left[k_{x}{\rm cos}(\theta_{p})+k_{y}{\rm sin}(\theta_{p})\right]\,,
\end{split}
\end{equation}
with $\bar{M}^{d}=M/k_{\rm F}^{2}$ and $\bar{M}^{p}=M/k_{\rm F}$. 
Here, $M$ is the strength of unconventional magnetic order, $k_{\rm F}$ is the Fermi wavevector, and $\theta_{j}$ is the angle between the $x$-axis and the lobe of $M^{j}_{\bm{k}}$ \cite{Bai_review24,FukayaCayaoReviewUMs,brekke24,hellenes2024P,sukhachov2025,PhysRevB.110.144412,Brekke23}. 
From $M^{d}_{\bm{k}}$ at $\theta_{d}=0$, the UM is the $d_{x^{2}-y^{2}}$-wave AM, while at $\theta_{d}=\pi/4$ we have a $d_{xy}$-wave AM. Also, using $M^{p}_{\bm{k}}$ at $\theta_{p}=0$, the UM is a $p_{x}$-wave magnet, while a $p_{y}$-wave magnet is achieved at $\theta_{p}=\pi/2$. 
The main property of UMs captured by Eqs.\,(\ref{UMs}) stems from     $M^{j}_{\bm{k}}$, which leads to an anisotropic spin split Fermi surfaces \cite{Bai_review24,FukayaCayaoReviewUMs}.  
These features are revealed in the spin-density $\rho^{j}_{z}(\bm{k},\omega)$, shown in Fig.\,\ref{Fig1}(b) (see End Matter S1.1).

\textit{Non-trivial light-matter coupling and Floquet engineering in UMs}.---We now inspect how light   affects the main properties of UMs described by Eq.\,(\ref{UMEq1}). We consider a spatially uniform time-periodic light drive  $\bm{E}(t)$ with frequency $\Omega=2\pi/T$ and period $T$.  This time-dependent field is taken into account by a minimal coupling substitution $\bm{k}\rightarrow\bm{k}+e\bm{A}(t)$, where $\bm{A}(t)$ is the vector potential $\bm{E}(t)=-\partial_{t}\bm{A}(t)$, with $e>0$ being the fundamental charge.
We focus on circularly (C) and linearly (L) polarized  drives described, respectively, by $A_{\rm C}(t)=A_{0}({\rm cos}\Omega t,\beta {\rm sin}\Omega t)$ and  $A_{\rm L}(t)=A_{0}{\rm cos}\Omega t({\rm cos}\phi,{\rm sin}\phi)$, with $A_{0}=E_{0}/\Omega$, $\beta=\pm$ defines the left- and right-handed circularly polarized light (CPL), and $\phi$ captures the spatial polarization direction. 
Thus, the   light drive renders the UM Hamiltonian time-periodic, taking the form $\hat{H}^{j}_{\bm{k}}(t) = H^{j}(\bm{k}) + V^{j}_{\bm{k}}(t)$. 
Here, $H^{j}(\bm{k})$ is given by  Eq.~(\ref{UMEq1}) with a renormalized chemical potential $\mu \rightarrow \mu - e^{2}|\bm{A}|^{2}/(2m)$, while $V^{j}_{\bm{k}}(t)$ captures the unique light-matter coupling in UMs (see End Matter S2).

To assess the effect of the light drives on UMs, and investigate their spin-dependent properties, we exploit the time-periodicity of the drive $\bm{A}(t+T)=\bm{A}(t)$; hence  $\hat{H}^{j}_{\bm{k}}(t+T)=\hat{H}^{j}_{\bm{k}}(t)$.  We then employ the Floquet theorem \cite{ASENS_1883_2_12__47_0,PhysRev.138.B979,PhysRevA.7.2203} and decompose the time-dependent Hamiltonian $\hat{H}^{j}_{\bm{k}}(t)$ and solutions of the Sch\"{o}dinger equation in series of the fundamental frequency of the drive $\Omega$ \cite{rudner2020band}. 
Thus, the time-dependent eigenvalue problem becomes,
\begin{equation}
\label{HFloquetEQ}
\sum_{m'} 
H_{n,m'}^{j}(\bm{k})\Phi_{n}^{m'}(\bm{k})=\varepsilon_{n}\Phi_{n}^{m}(\bm{k})\,,
\end{equation}    
where $n,m\in\mathbb{Z}$ are Floquet indices, $H^{j}_{n,m'}=(H^{j}_{0}-m'\hbar\Omega)\delta_{n,m'}+\sum_{n'}H^{j}_{n'}\delta_{n+n',m'}$, with $n'=\pm1,\pm2$ and the Hamiltonian harmonics given by $H^{j}_{n}(\bm{k})=(1/T)\int{dt}\hat{H}_{\bm{k}}^{j}(t){\rm exp}(in\Omega t)$. 
Note that $H^{j}_{n}(\bm{k})$ depends on the applied drive and on the type of UM, and read \cite{sm}
\begin{equation}
\label{Hncompo}
\begin{split}
H_{0}^{j}\left( \bm{k}\right)  &=H^{j}( \bm{k}) +\frac{%
(eA_{0}) ^{2}\bar{M}^{d}}{2}\cos(2\Theta_{d}) \sigma _{z}\delta _{j,d}\delta _{l,L} \,,\\
H_{+1}^{j}(\bm{k})  &=\frac{e\hbar ^{2}A_{0}k}{2m}\Big[
e^{i\beta \theta _{k}}\delta _{l,C}+\cos(\Theta_{k})
\delta _{l,L}\Big] \\
&+eA_{0}k\bar{M}^{d}\Big[
 e^{i\Theta_{dk}^{\beta}}\delta _{l,C}+\cos(\Theta_{kd}) \delta _{l,L}\Big] \sigma _{z}\delta _{j,d} \\
&+eA_{0}\bar{M}^{p}\Big[ e^{i\beta \theta _{p}}\delta _{l,C}+
\cos(\Theta _{p}) \delta _{l,L}\Big] \sigma _{z}\delta _{j,p} \\
H_{+2}^{j}\left( \bm{k}\right)  &=\frac{(\hbar eA_{0}) ^{2}%
}{8m}\delta _{l,L}+\frac{(eA_{0}) ^{2}\bar{M}^{d}}{4}%
\Big[ 2e^{2i\beta \theta _{d}}\delta _{l,C}\\
&+\cos \left( 2\Theta _{d}\right) \delta _{l,L}\Big] \delta _{j,d}\sigma _{z}\,,
\end{split}
\end{equation}
where $j=p,d$ indicates the type of UM and $l={\rm C(L)}$ labels C (L) polarized light.
Also, $\Theta_{i}=\theta _{i}-\phi$, with $i=d,p,k$, $\Theta_{kd}=\theta_{k}-2\theta _{d}+\phi$, $\Theta_{dk}^{\beta}=2\theta_{d}-\beta \theta _{k}$,   $\theta_k = \arctan(k_y/k_x)$. 
Here, $H_{-n}^{j}(\bm{k}) = [H_{+n}^{j}(\bm{k})]^\dagger$. 
We have thus reduced the time-dependent problem into a lattice in frequency space with diagonal elements $H_{0}^{j}-m\hbar\Omega$, where the nearest-neighbor (next-nearest-neighbor) Floquet bands are coupled by $H_{\pm1}^{j}$ ($H_{\pm2}^{j}$) by the emission/absorption of one (two) photons. Details on the derivation of Eqs.\,(\ref{Hncompo}) are presented in the Supplementary Material \cite{sm}.  A few remarks are important on Eq.\,(\ref{HFloquetEQ}) and Eq.\,(\ref{Hncompo}).  First, the terms with $|n|\geq3$  vanish.  Second, for a $d$-wave AM, there appears in the diagonal term $H_{0}^{d}$ a Zeeman-like field generated by a LPL  with a dependence on the relative direction between the AM lobe $\theta_{d}$ and the spatial light polarization $\phi$. 
Third, the one- and two-photon couplings between Floquet bands acquire nontrivial contribution due to the interplay between unconventional magnetism and light, see second and third elements (second element) in the expressions for $H^{j}_{\pm1}$ ($H^{j}_{\pm2}$) in Eqs.\,(\ref{Hncompo}). 
Interestingly, this nontrivial effect manifests in one- and two-photon coupling processes occurring in $d$-wave AMs under either CPL or LPL.  In contrast, the nontrivial effect in $p$-wave UMs only produces one-photon coupling processes.  These properties are reminiscent of the nontrivial light-matter coupling revealed by Eq.\,(\ref{VtUMs}) and further unveil the richness of driven UMs. 
 
\begin{figure}[!t]
\centering
\includegraphics[width=0.99\linewidth]{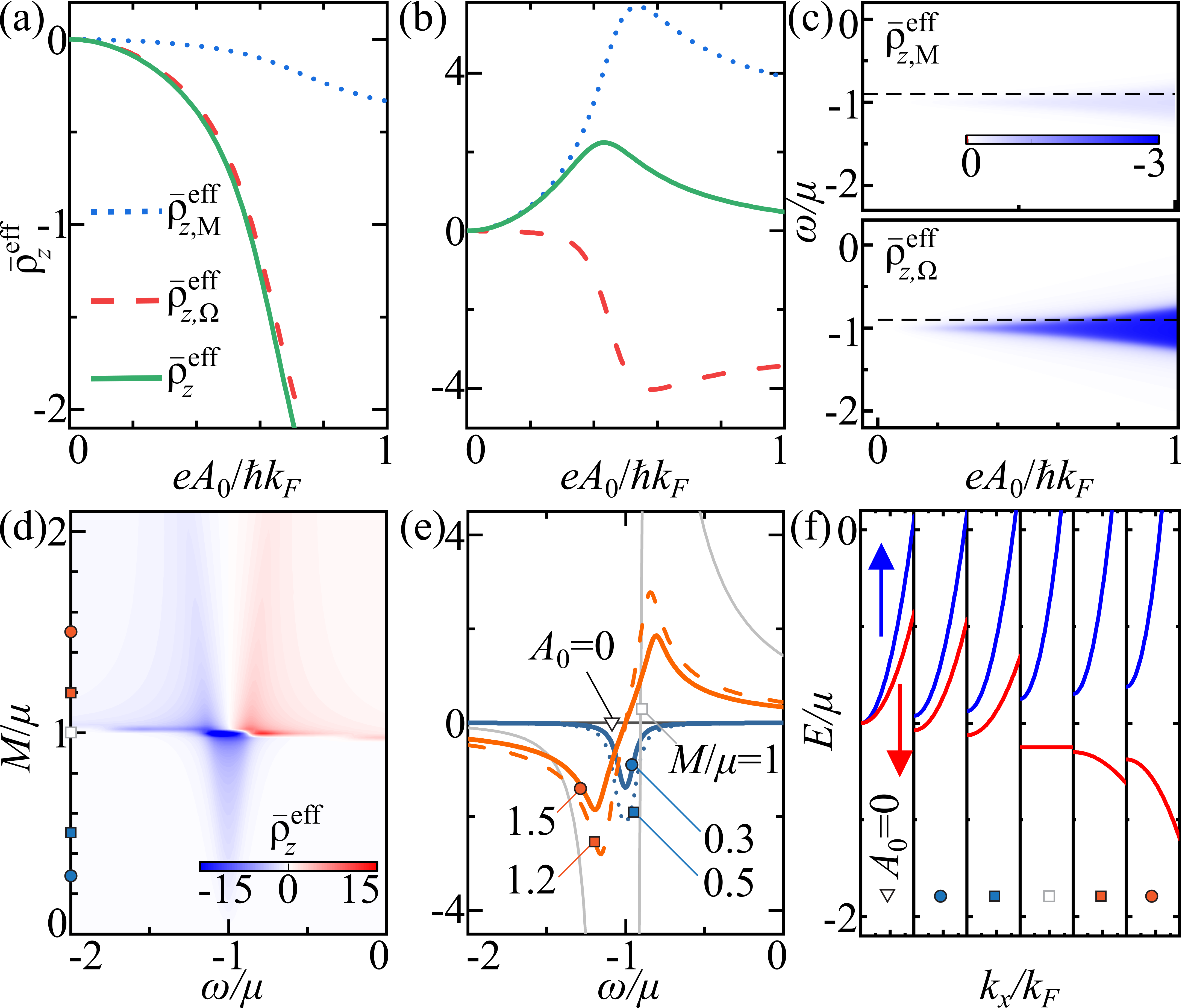}
\caption{(a,b) Integrated spin density $\bar{\rho}_z^\text{{eff}}$ (green curve) as a function of $A_{0}$ for  $M/\mu=0.5$ (a) and  $M/\mu=1.2$ (b), both at $\omega/\mu=-0.9$ for a $d_{x^{2}-y^{2}}$-wave AM under LPL. 
The dotted blue and dashed red curves show $\bar{\rho}_{z,M}^\text{{eff}}$ and   $\bar{\rho}_{z,\Omega}^\text{{eff}}$, respectively. 
(c) $\bar{\rho}_{z,M}^\text{{eff}}(\omega)$ and   $\bar{\rho}_{z,\Omega}^\text{{eff}}(\omega)$ as a function of $\omega$ and $A_{0}$ at $M/\mu=0.5$, where black dashed lines indicate $\omega/\mu=-0.9$. 
(d) Spin density $\bar{\rho}_z^\text{{eff}}$ as a function of $M$ and $\omega$ at $eA_{0}/(\hbar k_{\rm F})=0.5$, while (e) shows line cuts at distinct values of $M$ marked in (d) and depicted by blue, gray, and orange curves for $M/\mu<1$, $M/\mu=-1$, and $M/\mu>1$, respectively.  
(f) Energy versus $k_{x}$ at $k_{y}=0$ for the distinct values corresponding to the distinct values of $M$ in (e). 
The spin-up (spin-down) bands are denoted in blue (red). Parameters: $\mu=1$, $\hbar^2/(2m)=1$ and $k_{\rm F}= 1$, $\theta_d=0$ and $\phi=0$, $eA_0/(\hbar k_{\rm F})=0.5$.  
}
\label{figure2}
\end{figure}

To further inspect the effect of  light on UMs, we now assess the induced spin density in the high-frequency regime because it permits us to gain fundamental understanding analytically \cite{PhysRevB.84.235108,PhysRevB.79.081406,Fu2019Josephson,Fu2022Electrically,Lee2025Floquet,Fu2022The,Li2019Photon}.  
This situation is described by the effective Hamiltonian $H_{\rm eff}\approx H^{j}_{0}+\sum_{n}[H_{-n},H_{n}]/(n\Omega)$, where $H^{j}_{0}$ and $H_{n(-n)}$ are given by Eqs.\,(\ref{Hncompo}). 
Since $[H_{-n},H_{n}]=0$ for all types of considered UMs under LPL and CPL, light-induced effects appear from $H^{j}_{0}$ in the second term of Eq.\,(\ref{Hncompo}), which is nonzero only for $d$-wave AMs under LPL. Hence, under high-frequency LPL, we obtain an effective spin triplet density in $d$-wave AMs given by (see End Matter S2.1)
\begin{equation}
\label{rhoEffective}
\rho_{z}^{\rm eff}(\bm{k},\omega)=\rho_{z,{\rm M}}^{\rm eff}(\bm{k},\omega)+\rho_{z,\Omega}^{\rm eff}(\bm{k},\omega)\,,
\end{equation}
where $\rho_{z,{\rm M}}^{\rm eff}(\bm{k},\omega)=(2/\pi){\rm Im}[{M_{\bm{k}}^{d}}/{D_{\Omega}}]$ and $\rho_{z,\Omega}^{\rm eff}(\bm{k},\omega)=(2/\pi){\rm Im}[{M_{\Omega}^{d}}/{D_{\Omega}}]$; here $M_{\Omega}^{d}=\bar{M}^{d}e^{2}A^{2}_{0}{\rm cos}(2\Theta_{d})/2$, and $D_{\Omega}=(M_{\bm{k}}^{d}+M_{\Omega}^{d})^{2}-(\omega+i\eta-\xi_{\bm{k}})^{2}$. 
Thus, Eq.\,(\ref{rhoEffective}) implies that LPL induces a finite spin density ($\rho_{z,\Omega}^{\rm eff}$) in $d$-wave AMs, which does not vanish at the momenta where the part originated from the static density  ($\rho_{z,{\rm M}}^{\rm eff}$) does.  
To elucidate the effect of the LPL, we inspect the spin density integrated over momenta $\bar{\rho}_{z}^{\rm eff}(\omega)=[1/(2\pi)^{2}]\int d\bm{k}\rho_{z}^{\rm eff}(\bm{k},\omega)\equiv\bar{\rho}_{z,{\rm M}}^{\rm eff}(\omega)+\bar{\rho}_{z,\Omega}^{\rm eff}$. 
In the static regime ($A_{0}=0$), both components of  $\bar{\rho}_{z}^{\rm eff}(\omega)$ vanish regardless of  $M$, see   Fig.\,\ref{figure2} for a driven $d_{x^{2}-y^{2}}$-wave AM; here, the energy bands are   split at $\bm{k}\neq0$ [Fig.\,\ref{figure2}(f)].
Surprisingly, in the driven phase, both  $\bar{\rho}_{z,{\rm M}}^{\rm eff}$ and $\bar{\rho}_{z,\Omega}^{\rm eff}$ are finite, leading to a finite light-induced spin-triplet density $\bar{\rho}_{z}^{\rm eff}$ in $d$-wave AMs [Fig.\,\ref{figure2}].  Depending on   $M$ and   $\omega$, either $\bar{\rho}_{z,{\rm M}}^{\rm eff}$ or $\bar{\rho}_{z,\Omega}^{\rm eff}$ can have a dominant contribution. For $M/\mu<1$ and $\omega/\mu>-1$, the   spin density $\bar{\rho}_{z}^{\rm eff}$ is negative and its value entirely comes from $\bar{\rho}_{z,\Omega}^{\rm eff}$ [Fig.\,\ref{figure2}(a-c)].  Here, the   bands are split for all $\bm{k}$ due to the applied drive and the spin density develops a dip centred at $\omega=-\mu$, see dotted and solid blue curves in Fig.\,\ref{figure2}(d,e). 

At $M/\mu=1$, one of the spin bands becomes dispersionless, resembling a flat band [Fig.\,\ref{figure2}(f)], and leads to a large spin density due to high density of states [gray curve in Fig.\,\ref{figure2}(e)]. 
The spin density at $M/\mu=1$ acquires large values just above and below $\omega/\mu=-1$, forming a dip-peak structure that becomes more evident when $M$ takes larger values. In fact, for  $M/\mu>1$, the  two spin split bands exhibit opposite effective masses [Fig.\,\ref{figure2}(f).], which then makes $\bar{\rho}_{z,{\rm M}}^{\rm eff}$ to acquire positive values that are   comparable    to those of the negative $\bar{\rho}_{z,\Omega}^{\rm eff}$.  As a result, a finite spin density  $\bar{\rho}_{z}^{\rm eff}$ appears having a clear dip-peak structure, with a positive value (peak) when $\omega/\mu<-1$ and and negative value (dip) for  $\omega/\mu>-1$, see orange curves in Fig.\,\ref{figure2}(e); this leaves $\bar{\rho}_{z}^{\rm eff}=0$ at $\omega=-\mu$. For $M/\mu>1$, the dip is centred at $\omega=-\mu-M^d_\Omega$, while the peak is centred at $\omega=-\mu+M^d_\Omega$, where   $M_{\Omega}^{d}=\bar{M}^{d}e^{2}A^{2}_{0}/2$.  An important feature of the dip-peak structure seen for $M/\mu>1$ is that the separation between peak and dip   provides a direct way to extract the strength of the altermagnet field:   The energy difference $\delta\omega$ between the peak and dip can be measured by microwaves in electron spin resonance spectroscopy  \cite{wang2023single}, where  the altermagnet strength is identified as $M=\delta\omega/[(eA_0/\hbar k_F)^2 \cos(2\Theta_d)]$ at $\phi=0$. At  $\phi\neq0$,  $\bar{\rho}_{z}^{\text{eff}}(\bm{k})\sim\cos(2\theta_d - 2\phi)$, which implies that it is also possible to identify the altermagnet spin splitting and the type of AM,   e. g., via optical response. We have also verified that the  analysis presented above   holds for $d_{xy}$-wave AMs ($\theta_{d}=\pi/4$) under linearly polarized light, provided $\phi\neq n\pi/2$ \footnote{We note that  low frequency CPL and LPL drives can also induce a spin-triplet density in both $d$- and $p$-wave UMs; here, the Floquet bands play a role.} and also for driven higher angular momentum UMs \cite{sm}.

\textit{Floquet engineering spin-triplet Cooper pairs}.---Having shown the effect of light on UMs, here we assess its impact on UMs with conventional spin-singlet $s$-wave superconductivity [Fig.\,\ref{Fig1}(c)]. 
In particular, we address the generation and control of emergent Cooper pairs by light drives within the Floquet framework introduced in the previous section. 
Under a time-periodic drive, UMs with conventional superconductivity are modelled in Nambu space by \cite{sm} $\hat{H}_{\rm sc}^{j}(\bm{k},t)=H_{\rm sc}^{j}(\bm{k})+U^{j}_{\bm{k}}(t)$,  where $H_{\rm sc}^{j}(\bm{k})$ describes the static UM with spin-singlet $s$-wave superconductivity and $U^{j}_{\bm{k}}$ accounts for the time-dependent drive. 
The basis of $H_{\rm sc}^{j}(\bm{k})$ is  $\psi_{\bm{k}}=(c_{\bm{k}\uparrow},c_{\bm{k}\downarrow},c_{-\bm{k}\uparrow}^{\dagger},c_{-\bm{k}\downarrow}^{\dagger})^{\rm T}$, with T the transpose operation. 
Here, $H_{\rm sc}^{j}(\bm{k})=\xi_{\bm{k}}\tau_{z}+M^{d}_{\bm{k}}\sigma_{z}\tau_{z}+M^{p}_{\bm{k}}\sigma_{z}\tau_{0}-\Delta\sigma_{y}\tau_{y}$, with $\tau_{i}$ being the $i$-th Pauli matrix in Nambu space, while  $U^{j}_{\bm{k}}(t)$ is the Nambu light-matter coupling of $V^{j}_{\bm{k}}(t)$ in the previous section. We are interested in the emergent Cooper pairs in $\hat{H}_{\rm sc}^{j}(t)$, characterized by the anomalous Green's function  $\mathcal{F}_{\sigma_{1}\sigma_{2}}(\bm{k}_{1},\bm{k}_{2};t_{1},t_{2})=-i\langle\mathcal{T} c_{\bm{k}_{1}\sigma_{1}}(t_{1}) c_{\bm{k}_{2}\sigma_{2}}(t_{2})\rangle$; here $\mathcal{T}$ is the time-ordering operator \cite{zagoskin,mahan2013many}. To demonstrate the light-induced Cooper pairs, we focus on $d$-wave AMs under high-frequency linearly polarized light. 
Using the Floquet description of the previous section, the effective Hamiltonian is \cite{sm}
\begin{equation}
\label{HeffSCLPL}
H_{\rm eff}^{d}(\bm{k})=H_{\rm sc}^{d}(\bm{k})+M^{d}_{\Omega}\sigma_{z}\tau_{z}\,,
\end{equation}
where $H_{\rm sc}^{d}(\bm{k})$ is the static Hamiltonian and   $M^{d}_{\Omega}$ results from the light drive.  Then, we obtain the spin-singlet ($s$) and mixed spin-triplet ($t$) pair amplitudes from the anomalous Green's function associated with Eq.\,(\ref{HeffSCLPL}) \cite{sm},
\begin{equation}
\label{EqFtFs}
\begin{split}
F_{s}^{\rm eff}(\bm{k},z)&=\frac{\Delta \mathcal{P}(z,\Omega,\bm{k})}{[\mathcal{P}(z,\Omega,\bm{k})]^{2}+4z^{2}(M^{d}_{\bm{k}}+M^{d}_{\Omega})^{2}}\,,\\
F_{t}^{\rm eff}(\bm{k},z)&=\frac{2z\Delta (M^{d}_{\bm{k}}+M^{d}_{\Omega})}{[\mathcal{P}(z,\Omega,\bm{k})]^{2}+4z^{2}(M^{d}_{\bm{k}}+M^{d}_{\Omega})^{2}}\,.
\end{split}
\end{equation}
with $\mathcal{P}(z,\Omega,\bm{k})=[z^{2}-(M^{d}_{\bm{k}}+M^{d}_{\Omega})^{2}-\xi_{\bm{k}}^{2}-\Delta^{2}]$ is an even function of the complex frequency $z$ (End Matter S2.2).

The spin-singlet pairing $F_{s}^{\rm eff}$ naturally results from the parent conventional superconductor and has an even-frequency and even-parity symmetry, which fulfills the antisymmetry condition of Fermi-Dirac statistics \cite{RevModPhys.77.1321,tanaka2012symmetry,cayao2019odd,RevModPhys.91.045005,triola2020role,tanaka2024theory,Tanaka2007Theory,Berezinskii1974New}. 
The mixed spin-triplet component $F_{t}^{\rm eff}$ contains two contributions denoted by $F_{t,\rm M}^{\rm eff}$ and $F_{t,\Omega}^{\rm eff}$, respectively.
The first term results from the interplay between altermagnetism and the parent superconductor \cite{maeda2025pair}, while the second term emerges entirely due to the combined effect of light, altermagnetism, and superconductivity. 
The spin-triplet pairing $F_{t}^{\rm eff}$ belongs to the odd-frequency mixed even-parity class \cite{maeda2025pair,PhysRevB.111.064502,FukayaCayaoReviewUMs}, with its first and second parts having $d$-wave and $s$-wave parities, respectively \footnote{We have verified that light-driven unconventional magnets with unconventional superconductivity allow to engineer even more exotic Cooper pairs \cite{phjc}.}.  
The driven $d$-wave AM hence hosts light-induced spin-triplet $s$-wave components in addition to the $d$-wave pairs present in the static regime [Fig.\,\ref{Fig1}(d)], which is reminiscent of high-$T_{c}$ superconductors \cite{PhysRevB.48.437,PhysRevB.52.16208} and can be detected via quasiparticle interference \cite{hanaguri2007quasiparticle,sharma2020momentum,allan2012anisotropic}.

\begin{figure}[!t]
\centering
\includegraphics[width=1\linewidth]{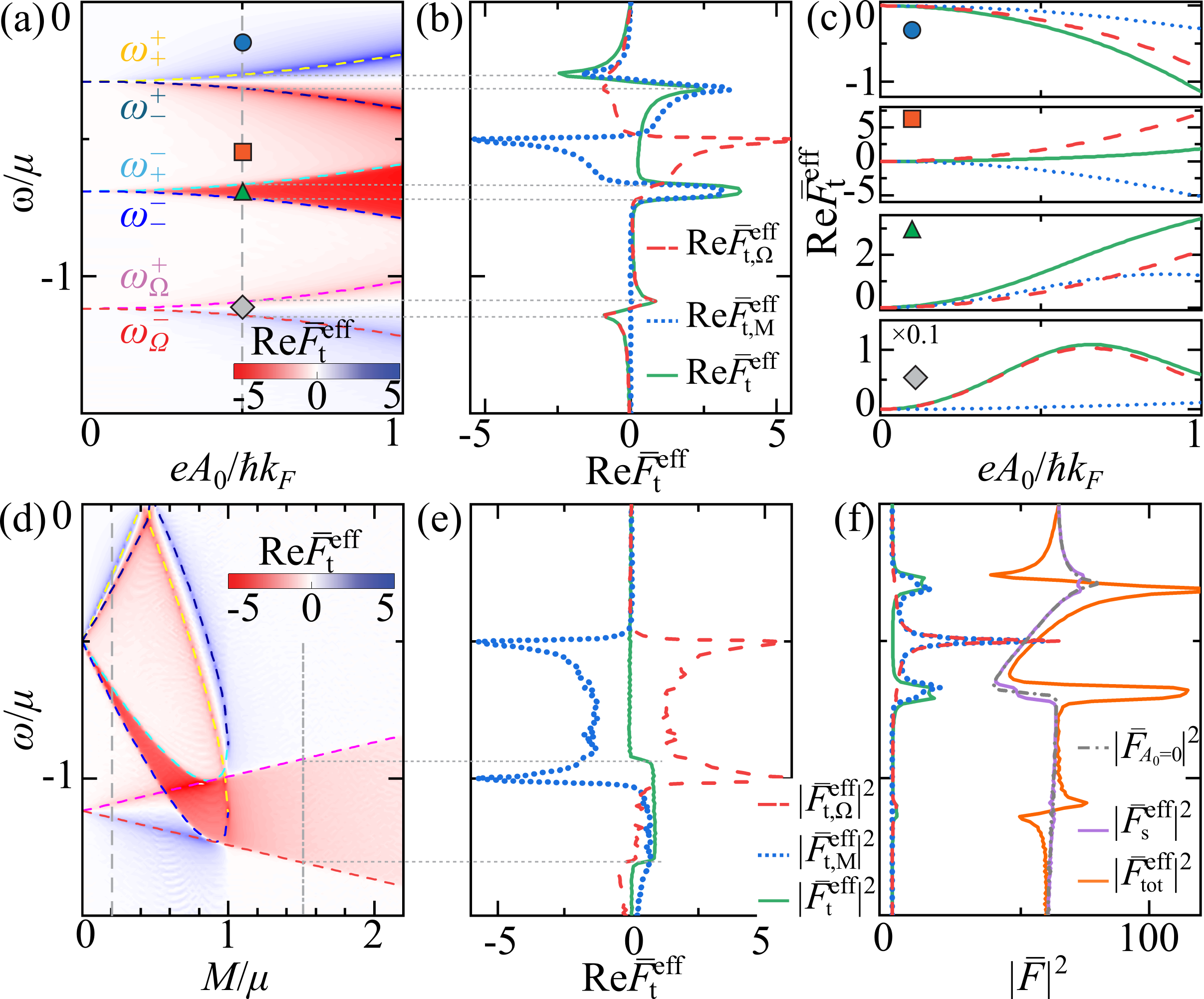}
\caption{
(a) Real part of the integrated spin-triplet pair amplitude ${\rm Re}\bar{F}^{\rm eff}_{t}$ as a function of $\omega$ and $A_{0}$ at $M/\mu=0.2$ for a $d_{x^{2}-y^{2}}$-wave AM with superconductivity under LPL. $\omega_{\pm}^{\pm}$ and $\omega_{\Omega}^{\pm}$ depicted by dashed curves are the negative band edges of the spectrum of Eq.\,(\ref{HeffSCLPL}). 
(b) Line cuts of (a) as a function of $\omega$ at $eA_{0}/(\hbar k_{\rm F})=0.5$, where dashed (dotted) curve shows $\bar{F}^{\rm eff}_{t,\rm M(\Omega)}$. 
(c) Real part of $\bar{F}^{\rm eff}_{t}$, $\bar{F}^{\rm eff}_{t,\rm M}$  and $\bar{F}^{\rm eff}_{t, \Omega}$ of (b) as a function of $A_{0}$ at $\omega$ indicated by  markers in (a). 
(d) ${\rm Re}\bar{F}^{\rm eff}_{t}$ as a function of $\omega$ and $M$  at $eA_{0}/(\hbar k_{\rm F})=0.5$. 
(e) Line cuts of (d) at $M/\mu=1.5$. 
(f) $|\bar{F}^{\rm eff}_{t}|^{2}$ and its singlet and triplet contributions as a function of $\omega$ at $M/\mu=0.2$ and $eA_{0}/(\hbar k_{\rm F})=0.5$. 
The gray curve shows $|\bar{F}^{\rm eff}_{t}|^{2}$ at $A_{0}=0$.  Parameters: $\Delta=0.5$, the rest same as in Fig.\,\ref{figure2}.
}
\label{Figure3}
\end{figure}
To visualize the light-induced spin-triplet pairing, Fig.\,\ref{Figure3} exhibits the integrated pair amplitude $\bar{F}_{t}^{\rm eff}(z)=(1/N)\int d\bm{k}F_{t}^{\rm eff}(\bm{k},z)=\bar{F}_{t,\rm M}^{\rm eff}+\bar{F}_{t,\Omega}^{\rm eff}$ as well as its components $\bar{F}_{t,\rm M (\Omega)}^{\rm eff}$ for a $d_{x^{2}-y^{2}}$-AM. 
Fig.\,\ref{Figure3}(a) shows  ${\rm Re}\bar{F}_{t}^{\rm eff}$ as a function of $\omega$ and $A_{0}$ at $M/\mu=0.2$, which demonstrates that the drive induces a finite spin-triplet pairing. It also
 exhibits a peak-dip structure at the edges of the negative energy bands here denoted by $\omega_{\pm}^{\pm}=\pm M\pm M_{\Omega}^{d}-\Delta\sqrt{1-(M/\mu)^{2}}$ and  $\omega_{\Omega}^{\pm}=\pm M_{\Omega}^{d}-\sqrt{\mu^{2}-\Delta^{2}}$ \footnote{We note that the same conclusions can be obtained from the imaginary part ${\rm Im}\bar{F}_{t}^{\rm eff}$ demonstrating peak-dip structures and competition between ${\rm Im}\bar{F}_{t,\rm M}^{\rm eff}$ and ${\rm Im}\bar{F}_{t,\Omega}^{\rm eff}$.}.
In the static regime, the energy spectrum of Eq.\,(\ref{HeffSCLPL}) becomes gapless when $M>M_{\rm c}\equiv\Delta/\sqrt{1+(\Delta/\mu)}$, where $\omega_{+}^{+}=0$.  
In the driven phase, the gap closing is modified by the light-induced spin splitting $M_{\Omega}^{d}$, opening a gap at $\bm{k}=0$ of size $|\omega_{\Omega}^{+}-\omega_{\Omega}^{-}|$. 
To resolve the peak-dip structure, Fig.\,\ref{Figure3}(b) displays the frequency dependence of ${\rm Re}\bar{F}_{t}^{\rm eff}$ and ${\rm Re}\bar{F}_{t,\rm M(\Omega)}^{\rm eff}$ at fixed $A_{0}\neq0$. 
For $\omega\in[\omega_{+}^{+},0]$, both ${\rm Re}\bar{F}_{t,\rm M}^{\rm eff}$ and ${\rm Re}\bar{F}_{t,\Omega}^{\rm eff}$ exhibit equal sign and enhance the total spin-triplet pairing within $\Delta$ corresponding to Fig.\,\ref{Figure3}(c). 
As $\omega$ enters the interval $[\omega_{-}^{+},\omega_{+}^{+}]$, $\bar{F}^{\rm eff}_{t, \rm M}$ changes sign, inducing a sharp peak-dip structure at $\omega=\omega_{+}^{\pm}$. 
For $\omega\in[\omega_{+}^{-},\omega_{-}^{+}]$, both  components develop  opposite signs, suppressing ${\rm Re}\bar{F}^{\rm eff}_{t}$ [Fig. \ref{Figure3}(b,c)]. 
At $\omega=\Delta$, ${\rm Re}\bar{F}^{\rm eff}_{t, \rm M}$ and ${\rm Re}\bar{F}^{\rm eff}_{t,\Omega}$ have opposite signs, giving rise to a vanishing spin triplet pairing. 
A constructive interference resumes for $\omega\in[\omega_{-}^{-},\omega_{+}^{-} ]$ before the spin-triplet pairing is suppressed again for $\omega < \omega_{-}^{-}$.   Interestingly, another sharp peak-dip structure  appears at $\omega = \omega_{\Omega}^{\pm}$, where Re$\bar{F}^{\rm eff}_{t}$ is dominated by the light-induced term ${\rm Re}\bar{F}^{\rm eff}_{t,\Omega}$ [Fig. \ref{Figure3}(c)]. 
Further insights are obtained from Fig.\,\ref{Figure3}(d), where ${\rm Re}\bar{F}^{\rm eff}_{t}$ as a function of $\omega$ and $M$ reflects a large spin-triplet pairing when $M<\mu$ for a broad range of $\omega$.  For $M>\mu$, however, a finite spin-triplet emerges for $\omega\in[\omega_{\Omega}^{-},\omega_{\Omega}^{+}]$ with a roughly constant (plateau) value [Fig.\,\ref{Figure3}(e)].

The peak-dip structure for $M/\mu < 1$ and the plateau for $M/\mu > 1$ in the spin-triplet pair amplitude $\bar{F}^{\rm eff}_{t}$ can be detected via Andreev conductance $G_{\rm A} \propto |\bar{F}^{\rm eff}_{\rm tot}(\omega)|^2 =| \bar{F}^{\rm eff}_{s}(\omega) + \bar{F}^{\rm eff}_{t}(\omega)|^2$  \cite{kashiwaya2000,PhysRevB.54.7366,PhysRevB.93.201402,PhysRevB.111.024507,PhysRevB.109.205406}, see Fig. \ref{Figure3}(f). 
While $|\bar{F}^{\rm eff}_{\rm tot}(\omega)|$ is dominated by the spin-singlet term, it has two clear peak-dip features due to the spin-triplet part.  The first peak-dip signal occurs at $\omega\sim \omega_{\pm}^{+}$ and is mainly due to $|\bar{F}^{\rm eff}_{t,\rm M}|$, while the second peak-dip feature at $\omega\sim \omega_{\Omega}^{\pm}$ is dominated by $|\bar{F}^{\rm eff}_{t,\Omega}|$. From the peak-dip energy difference, $\delta \omega = | \omega_{+}^{+} - \omega_{-}^{+} | = | \omega_{\Omega}^{+} - \omega_{\Omega}^{-} |$, the altermagnetic strength is extracted as $M = \delta \omega / [ ( eA_0 / k_F )^2 \cos(2\Theta_d) ]$, in line with the normal state spin density. These phenomena remain in higher angular momentum UMs under LPL, but with broader parities of Cooper pairs \cite{sm}. Moreover, while spin-triplet odd-frequency pairs appear only in high-$\Omega$ driven $d$-wave UMs, we have confirmed that both odd- and even-frequency spin-triplet pairs emerge in all driven UMs at low $\Omega$, see End Matter S2.3.

In conclusion, the intrinsic nonrelativistic spin splitting of unconventional magnets enables nontrivial light-matter interactions and Floquet engineering of spin-triplet states. 
High-frequency linearly polarized light induces spin-triplet spin densities in $d$-wave altermagnets and generates tunable spin-triplet odd-frequency Cooper pairs with $d$- and $s$-wave symmetries, which provide spectroscopic signatures sensitive to the symmetry and strength of   altermagnetism. These emergent effects remain in higher-order unconventional magnets, broadening the parities of Cooper pairs \cite{sm}.
Low-frequency driving enables mixed Floquet sideband opening additional spin-triplet channels absent in static or high-frequency regimes. Given the recent advances,  in Ref.\,\cite{sm}, we estimate that the found light-induced spin-triplet states are very likely to appear under current Floquet experimental conditions \cite{Matsuda2020Room,Zhang2023Light,Shan2021Giant,mciver2020light,Luo2021LightInduced,aeschlimann2021survival,
wang2013observation,
Merboldt2025Observation}, and expected to be robust against unavoidable screening effects \cite{Merboldt2025Observation}. Our results establish unconventional magnets under periodic driving as a promising platform for engineering nonequilibrium spin-triplet superconductivity.

\begin{acknowledgments}
\emph{Note added.} 
During the final stages of preparing this manuscript, we became aware of Refs.\, \cite{Ghorashi2025Dynamical,Yarmohammadi2025Anisotropic,yokoyama2025FloquetSC}, which, although they use similar techniques, consider other effects and address distinct properties.

 P.-H. Fu acknowledges support from W. Xu. 
    We thank M. Hirschberger for drawing our attention to thin flakes of an unconventional magnet realized in NiI$_2$ \cite{Song2025Electrical} and Gd$_3$Ru$_4$Al$_{12}$ \cite{Yamada2025Metallic}. S. M. and J. C. acknowledge financial support from the Carl Trygger’s Foundation (Grant No. 22: 2093), the Sweden-Japan Foundation (Grant No. BA24-0003), the G\"{o}ran Gustafsson Foundation (Grant No. 2216), and the Swedish Research Council (Vetenskapsr\aa det Grant No.~2021-04121). 
    J.-F. Liu acknowledges financial support from the National Natural Science Foundation of China (Grant No. 12174077). 
    Y. T. acknowledges financial support from JSPS with Grants-in-Aid for Scientific research (KAKENHI Grants No. 23K17668  and 24K00583).
\end{acknowledgments}

\onecolumngrid
\newpage
\begin{center}
\textbf{\textsc{END MATTER}}
\end{center}
\vspace{1em}
\twocolumngrid

\textit{\textbf{S1.} Spin-triplet density and spin-triplet Cooper pairs in the static regime}.---
We present the spin density and spin-triplet pair amplitudes corresponding to the static states shown in Figs. \ref{Fig1}(b) and \ref{Fig1}(c). 

\textit{\textbf{S1.1. }Spin-triplet density}.---
The spin density can be obtained from the retarded Green’s function, $G^{r}(\bm{k},\omega) = [\omega + i\eta - H^{j}(\bm{k})]^{-1}$, as $\rho_{z}(\bm{k},\omega) = (-1/\pi)\,{\rm Im\,Tr}\big[G^{r}(\bm{k},\omega)\sigma_{z}\big]$~\cite{zagoskin,mahan2013many}. This yields
\begin{equation}
\label{rhz_static}
\rho^{j}_{z}(\bm{k},\omega) = \frac{2}{\pi}\,{\rm Im} \left[\frac{M^{j}_{\bm{k}}}{\big(M^{j}_{\bm{k}}\big)^{2} - (\omega + i\eta - \xi_{\bm{k}})^{2}}\right]\,.
\end{equation}
The spin density along the $z$-axis is thus governed by $M^{j}_{\bm{k}}$, allowing direct access to the $d$-wave and $p$-wave parities of UMs [Fig.~\ref{Fig1}(b)].

\textit{\textbf{S1.2. }Spin triplet Cooper pairs.}---
To analyze the Cooper pair amplitude [Fig.~\ref{Fig1}(d)] in UMs proximitized by conventional spin-singlet $s$-wave superconductors, we employ the Bogoliubov–de Gennes (BdG) formalism in the Nambu basis $\psi_{\bm{k}} = (c_{\bm{k}\uparrow}, c_{\bm{k}\downarrow}, c_{-\bm{k}\uparrow}^{\dagger}, c_{-\bm{k}\downarrow}^{\dagger})^{\rm T}$. 
The resulting Hamiltonian reads
\begin{equation}
H^j_{\rm sc} = \xi_{\bm{k}} \tau_z \sigma_z - \Delta \tau_y \sigma_y 
+ M^{j}_{\bm{k}} \left( \delta_{j,d} \tau_z \sigma_z + \delta_{j,p} \tau_0 \sigma_z \right)\,,
\label{eq_hqbdg}
\end{equation}
where the last term reflects the momentum parity of the UM: even ($d$-wave) or odd ($p$-wave). 
This Hamiltonian structure encodes the influence of the UM's momentum parity on the induced pairing symmetry. 
The corresponding BdG spectrum is given by
\begin{equation}
E^{j}_{\nu,\gamma}(\bm{k}) = \nu M^{d}_{\bm{k}} \delta_{j,d} 
+ \gamma \sqrt{\big( \xi_{\bm{k}} + \nu M^{p}_{\bm{k}} \delta_{j,p} \big)^2 + \Delta^2}\,,
\label{eq_Eqbdg}
\end{equation}
where $\nu, \gamma = \pm 1$ label the spin and particle-hole branches, respectively. 
This expression highlights the contrasting roles of $d$-wave and $p$-wave magnetism on the $s$-wave superconducting states.

We focus on the Cooper pair amplitude encoded in the anomalous (off-diagonal) components of the Green’s function of the BdG Hamiltonian. 
The full Green’s function is given by $\mathcal{G}(z,\bm{k}) = [z - H^j_{\rm sc}]^{-1}$, with $z = \omega + i\eta$ a complex frequency and $\eta$ a positive infinitesimal. 
In Nambu space, $\mathcal{G}(z,\bm{k})$ takes the form~\cite{RevModPhys.91.045005,maeda2025pair}
\begin{equation}
\label{eq_Greenfunction}
\mathcal{G}(z,\bm{k}) =
\begin{pmatrix}
G_0(z,\bm{k}) & F(z,\bm{k}) \\
\bar{F}(z,\bm{k}) & \bar{G}_0(z,\bm{k})
\end{pmatrix},
\end{equation}
where the diagonal (off-diagonal) blocks denote the normal (anomalous) Green’s functions
The pairing correlations encoded in $F(z,\bm{k})$ can be written in spin space as
\begin{equation}
F(z,\bm{k}) =
\begin{pmatrix}
F_{\uparrow\uparrow}(z,\bm{k}) & F_{\uparrow\downarrow}(z,\bm{k}) \\
F_{\downarrow\uparrow}(z,\bm{k}) & F_{\downarrow\downarrow}(z,\bm{k})
\end{pmatrix},
\end{equation}
where $F_{\sigma_1\sigma_2}(z,\bm{k})$ represents the pair amplitude between electrons with spins $\sigma_1$ and $\sigma_2$.

The symmetry of $F_{\sigma_1\sigma_2}(z,\bm{k})$ under exchanges of frequency, momentum, and spin determines the nature of the pairing. Fermi-Dirac statistics imposes the antisymmetry condition $F_{\sigma_1\sigma_2}(z,\bm{k}) = -F_{\sigma_2\sigma_1}(-z,-\bm{k})$, meaning the full exchange must yield a minus sign, while individual components may be even or odd. Decomposing the spin structure, the pair amplitude reads \cite{RevModPhys.91.045005,tanaka2012symmetry,cayao2019odd,triola2020role,tanaka2024theory}
\begin{equation}
F(z,\bm{k}) = \big[F_s(z,\bm{k}) \sigma_0 + \bm{F}_t(z,\bm{k}) \cdot \bm{\sigma} \big] i\sigma_y\,,
\end{equation}
where $F_s(z,\bm{k}) =  (F_{\uparrow\downarrow} - F_{\downarrow\uparrow})/2$ is the spin-singlet component, and $\bm{F}_t = (F_t^x, F_t^y, F_t^z)$ denotes the spin-triplet vector with 
$F_t^x = (-F_{\uparrow\uparrow} + F_{\downarrow\downarrow})/2$,
$F_t^y = (F_{\uparrow\uparrow} + F_{\downarrow\downarrow})/(2i)$, and
$F_t^z =  (F_{\uparrow\downarrow} + F_{\downarrow\uparrow})/2$.

For UMs proximitized by spin-singlet $s$-wave superconductors, equal-spin pairing is absent, i.e., $F_{\uparrow\uparrow} = F_{\downarrow\downarrow} = 0$. Thus, only $F_t^z(z,\bm{k})$ remains finite. From Eq.~(\ref{eq_hqbdg}), the spin-triplet and spin-singlet pair amplitudes are
\begin{align}
F_t(z,\bm{k}) &= \frac{4\Delta M^{j}_{\bm{k}} (z \delta_{j,d} - \xi_{\bm{k}} \delta_{j,p})}{\prod_{\nu,\gamma}(z - E^{j}_{\nu,\gamma})}, \label{eq_fst} \\
F_s(z,\bm{k}) &= \frac{2\Delta \left[z^2 - \xi_{\bm{k}}^2 - \Delta^2 + M^{j}_{\bm{k}} (\delta_{j,d} - \delta_{j,p})\right]}{\prod_{\nu,\gamma}(z - E^{j}_{\nu,\gamma})}. \label{eq_fss}
\end{align} 
Although the parent superconductor is an isotropic spin-singlet $s$-wave, the UM induces anisotropic spin-triplet components $F_t(z,\bm{k}) \propto M^j_{\bm{k}}$, exhibiting a momentum parity inherited from UM and vanishing  along spin-degenerate directions [Fig.~\ref{Fig1}(d)] \cite{maeda2025pair}. 

\textit{\textbf{S2. }Light-matter interaction in UMs.}---
In this section, we introduce the time-dependent Hamiltonian $\hat{H}^{j}_{\bm{k}}(t)$ for UMs under periodic driving, which relates to Eq.~(\ref{Hncompo}) via the Fourier transform $H^{j}_{n}(\bm{k}) = (1/T) \int dt\, \hat{H}^{j}_{\bm{k}}(t) e^{in\Omega t}$. 
Applying the minimal coupling substitution $\bm{k} \rightarrow \bm{k} + e\bm{A}(t)$ to Eq.~(\ref{UMEq1}), we obtain the time-periodic Hamiltonian under a driving field $\bm{A}(t) = (A_x(t), A_y(t))$ as
\begin{equation}
\hat{H}^{j}_{\bm{k}}(t)=H^{j}(\bm{k}) +V^{j}_{\bm{k}}(t)\,,
\end{equation}
where $H^{j}(\bm{k})$ is given by Eqs.\,(\ref{UMEq1}) with a renormalized chemical potential $\mu\rightarrow \mu-e^{2}|\bm{A}|^{2}/(2m)$ and  $V^{j}_{\bm{k}}(t)$ reads
\begin{align}
\label{VtUMs}
V^{d}_{\bm{k}}(t) &= \frac{\hbar^{2}}{m} \bm{k} \cdot \bm{A}(t) 
\notag \\
&\quad + 2 \bar{M}^{d} \big[ \bar{\bm{k}} \cdot \bm{A}(t) \cos(2\theta_{d}) 
+ [\bar{\bm{k}} \times \bm{A}(t)]_{z} \sin(2\theta_{d}) \big] \sigma_{z} \notag \\
&\quad + \bar{M}^{d} \big[ A_{x^2 - y^2}(t) \cos(2\theta_{d}) 
+ 2 A_{xy}(t) \sin(2\theta_{d}) \big] \sigma_{z}, \notag \\
V^{p}_{\bm{k}}(t) &= \frac{\hbar^{2}}{m} \bm{k} \cdot \bm{A}(t) 
+ \bar{M}^{p} (\bm{\alpha} \cdot \bm{A}(t)) \sigma_{z}.
\end{align}
where $\bar{\bm{k}}=(k_{x},-k_{y})$, $A_{x^{2}-y^{2}}(t)=A_{x}^{2}(t)-A_{y}^{2}(t)$, $A_{xy}(t)=A_{x}(t)A_{y}(t)$, ${\bm{\alpha}}=({\rm cos}\theta_{p},{\rm sin}\theta_{p})$.  Thus, Eqs.\,(\ref{VtUMs}) uncover the effect of the time-periodic drive $\bm{A}(t)$ on UMs. 
The first term in $V^{d}_{\bm{k}}(t)$ and $V^{p}_{\bm{k}}(t)$ is a trivial term arising due to the parabolic band, while the remaining parts are proportional to $\sigma_{z}$ and depend on $\bm{A}(t)$, reflecting the effect of   light   on UMs. 
For $d$-wave AMs, the second term of $V^{d}_{\bm{k}}(t)$ shows that the time-dependent light drive directly affects the altermagnetic field by linearly coupling to momentum akin to common SOC and that depends on the type of $d$-wave AM, hence causing a nontrivial light-matter interaction. 
Also, the third term of $V^{d}_{\bm{k}}(t)$ manifests that the time-dependent drive also affects the altermagnetic field in the form of a Zeeman-like effect, which is independent of momentum but still distinct for each type of $d$-wave AM. 
For $p$-wave UMs, the second term of $V^{p}_{\bm{k}}(t)$ is a momentum-independent Zeeman-like contribution of the light drive, which, however, strongly depends on the type of $p$-wave UM.

\textit{\textbf{S2.1. }High-frequency driven UMs in the normal state.}---
In the LPL-driven UMs, the spin density [Eq.~(\ref{rhoEffective})] is obtained by replacing $M_{\bm{k}}^{d} \to M_{\bm{k}}^{d} + M_{\Omega}^{d}$ in Eq.~(\ref{rhz_static}).

\begin{figure}[!t]
\centering
\includegraphics[width=0.95\linewidth]{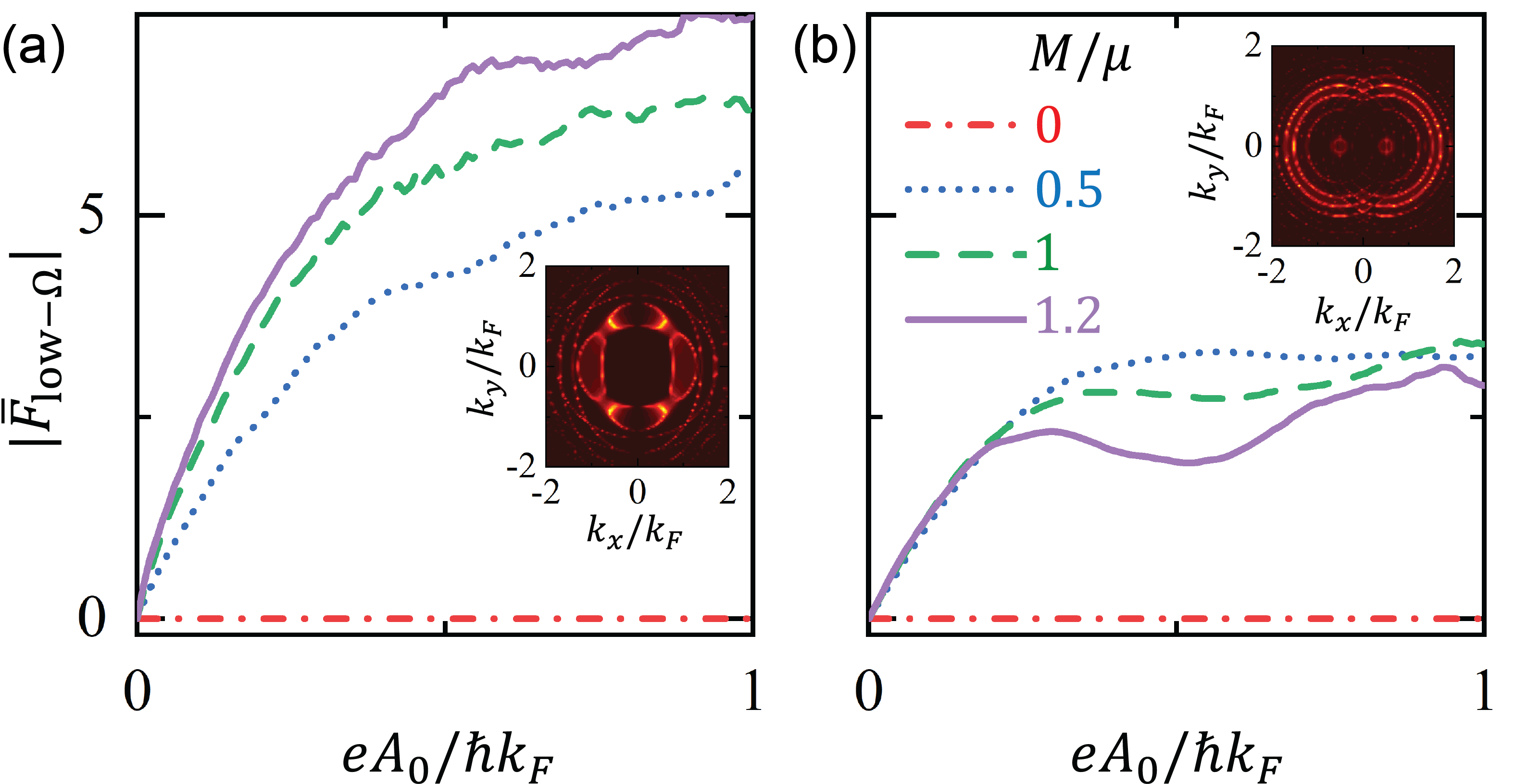}
\caption{(a) Spin-triplet odd-Floquet odd-frequency odd-parity pair amplitude pair amplitude $\bar{F}_{\rm low- \Omega}$ integrated in $\bm{k}$ as a function of $A_{0}$ for distinct $M$   in a $d_{x^{2}-y^{2}}$-wave AM  with conventional superconductivity under low-$\Omega$ LPL. 
(b) Same as in (a) but for an emergent spin-triplet odd-Floquet even-frequency even-parity pairing in a $p_{x}$-wave UM. The insets show the momentum dependence of $F_{\rm low- \Omega}$.  Parameters: $\phi=0$, $\Delta=0.5$, and $\omega = \Omega/2$, while the rest same as in Fig.\,\ref{Figure3}. 
}
\label{Figure4}
\end{figure}

\textit{\textbf{S2.2. }High-frequency driven UMs with conventional superconductivity.}---The spin-triplet density and spin-triplet pairing can be obtained from the diagonal and off-diagonal parts of $G_{\rm eff}^{d}(\bm{k},\omega)=[z-H_{\rm eff}^{d}(\bm{k})]^{-1}$, with $H_{\rm eff}^{d}(\bm{k})$ given by Eq.\,(\ref{HeffSCLPL}); see also S1.1 and Eqs.\,(\ref{eq_Greenfunction}). We find that the spin-triplet pairing is accompanied by a spin-triplet spin density along $z$  given by
\begin{equation}
\rho_{z,{\rm sc}}^{\rm eff}(\bm{k},\omega)=\frac{2}{\pi}{\rm Im}\bigg[\frac{\mathcal{Q}(\omega,\Omega,\bm{k})}{2\Delta (\omega+i\eta)}F^{\rm eff}_{t}(\bm{k},\omega+i\eta)\bigg]\,,
\end{equation}
where  $\mathcal{Q}(\omega,\Omega,\bm{k})=(\omega+i\eta+\xi_{\bm{k}})^{2}-(M^{d}_{\bm{k}}+M^{d}_{\Omega})^{2}+\Delta^{2}$,   $F^{\rm eff}_{t}(\bm{k},\omega+i\eta)$ is given by Eq. (\ref{EqFtFs}). Note that the pair amplitudes [Eqs.~(\ref{EqFtFs})] can be obtained by replacing $M_{\bm{k}}^{d} \to M_{\bm{k}}^{d} + M_{\Omega}^{d}$ in (\ref{eq_fst}), and (\ref{eq_fss}).  In the normal state, $\rho_{z,{\rm sc}}^{\rm eff}$ reduces to the normal state expression given by Eq.\,(\ref{rhoEffective}).  In the superconducting state, the spin density is tied to the emergent spin-triplet pairing.  While this relationship already happens in the static phase, therein the spin density (and hence the spin-triplet pairing) vanishes when integrating over $\bm{k}$, as we discussed before. Hence, a finite spin-triplet pairing appears only due to the light drive.

\textit{\textbf{S2.3. } Low-frequency driven UMs with conventional superconductivity.}---
While spin-triplet odd-frequency pairing appears only in high-frequency driven $d$-wave UMs, both odd- and even-frequency spin-triplet pairs emerge in all driven UMs at low $\Omega$.
The low-frequency pairs $\bar F_{\text{low-}\Omega}$ originate from the Floquet anomalous Green's functions $F_{n,m}^{\sigma_{1}\sigma_{2}}(\bm{k}_1, \bm{k}_2, z)$, which represent   electron pairing   between  the   Floquet bands ($n,m$) via the    absorption/emission of $|n-m|$ photons. Here, $F_{n,m}^{\sigma_{1}\sigma_{2}}(\bm{k}_1, \bm{k}_2, z)$ obey Fermi-Dirac statistics \cite{PhysRevB.103.104505,PhysRevB.109.134517},  enabling the emergence of $\bar F_{\text{low-}\Omega}$ with multiple symmetries, see Section IIC in \cite{sm} for more details. 
Fig.\,\ref{Figure4}(a) shows the magnitude of the spin-triplet odd-Floquet odd-frequency odd-parity pairing integrated in $\bm{k}$ as a function of $A_{0}$ for distinct $M$ in a $d_{x^{2}-y^{2}}$-wave UM under LPL. Also, Fig.\,\ref{Figure4}(b)   demonstrates the emergent pairing with spin-triplet odd-Floquet even-frequency even-parity  pairs in a $p$-wave UM under LPL. Both cases  unveil the role of $M$ for the spin-triplet character of the pairing, while $A_0$ enables its presence between Floquet bands and the $d_{x^2-y^2}$-wave parity symmetry [inset of Fig.\,\ref{Figure4}(a)].  Thus, low-$\Omega$ light drives offer a broad landscape for engineering Cooper pairs in UMs.


\begin{thebibliography}{139}%
\makeatletter
\providecommand \@ifxundefined [1]{%
 \@ifx{#1\undefined}
}%
\providecommand \@ifnum [1]{%
 \ifnum #1\expandafter \@firstoftwo
 \else \expandafter \@secondoftwo
 \fi
}%
\providecommand \@ifx [1]{%
 \ifx #1\expandafter \@firstoftwo
 \else \expandafter \@secondoftwo
 \fi
}%
\providecommand \natexlab [1]{#1}%
\providecommand \enquote  [1]{``#1''}%
\providecommand \bibnamefont  [1]{#1}%
\providecommand \bibfnamefont [1]{#1}%
\providecommand \citenamefont [1]{#1}%
\providecommand \href@noop [0]{\@secondoftwo}%
\providecommand \href [0]{\begingroup \@sanitize@url \@href}%
\providecommand \@href[1]{\@@startlink{#1}\@@href}%
\providecommand \@@href[1]{\endgroup#1\@@endlink}%
\providecommand \@sanitize@url [0]{\catcode `\\12\catcode `\$12\catcode
  `\&12\catcode `\#12\catcode `\^12\catcode `\_12\catcode `\%12\relax}%
\providecommand \@@startlink[1]{}%
\providecommand \@@endlink[0]{}%
\providecommand \url  [0]{\begingroup\@sanitize@url \@url }%
\providecommand \@url [1]{\endgroup\@href {#1}{\urlprefix }}%
\providecommand \urlprefix  [0]{URL }%
\providecommand \Eprint [0]{\href }%
\providecommand \doibase [0]{https://doi.org/}%
\providecommand \selectlanguage [0]{\@gobble}%
\providecommand \bibinfo  [0]{\@secondoftwo}%
\providecommand \bibfield  [0]{\@secondoftwo}%
\providecommand \translation [1]{[#1]}%
\providecommand \BibitemOpen [0]{}%
\providecommand \bibitemStop [0]{}%
\providecommand \bibitemNoStop [0]{.\EOS\space}%
\providecommand \EOS [0]{\spacefactor3000\relax}%
\providecommand \BibitemShut  [1]{\csname bibitem#1\endcsname}%
\let\auto@bib@innerbib\@empty
%</preamble>
\bibitem [{\citenamefont {Cayssol}\ \emph {et~al.}(2013)\citenamefont
  {Cayssol}, \citenamefont {D{\'o}ra}, \citenamefont {Simon},\ and\
  \citenamefont {Moessner}}]{cayssol2013floquet}%
  \BibitemOpen
  \bibfield  {author} {\bibinfo {author} {\bibfnamefont {J.}~\bibnamefont
  {Cayssol}}, \bibinfo {author} {\bibfnamefont {B.}~\bibnamefont {D{\'o}ra}},
  \bibinfo {author} {\bibfnamefont {F.}~\bibnamefont {Simon}},\ and\ \bibinfo
  {author} {\bibfnamefont {R.}~\bibnamefont {Moessner}},\ }\bibfield  {title}
  {\bibinfo {title} {Floquet topological insulators},\ }\href
  {https://doi.org/10.1002/pssr.201206451} {\bibfield  {journal} {\bibinfo
  {journal} {Phys. Status Solidi RRL}\ }\textbf {\bibinfo {volume} {7}},\
  \bibinfo {pages} {101} (\bibinfo {year} {2013})}\BibitemShut {NoStop}%
\bibitem [{\citenamefont {Bukov}\ \emph {et~al.}(2015)\citenamefont {Bukov},
  \citenamefont {D'Alessio},\ and\ \citenamefont
  {Polkovnikov}}]{bukov2015universal}%
  \BibitemOpen
  \bibfield  {author} {\bibinfo {author} {\bibfnamefont {M.}~\bibnamefont
  {Bukov}}, \bibinfo {author} {\bibfnamefont {L.}~\bibnamefont {D'Alessio}},\
  and\ \bibinfo {author} {\bibfnamefont {A.}~\bibnamefont {Polkovnikov}},\
  }\bibfield  {title} {\bibinfo {title} {Universal high-frequency behavior of
  periodically driven systems: from dynamical stabilization to {F}loquet
  engineering},\ }\href {https://doi.org/10.1080/00018732.2015.1055918}
  {\bibfield  {journal} {\bibinfo  {journal} {Adv. Phys.}\ }\textbf {\bibinfo
  {volume} {64}},\ \bibinfo {pages} {139} (\bibinfo {year} {2015})}\BibitemShut
  {NoStop}%
\bibitem [{\citenamefont {Oka}\ and\ \citenamefont
  {Kitamura}(2019)}]{OkaReview2019}%
  \BibitemOpen
  \bibfield  {author} {\bibinfo {author} {\bibfnamefont {T.}~\bibnamefont
  {Oka}}\ and\ \bibinfo {author} {\bibfnamefont {S.}~\bibnamefont {Kitamura}},\
  }\bibfield  {title} {\bibinfo {title} {Floquet engineering of quantum
  materials},\ }\href
  {https://doi.org/10.1146/annurev-conmatphys-031218-013423} {\bibfield
  {journal} {\bibinfo  {journal} {Annu. Rev. Condens. Matter Phys.}\ }\textbf
  {\bibinfo {volume} {10}},\ \bibinfo {pages} {387} (\bibinfo {year}
  {2019})}\BibitemShut {NoStop}%
\bibitem [{\citenamefont {Giovannini}\ and\ \citenamefont
  {H\"{u}bener}(2019)}]{Giovannini_2019}%
  \BibitemOpen
  \bibfield  {author} {\bibinfo {author} {\bibfnamefont {U.~D.}\ \bibnamefont
  {Giovannini}}\ and\ \bibinfo {author} {\bibfnamefont {H.}~\bibnamefont
  {H\"{u}bener}},\ }\bibfield  {title} {\bibinfo {title} {Floquet analysis of
  excitations in materials},\ }\href {https://doi.org/10.1088/2515-7639/ab387b}
  {\bibfield  {journal} {\bibinfo  {journal} {J. Phys. Mater.}\ }\textbf
  {\bibinfo {volume} {3}},\ \bibinfo {pages} {012001} (\bibinfo {year}
  {2019})}\BibitemShut {NoStop}%
\bibitem [{\citenamefont {Rudner}\ and\ \citenamefont
  {Lindner}(2020)}]{rudner2020band}%
  \BibitemOpen
  \bibfield  {author} {\bibinfo {author} {\bibfnamefont {M.~S.}\ \bibnamefont
  {Rudner}}\ and\ \bibinfo {author} {\bibfnamefont {N.~H.}\ \bibnamefont
  {Lindner}},\ }\bibfield  {title} {\bibinfo {title} {Band structure
  engineering and non-equilibrium dynamics in {F}loquet topological
  insulators},\ }\href {https://doi.org/10.1038/s42254-020-0170-z} {\bibfield
  {journal} {\bibinfo  {journal} {Nat. Rev. Phys.}\ }\textbf {\bibinfo {volume}
  {2}},\ \bibinfo {pages} {229} (\bibinfo {year} {2020})}\BibitemShut {NoStop}%
\bibitem [{\citenamefont {Francesconi}\ \emph {et~al.}(2020)\citenamefont
  {Francesconi}, \citenamefont {Baboux}, \citenamefont {Raymond}, \citenamefont
  {Fabre}, \citenamefont {Boucher}, \citenamefont {Lemaître}, \citenamefont
  {Milman}, \citenamefont {Amanti},\ and\ \citenamefont
  {Ducci}}]{francesconi2020engineering}%
  \BibitemOpen
  \bibfield  {author} {\bibinfo {author} {\bibfnamefont {S.}~\bibnamefont
  {Francesconi}}, \bibinfo {author} {\bibfnamefont {F.}~\bibnamefont {Baboux}},
  \bibinfo {author} {\bibfnamefont {A.}~\bibnamefont {Raymond}}, \bibinfo
  {author} {\bibfnamefont {N.}~\bibnamefont {Fabre}}, \bibinfo {author}
  {\bibfnamefont {G.}~\bibnamefont {Boucher}}, \bibinfo {author} {\bibfnamefont
  {A.}~\bibnamefont {Lemaître}}, \bibinfo {author} {\bibfnamefont
  {P.}~\bibnamefont {Milman}}, \bibinfo {author} {\bibfnamefont {M.~I.}\
  \bibnamefont {Amanti}},\ and\ \bibinfo {author} {\bibfnamefont
  {S.}~\bibnamefont {Ducci}},\ }\bibfield  {title} {\bibinfo {title}
  {Engineering two-photon wavefunction and exchange statistics in a
  semiconductor chip},\ }\href {https://doi.org/10.1364/OPTICA.379477}
  {\bibfield  {journal} {\bibinfo  {journal} {Optica}\ }\textbf {\bibinfo
  {volume} {7}},\ \bibinfo {pages} {316} (\bibinfo {year} {2020})}\BibitemShut
  {NoStop}%
\bibitem [{\citenamefont {Floquet}(1883)}]{ASENS_1883_2_12__47_0}%
  \BibitemOpen
  \bibfield  {author} {\bibinfo {author} {\bibfnamefont {G.}~\bibnamefont
  {Floquet}},\ }\bibfield  {title} {\bibinfo {title} {Sur les \'equations
  diff\'erentielles lin\'eaires \`a coefficients p\'eriodiques},\ }\href
  {https://doi.org/10.24033/asens.220} {\bibfield  {journal} {\bibinfo
  {journal} {Annales scientifiques de l'\'Ecole Normale Sup\'erieure}\ }\textbf
  {\bibinfo {volume} {12}},\ \bibinfo {pages} {47} (\bibinfo {year}
  {1883})}\BibitemShut {NoStop}%
\bibitem [{\citenamefont {Shirley}(1965)}]{PhysRev.138.B979}%
  \BibitemOpen
  \bibfield  {author} {\bibinfo {author} {\bibfnamefont {J.~H.}\ \bibnamefont
  {Shirley}},\ }\bibfield  {title} {\bibinfo {title} {Solution of the
  schr\"odinger equation with a hamiltonian periodic in time},\ }\href
  {https://doi.org/10.1103/PhysRev.138.B979} {\bibfield  {journal} {\bibinfo
  {journal} {Phys. Rev.}\ }\textbf {\bibinfo {volume} {138}},\ \bibinfo {pages}
  {B979} (\bibinfo {year} {1965})}\BibitemShut {NoStop}%
\bibitem [{\citenamefont {Sambe}(1973)}]{PhysRevA.7.2203}%
  \BibitemOpen
  \bibfield  {author} {\bibinfo {author} {\bibfnamefont {H.}~\bibnamefont
  {Sambe}},\ }\bibfield  {title} {\bibinfo {title} {Steady states and
  quasienergies of a quantum-mechanical system in an oscillating field},\
  }\href {https://doi.org/10.1103/PhysRevA.7.2203} {\bibfield  {journal}
  {\bibinfo  {journal} {Phys. Rev. A}\ }\textbf {\bibinfo {volume} {7}},\
  \bibinfo {pages} {2203} (\bibinfo {year} {1973})}\BibitemShut {NoStop}%
\bibitem [{\citenamefont {Oka}\ and\ \citenamefont
  {Aoki}(2009)}]{PhysRevB.79.081406}%
  \BibitemOpen
  \bibfield  {author} {\bibinfo {author} {\bibfnamefont {T.}~\bibnamefont
  {Oka}}\ and\ \bibinfo {author} {\bibfnamefont {H.}~\bibnamefont {Aoki}},\
  }\bibfield  {title} {\bibinfo {title} {Photovoltaic {H}all effect in
  graphene},\ }\href {https://doi.org/10.1103/PhysRevB.79.081406} {\bibfield
  {journal} {\bibinfo  {journal} {Phys. Rev. B}\ }\textbf {\bibinfo {volume}
  {79}},\ \bibinfo {pages} {081406} (\bibinfo {year} {2009})}\BibitemShut
  {NoStop}%
\bibitem [{\citenamefont {Kitagawa}\ \emph {et~al.}(2010)\citenamefont
  {Kitagawa}, \citenamefont {Berg}, \citenamefont {Rudner},\ and\ \citenamefont
  {Demler}}]{PhysRevB.82.235114}%
  \BibitemOpen
  \bibfield  {author} {\bibinfo {author} {\bibfnamefont {T.}~\bibnamefont
  {Kitagawa}}, \bibinfo {author} {\bibfnamefont {E.}~\bibnamefont {Berg}},
  \bibinfo {author} {\bibfnamefont {M.}~\bibnamefont {Rudner}},\ and\ \bibinfo
  {author} {\bibfnamefont {E.}~\bibnamefont {Demler}},\ }\bibfield  {title}
  {\bibinfo {title} {Topological characterization of periodically driven
  quantum systems},\ }\href {https://doi.org/10.1103/PhysRevB.82.235114}
  {\bibfield  {journal} {\bibinfo  {journal} {Phys. Rev. B}\ }\textbf {\bibinfo
  {volume} {82}},\ \bibinfo {pages} {235114} (\bibinfo {year}
  {2010})}\BibitemShut {NoStop}%
\bibitem [{\citenamefont {Lindner}\ \emph {et~al.}(2011)\citenamefont
  {Lindner}, \citenamefont {Refael},\ and\ \citenamefont
  {Galitski}}]{lindner2011floquet}%
  \BibitemOpen
  \bibfield  {author} {\bibinfo {author} {\bibfnamefont {N.~H.}\ \bibnamefont
  {Lindner}}, \bibinfo {author} {\bibfnamefont {G.}~\bibnamefont {Refael}},\
  and\ \bibinfo {author} {\bibfnamefont {V.}~\bibnamefont {Galitski}},\
  }\bibfield  {title} {\bibinfo {title} {Floquet topological insulator in
  semiconductor quantum wells},\ }\href {https://doi.org/10.1038/nphys1926}
  {\bibfield  {journal} {\bibinfo  {journal} {Nat. Phys.}\ }\textbf {\bibinfo
  {volume} {7}},\ \bibinfo {pages} {490} (\bibinfo {year} {2011})}\BibitemShut
  {NoStop}%
\bibitem [{\citenamefont {Klinovaja}\ \emph {et~al.}(2016)\citenamefont
  {Klinovaja}, \citenamefont {Stano},\ and\ \citenamefont
  {Loss}}]{PhysRevLett.116.176401}%
  \BibitemOpen
  \bibfield  {author} {\bibinfo {author} {\bibfnamefont {J.}~\bibnamefont
  {Klinovaja}}, \bibinfo {author} {\bibfnamefont {P.}~\bibnamefont {Stano}},\
  and\ \bibinfo {author} {\bibfnamefont {D.}~\bibnamefont {Loss}},\ }\bibfield
  {title} {\bibinfo {title} {Topological {F}loquet phases in driven coupled
  {R}ashba nanowires},\ }\href {https://doi.org/10.1103/PhysRevLett.116.176401}
  {\bibfield  {journal} {\bibinfo  {journal} {Phys. Rev. Lett.}\ }\textbf
  {\bibinfo {volume} {116}},\ \bibinfo {pages} {176401} (\bibinfo {year}
  {2016})}\BibitemShut {NoStop}%
\bibitem [{\citenamefont {Faisal}\ and\ \citenamefont
  {Kami\ifmmode~\acute{n}\else \'{n}\fi{}ski}(1997)}]{PhysRevA.56.748}%
  \BibitemOpen
  \bibfield  {author} {\bibinfo {author} {\bibfnamefont {F.~H.~M.}\
  \bibnamefont {Faisal}}\ and\ \bibinfo {author} {\bibfnamefont {J.~Z.}\
  \bibnamefont {Kami\ifmmode~\acute{n}\else \'{n}\fi{}ski}},\ }\bibfield
  {title} {\bibinfo {title} {Floquet-{B}loch theory of high-harmonic generation
  in periodic structures},\ }\href {https://doi.org/10.1103/PhysRevA.56.748}
  {\bibfield  {journal} {\bibinfo  {journal} {Phys. Rev. A}\ }\textbf {\bibinfo
  {volume} {56}},\ \bibinfo {pages} {748} (\bibinfo {year} {1997})}\BibitemShut
  {NoStop}%
\bibitem [{\citenamefont {Kitagawa}\ \emph {et~al.}(2011)\citenamefont
  {Kitagawa}, \citenamefont {Oka}, \citenamefont {Brataas}, \citenamefont
  {Fu},\ and\ \citenamefont {Demler}}]{PhysRevB.84.235108}%
  \BibitemOpen
  \bibfield  {author} {\bibinfo {author} {\bibfnamefont {T.}~\bibnamefont
  {Kitagawa}}, \bibinfo {author} {\bibfnamefont {T.}~\bibnamefont {Oka}},
  \bibinfo {author} {\bibfnamefont {A.}~\bibnamefont {Brataas}}, \bibinfo
  {author} {\bibfnamefont {L.}~\bibnamefont {Fu}},\ and\ \bibinfo {author}
  {\bibfnamefont {E.}~\bibnamefont {Demler}},\ }\bibfield  {title} {\bibinfo
  {title} {Transport properties of nonequilibrium systems under the application
  of light: Photoinduced quantum {H}all insulators without {L}andau levels},\
  }\href {https://doi.org/10.1103/PhysRevB.84.235108} {\bibfield  {journal}
  {\bibinfo  {journal} {Phys. Rev. B}\ }\textbf {\bibinfo {volume} {84}},\
  \bibinfo {pages} {235108} (\bibinfo {year} {2011})}\BibitemShut {NoStop}%
\bibitem [{\citenamefont {Rechtsman}\ \emph {et~al.}(2013)\citenamefont
  {Rechtsman}, \citenamefont {Zeuner}, \citenamefont {Plotnik}, \citenamefont
  {Lumer}, \citenamefont {Podolsky}, \citenamefont {Dreisow}, \citenamefont
  {Nolte}, \citenamefont {Segev},\ and\ \citenamefont
  {Szameit}}]{rechtsman2013photonic}%
  \BibitemOpen
  \bibfield  {author} {\bibinfo {author} {\bibfnamefont {M.~C.}\ \bibnamefont
  {Rechtsman}}, \bibinfo {author} {\bibfnamefont {J.~M.}\ \bibnamefont
  {Zeuner}}, \bibinfo {author} {\bibfnamefont {Y.}~\bibnamefont {Plotnik}},
  \bibinfo {author} {\bibfnamefont {Y.}~\bibnamefont {Lumer}}, \bibinfo
  {author} {\bibfnamefont {D.}~\bibnamefont {Podolsky}}, \bibinfo {author}
  {\bibfnamefont {F.}~\bibnamefont {Dreisow}}, \bibinfo {author} {\bibfnamefont
  {S.}~\bibnamefont {Nolte}}, \bibinfo {author} {\bibfnamefont
  {M.}~\bibnamefont {Segev}},\ and\ \bibinfo {author} {\bibfnamefont
  {A.}~\bibnamefont {Szameit}},\ }\bibfield  {title} {\bibinfo {title}
  {Photonic {F}loquet topological insulators},\ }\href
  {https://doi.org/10.1038/nature12066} {\bibfield  {journal} {\bibinfo
  {journal} {Nature}\ }\textbf {\bibinfo {volume} {496}},\ \bibinfo {pages}
  {196} (\bibinfo {year} {2013})}\BibitemShut {NoStop}%
\bibitem [{\citenamefont {Fregoso}\ \emph {et~al.}(2013)\citenamefont
  {Fregoso}, \citenamefont {Wang}, \citenamefont {Gedik},\ and\ \citenamefont
  {Galitski}}]{PhysRevB.88.155129}%
  \BibitemOpen
  \bibfield  {author} {\bibinfo {author} {\bibfnamefont {B.~M.}\ \bibnamefont
  {Fregoso}}, \bibinfo {author} {\bibfnamefont {Y.~H.}\ \bibnamefont {Wang}},
  \bibinfo {author} {\bibfnamefont {N.}~\bibnamefont {Gedik}},\ and\ \bibinfo
  {author} {\bibfnamefont {V.}~\bibnamefont {Galitski}},\ }\bibfield  {title}
  {\bibinfo {title} {Driven electronic states at the surface of a topological
  insulator},\ }\href {https://doi.org/10.1103/PhysRevB.88.155129} {\bibfield
  {journal} {\bibinfo  {journal} {Phys. Rev. B}\ }\textbf {\bibinfo {volume}
  {88}},\ \bibinfo {pages} {155129} (\bibinfo {year} {2013})}\BibitemShut
  {NoStop}%
\bibitem [{\citenamefont {Wang}\ \emph {et~al.}(2013)\citenamefont {Wang},
  \citenamefont {Steinberg}, \citenamefont {Jarillo-Herrero},\ and\
  \citenamefont {Gedik}}]{wang2013observation}%
  \BibitemOpen
  \bibfield  {author} {\bibinfo {author} {\bibfnamefont {Y.~H.}\ \bibnamefont
  {Wang}}, \bibinfo {author} {\bibfnamefont {H.}~\bibnamefont {Steinberg}},
  \bibinfo {author} {\bibfnamefont {P.}~\bibnamefont {Jarillo-Herrero}},\ and\
  \bibinfo {author} {\bibfnamefont {N.}~\bibnamefont {Gedik}},\ }\bibfield
  {title} {\bibinfo {title} {Observation of {F}loquet–{B}loch states on the
  surface of a topological insulator},\ }\href
  {https://doi.org/10.1126/science.1239834} {\bibfield  {journal} {\bibinfo
  {journal} {Science}\ }\textbf {\bibinfo {volume} {342}},\ \bibinfo {pages}
  {453} (\bibinfo {year} {2013})}\BibitemShut {NoStop}%
\bibitem [{\citenamefont {McIver}\ \emph {et~al.}(2020)\citenamefont {McIver},
  \citenamefont {Schulte}, \citenamefont {Stein}, \citenamefont {Matsuyama},
  \citenamefont {Jotzu}, \citenamefont {Meier},\ and\ \citenamefont
  {Cavalleri}}]{mciver2020light}%
  \BibitemOpen
  \bibfield  {author} {\bibinfo {author} {\bibfnamefont {J.~W.}\ \bibnamefont
  {McIver}}, \bibinfo {author} {\bibfnamefont {B.}~\bibnamefont {Schulte}},
  \bibinfo {author} {\bibfnamefont {F.~U.}\ \bibnamefont {Stein}}, \bibinfo
  {author} {\bibfnamefont {T.}~\bibnamefont {Matsuyama}}, \bibinfo {author}
  {\bibfnamefont {G.}~\bibnamefont {Jotzu}}, \bibinfo {author} {\bibfnamefont
  {G.}~\bibnamefont {Meier}},\ and\ \bibinfo {author} {\bibfnamefont
  {A.}~\bibnamefont {Cavalleri}},\ }\bibfield  {title} {\bibinfo {title}
  {Light‑induced anomalous {H}all effect in graphene},\ }\href
  {https://doi.org/10.1038/s41567-019-0698-y} {\bibfield  {journal} {\bibinfo
  {journal} {Nat. Phys.}\ }\textbf {\bibinfo {volume} {16}},\ \bibinfo {pages}
  {38} (\bibinfo {year} {2020})}\BibitemShut {NoStop}%
\bibitem [{\citenamefont {Aeschlimann}\ \emph {et~al.}(2021)\citenamefont
  {Aeschlimann}, \citenamefont {Sato}, \citenamefont {Krause}, \citenamefont
  {Chávez-Cervantes}, \citenamefont {Giovannini}, \citenamefont {Hübener},
  \citenamefont {Forti}, \citenamefont {Coletti}, \citenamefont {Hanff},
  \citenamefont {Rossnagel}, \citenamefont {Rubio},\ and\ \citenamefont
  {Gierz}}]{aeschlimann2021survival}%
  \BibitemOpen
  \bibfield  {author} {\bibinfo {author} {\bibfnamefont {S.}~\bibnamefont
  {Aeschlimann}}, \bibinfo {author} {\bibfnamefont {S.~A.}\ \bibnamefont
  {Sato}}, \bibinfo {author} {\bibfnamefont {R.}~\bibnamefont {Krause}},
  \bibinfo {author} {\bibfnamefont {M.}~\bibnamefont {Chávez-Cervantes}},
  \bibinfo {author} {\bibfnamefont {U.~D.}\ \bibnamefont {Giovannini}},
  \bibinfo {author} {\bibfnamefont {H.}~\bibnamefont {Hübener}}, \bibinfo
  {author} {\bibfnamefont {S.}~\bibnamefont {Forti}}, \bibinfo {author}
  {\bibfnamefont {C.}~\bibnamefont {Coletti}}, \bibinfo {author} {\bibfnamefont
  {K.}~\bibnamefont {Hanff}}, \bibinfo {author} {\bibfnamefont
  {K.}~\bibnamefont {Rossnagel}}, \bibinfo {author} {\bibfnamefont
  {A.}~\bibnamefont {Rubio}},\ and\ \bibinfo {author} {\bibfnamefont
  {I.}~\bibnamefont {Gierz}},\ }\bibfield  {title} {\bibinfo {title} {Survival
  of {F}loquet–{B}loch states in the presence of scattering},\ }\href
  {https://doi.org/10.1021/acs.nanolett.1c00801} {\bibfield  {journal}
  {\bibinfo  {journal} {Nano Lett.}\ }\textbf {\bibinfo {volume} {21}},\
  \bibinfo {pages} {5028} (\bibinfo {year} {2021})}\BibitemShut {NoStop}%
\bibitem [{\citenamefont {Mahmood}\ \emph {et~al.}(2016)\citenamefont
  {Mahmood}, \citenamefont {Chan}, \citenamefont {Alpichshev}, \citenamefont
  {Gardner}, \citenamefont {Lee}, \citenamefont {Lee},\ and\ \citenamefont
  {Gedik}}]{mahmood2016selective}%
  \BibitemOpen
  \bibfield  {author} {\bibinfo {author} {\bibfnamefont {F.}~\bibnamefont
  {Mahmood}}, \bibinfo {author} {\bibfnamefont {C.-K.}\ \bibnamefont {Chan}},
  \bibinfo {author} {\bibfnamefont {Z.}~\bibnamefont {Alpichshev}}, \bibinfo
  {author} {\bibfnamefont {D.~R.}\ \bibnamefont {Gardner}}, \bibinfo {author}
  {\bibfnamefont {Y.~S.}\ \bibnamefont {Lee}}, \bibinfo {author} {\bibfnamefont
  {P.~A.}\ \bibnamefont {Lee}},\ and\ \bibinfo {author} {\bibfnamefont
  {N.}~\bibnamefont {Gedik}},\ }\bibfield  {title} {\bibinfo {title} {Selective
  scattering between {F}loquet–{B}loch and {V}olkov states in a topological
  insulator},\ }\href {https://doi.org/10.1038/nphys3609} {\bibfield  {journal}
  {\bibinfo  {journal} {Nat. Phys.}\ }\textbf {\bibinfo {volume} {12}},\
  \bibinfo {pages} {306} (\bibinfo {year} {2016})}\BibitemShut {NoStop}%
\bibitem [{\citenamefont {Galitski}\ and\ \citenamefont
  {Spielman}(2013)}]{galitski2013spin}%
  \BibitemOpen
  \bibfield  {author} {\bibinfo {author} {\bibfnamefont {V.}~\bibnamefont
  {Galitski}}\ and\ \bibinfo {author} {\bibfnamefont {I.~B.}\ \bibnamefont
  {Spielman}},\ }\bibfield  {title} {\bibinfo {title} {Spin--orbit coupling in
  quantum gases},\ }\href {https://doi.org/10.1038/nature11841} {\bibfield
  {journal} {\bibinfo  {journal} {Nature}\ }\textbf {\bibinfo {volume} {494}},\
  \bibinfo {pages} {49} (\bibinfo {year} {2013})}\BibitemShut {NoStop}%
\bibitem [{\citenamefont {Manchon}\ \emph {et~al.}(2015)\citenamefont
  {Manchon}, \citenamefont {Koo}, \citenamefont {Nitta}, \citenamefont
  {Frolov},\ and\ \citenamefont {Duine}}]{manchon2015new}%
  \BibitemOpen
  \bibfield  {author} {\bibinfo {author} {\bibfnamefont {A.}~\bibnamefont
  {Manchon}}, \bibinfo {author} {\bibfnamefont {H.}~\bibnamefont {Koo}},
  \bibinfo {author} {\bibfnamefont {J.}~\bibnamefont {Nitta}}, \bibinfo
  {author} {\bibfnamefont {S.}~\bibnamefont {Frolov}},\ and\ \bibinfo {author}
  {\bibfnamefont {R.}~\bibnamefont {Duine}},\ }\bibfield  {title} {\bibinfo
  {title} {New perspectives for {R}ashba spin–orbit coupling},\ }\href
  {https://doi.org/10.1038/nmat4360} {\bibfield  {journal} {\bibinfo  {journal}
  {Nat. Mater.}\ }\textbf {\bibinfo {volume} {14}},\ \bibinfo {pages} {871}
  (\bibinfo {year} {2015})}\BibitemShut {NoStop}%
\bibitem [{\citenamefont {Bai}\ \emph {et~al.}(2024)\citenamefont {Bai},
  \citenamefont {Feng}, \citenamefont {Liu}, \citenamefont {\v{S}mejkal},
  \citenamefont {Mokrousov},\ and\ \citenamefont {Yao}}]{Bai_review24}%
  \BibitemOpen
  \bibfield  {author} {\bibinfo {author} {\bibfnamefont {L.}~\bibnamefont
  {Bai}}, \bibinfo {author} {\bibfnamefont {W.}~\bibnamefont {Feng}}, \bibinfo
  {author} {\bibfnamefont {S.}~\bibnamefont {Liu}}, \bibinfo {author}
  {\bibfnamefont {L.}~\bibnamefont {\v{S}mejkal}}, \bibinfo {author}
  {\bibfnamefont {Y.}~\bibnamefont {Mokrousov}},\ and\ \bibinfo {author}
  {\bibfnamefont {Y.}~\bibnamefont {Yao}},\ }\bibfield  {title} {\bibinfo
  {title} {Altermagnetism: Exploring new frontiers in magnetism and
  spintronics},\ }\href
  {https://advanced.onlinelibrary.wiley.com/doi/abs/10.1002/adfm.202409327}
  {\bibfield  {journal} {\bibinfo  {journal} {Advanced Functional Materials}\
  }\textbf {\bibinfo {volume} {34}},\ \bibinfo {pages} {2409327} (\bibinfo
  {year} {2024})}\BibitemShut {NoStop}%
\bibitem [{\citenamefont {Jungwirth}\ \emph
  {et~al.}(2024{\natexlab{a}})\citenamefont {Jungwirth}, \citenamefont
  {Fernandes}, \citenamefont {Fradkin}, \citenamefont {MacDonald},
  \citenamefont {Sinova},\ and\ \citenamefont {Smejkal}}]{jungwirth2024He3Ams}%
  \BibitemOpen
  \bibfield  {author} {\bibinfo {author} {\bibfnamefont {T.}~\bibnamefont
  {Jungwirth}}, \bibinfo {author} {\bibfnamefont {R.~M.}\ \bibnamefont
  {Fernandes}}, \bibinfo {author} {\bibfnamefont {E.}~\bibnamefont {Fradkin}},
  \bibinfo {author} {\bibfnamefont {A.~H.}\ \bibnamefont {MacDonald}}, \bibinfo
  {author} {\bibfnamefont {J.}~\bibnamefont {Sinova}},\ and\ \bibinfo {author}
  {\bibfnamefont {L.}~\bibnamefont {Smejkal}},\ }\bibfield  {title} {\bibinfo
  {title} {From supefluid $^3${H}e to altermagnets},\ }\href
  {https://arxiv.org/abs/2411.00717} {\bibfield  {journal} {\bibinfo  {journal}
  {arXiv: 2411.00717}\ } (\bibinfo {year} {2024}{\natexlab{a}})}\BibitemShut
  {NoStop}%
\bibitem [{\citenamefont {Jungwirth}\ \emph
  {et~al.}(2024{\natexlab{b}})\citenamefont {Jungwirth}, \citenamefont
  {Fernandes}, \citenamefont {Sinova},\ and\ \citenamefont
  {\v{S}mejkal}}]{jungwirth2024}%
  \BibitemOpen
  \bibfield  {author} {\bibinfo {author} {\bibfnamefont {T.}~\bibnamefont
  {Jungwirth}}, \bibinfo {author} {\bibfnamefont {R.~M.}\ \bibnamefont
  {Fernandes}}, \bibinfo {author} {\bibfnamefont {J.}~\bibnamefont {Sinova}},\
  and\ \bibinfo {author} {\bibfnamefont {L.}~\bibnamefont {\v{S}mejkal}},\
  }\bibfield  {title} {\bibinfo {title} {Altermagnets and beyond: Nodal
  magnetically-ordered phases},\ }\href {https://arxiv.org/abs/2409.10034}
  {\bibfield  {journal} {\bibinfo  {journal} {arXiv: 2409.10034}\ } (\bibinfo
  {year} {2024}{\natexlab{b}})}\BibitemShut {NoStop}%
\bibitem [{\citenamefont {Fukaya}\ \emph
  {et~al.}(2025{\natexlab{a}})\citenamefont {Fukaya}, \citenamefont {Lu},
  \citenamefont {Yada}, \citenamefont {Tanaka},\ and\ \citenamefont
  {Cayao}}]{FukayaCayaoReviewUMs}%
  \BibitemOpen
  \bibfield  {author} {\bibinfo {author} {\bibfnamefont {Y.}~\bibnamefont
  {Fukaya}}, \bibinfo {author} {\bibfnamefont {B.}~\bibnamefont {Lu}}, \bibinfo
  {author} {\bibfnamefont {K.}~\bibnamefont {Yada}}, \bibinfo {author}
  {\bibfnamefont {Y.}~\bibnamefont {Tanaka}},\ and\ \bibinfo {author}
  {\bibfnamefont {J.}~\bibnamefont {Cayao}},\ }\bibfield  {title} {\bibinfo
  {title} {Superconducting phenomena in systems with unconventional magnets},\
  }\href {https://doi.org/10.1088/1361-648X/adf1cf} {\bibfield  {journal}
  {\bibinfo  {journal} {J. Phys.: Condens. Matter}\ }\textbf {\bibinfo {volume}
  {37}},\ \bibinfo {pages} {313003} (\bibinfo {year}
  {2025}{\natexlab{a}})}\BibitemShut {NoStop}%
\bibitem [{\citenamefont {\ifmmode~\check{S}\else \v{S}\fi{}mejkal}\ \emph
  {et~al.}(2022{\natexlab{a}})\citenamefont {\ifmmode~\check{S}\else
  \v{S}\fi{}mejkal}, \citenamefont {Sinova},\ and\ \citenamefont
  {Jungwirth}}]{landscape22}%
  \BibitemOpen
  \bibfield  {author} {\bibinfo {author} {\bibfnamefont {L.}~\bibnamefont
  {\ifmmode~\check{S}\else \v{S}\fi{}mejkal}}, \bibinfo {author} {\bibfnamefont
  {J.}~\bibnamefont {Sinova}},\ and\ \bibinfo {author} {\bibfnamefont
  {T.}~\bibnamefont {Jungwirth}},\ }\bibfield  {title} {\bibinfo {title}
  {Emerging research landscape of altermagnetism},\ }\href
  {https://doi.org/10.1103/PhysRevX.12.040501} {\bibfield  {journal} {\bibinfo
  {journal} {Phys. Rev. X}\ }\textbf {\bibinfo {volume} {12}},\ \bibinfo
  {pages} {040501} (\bibinfo {year} {2022}{\natexlab{a}})}\BibitemShut
  {NoStop}%
\bibitem [{\citenamefont {Mazin}(2022)}]{MazinPRX22}%
  \BibitemOpen
  \bibfield  {author} {\bibinfo {author} {\bibfnamefont {I.}~\bibnamefont
  {Mazin}},\ }\bibfield  {title} {\bibinfo {title} {Editorial:
  Altermagnetism---a new punch line of fundamental magnetism},\ }\href
  {https://doi.org/10.1103/PhysRevX.12.040002} {\bibfield  {journal} {\bibinfo
  {journal} {Phys. Rev. X}\ }\textbf {\bibinfo {volume} {12}},\ \bibinfo
  {pages} {040002} (\bibinfo {year} {2022})}\BibitemShut {NoStop}%
\bibitem [{\citenamefont {Brekke}\ \emph {et~al.}(2024)\citenamefont {Brekke},
  \citenamefont {Sukhachov}, \citenamefont {Giil}, \citenamefont {Brataas},\
  and\ \citenamefont {Linder}}]{brekke24}%
  \BibitemOpen
  \bibfield  {author} {\bibinfo {author} {\bibfnamefont {B.}~\bibnamefont
  {Brekke}}, \bibinfo {author} {\bibfnamefont {P.}~\bibnamefont {Sukhachov}},
  \bibinfo {author} {\bibfnamefont {H.~G.}\ \bibnamefont {Giil}}, \bibinfo
  {author} {\bibfnamefont {A.}~\bibnamefont {Brataas}},\ and\ \bibinfo {author}
  {\bibfnamefont {J.}~\bibnamefont {Linder}},\ }\bibfield  {title} {\bibinfo
  {title} {Minimal models and transport properties of unconventional $p$-wave
  magnets},\ }\href {https://doi.org/10.1103/PhysRevLett.133.236703} {\bibfield
   {journal} {\bibinfo  {journal} {Phys. Rev. Lett.}\ }\textbf {\bibinfo
  {volume} {133}},\ \bibinfo {pages} {236703} (\bibinfo {year}
  {2024})}\BibitemShut {NoStop}%
\bibitem [{\citenamefont {Hellenes}\ \emph {et~al.}(2024)\citenamefont
  {Hellenes}, \citenamefont {Jungwirth}, \citenamefont {Jaeschke-Ubiergo},
  \citenamefont {Chakraborty}, \citenamefont {Sinova},\ and\ \citenamefont
  {\v{S}mejkal}}]{hellenes2024P}%
  \BibitemOpen
  \bibfield  {author} {\bibinfo {author} {\bibfnamefont {A.~B.}\ \bibnamefont
  {Hellenes}}, \bibinfo {author} {\bibfnamefont {T.}~\bibnamefont {Jungwirth}},
  \bibinfo {author} {\bibfnamefont {R.}~\bibnamefont {Jaeschke-Ubiergo}},
  \bibinfo {author} {\bibfnamefont {A.}~\bibnamefont {Chakraborty}}, \bibinfo
  {author} {\bibfnamefont {J.}~\bibnamefont {Sinova}},\ and\ \bibinfo {author}
  {\bibfnamefont {L.}~\bibnamefont {\v{S}mejkal}},\ }\bibfield  {title}
  {\bibinfo {title} {P-wave magnets},\ }\href
  {https://arxiv.org/abs/2309.01607} {\bibfield  {journal} {\bibinfo  {journal}
  {arXiv:2309.01607}\ } (\bibinfo {year} {2024})}\BibitemShut {NoStop}%
\bibitem [{\citenamefont {Naka}\ \emph {et~al.}(2019)\citenamefont {Naka},
  \citenamefont {Hayami}, \citenamefont {Kusunose}, \citenamefont {Yanagi},
  \citenamefont {Motome},\ and\ \citenamefont {Seo}}]{NakaNatCommun2019}%
  \BibitemOpen
  \bibfield  {author} {\bibinfo {author} {\bibfnamefont {M.}~\bibnamefont
  {Naka}}, \bibinfo {author} {\bibfnamefont {S.}~\bibnamefont {Hayami}},
  \bibinfo {author} {\bibfnamefont {H.}~\bibnamefont {Kusunose}}, \bibinfo
  {author} {\bibfnamefont {Y.}~\bibnamefont {Yanagi}}, \bibinfo {author}
  {\bibfnamefont {Y.}~\bibnamefont {Motome}},\ and\ \bibinfo {author}
  {\bibfnamefont {H.}~\bibnamefont {Seo}},\ }\bibfield  {title} {\bibinfo
  {title} {Spin current generation in organic antiferromagnets},\ }\href
  {https://doi.org/10.1038/s41467-019-12229-y} {\bibfield  {journal} {\bibinfo
  {journal} {Nat. Commun.}\ }\textbf {\bibinfo {volume} {10}},\ \bibinfo
  {pages} {4305} (\bibinfo {year} {2019})}\BibitemShut {NoStop}%
\bibitem [{\citenamefont {Hayami}\ \emph {et~al.}(2019)\citenamefont {Hayami},
  \citenamefont {Yanagi},\ and\ \citenamefont {Kusunose}}]{Hayami19}%
  \BibitemOpen
  \bibfield  {author} {\bibinfo {author} {\bibfnamefont {S.}~\bibnamefont
  {Hayami}}, \bibinfo {author} {\bibfnamefont {Y.}~\bibnamefont {Yanagi}},\
  and\ \bibinfo {author} {\bibfnamefont {H.}~\bibnamefont {Kusunose}},\
  }\bibfield  {title} {\bibinfo {title} {Momentum-dependent spin splitting by
  collinear antiferromagnetic ordering},\ }\href
  {https://doi.org/10.7566/JPSJ.88.123702} {\bibfield  {journal} {\bibinfo
  {journal} {J. Phys. Soc. Jpn.}\ }\textbf {\bibinfo {volume} {88}},\ \bibinfo
  {pages} {123702} (\bibinfo {year} {2019})}\BibitemShut {NoStop}%
\bibitem [{\citenamefont {Naka}\ \emph {et~al.}(2020)\citenamefont {Naka},
  \citenamefont {Hayami}, \citenamefont {Kusunose}, \citenamefont {Yanagi},
  \citenamefont {Motome},\ and\ \citenamefont {Seo}}]{NakaPRB2020}%
  \BibitemOpen
  \bibfield  {author} {\bibinfo {author} {\bibfnamefont {M.}~\bibnamefont
  {Naka}}, \bibinfo {author} {\bibfnamefont {S.}~\bibnamefont {Hayami}},
  \bibinfo {author} {\bibfnamefont {H.}~\bibnamefont {Kusunose}}, \bibinfo
  {author} {\bibfnamefont {Y.}~\bibnamefont {Yanagi}}, \bibinfo {author}
  {\bibfnamefont {Y.}~\bibnamefont {Motome}},\ and\ \bibinfo {author}
  {\bibfnamefont {H.}~\bibnamefont {Seo}},\ }\bibfield  {title} {\bibinfo
  {title} {Anomalous {H}all effect in $\ensuremath{\kappa}$-type organic
  antiferromagnets},\ }\href {https://doi.org/10.1103/PhysRevB.102.075112}
  {\bibfield  {journal} {\bibinfo  {journal} {Phys. Rev. B}\ }\textbf {\bibinfo
  {volume} {102}},\ \bibinfo {pages} {075112} (\bibinfo {year}
  {2020})}\BibitemShut {NoStop}%
\bibitem [{\citenamefont {Hayami}\ \emph
  {et~al.}(2020{\natexlab{a}})\citenamefont {Hayami}, \citenamefont {Yanagi},\
  and\ \citenamefont {Kusunose}}]{Hayami20}%
  \BibitemOpen
  \bibfield  {author} {\bibinfo {author} {\bibfnamefont {S.}~\bibnamefont
  {Hayami}}, \bibinfo {author} {\bibfnamefont {Y.}~\bibnamefont {Yanagi}},\
  and\ \bibinfo {author} {\bibfnamefont {H.}~\bibnamefont {Kusunose}},\
  }\bibfield  {title} {\bibinfo {title} {Bottom-up design of spin-split and
  reshaped electronic band structures in antiferromagnets without spin-orbit
  coupling: Procedure on the basis of augmented multipoles},\ }\href
  {https://doi.org/10.1103/PhysRevB.102.144441} {\bibfield  {journal} {\bibinfo
   {journal} {Phys. Rev. B}\ }\textbf {\bibinfo {volume} {102}},\ \bibinfo
  {pages} {144441} (\bibinfo {year} {2020}{\natexlab{a}})}\BibitemShut
  {NoStop}%
\bibitem [{\citenamefont {Hayami}\ \emph
  {et~al.}(2020{\natexlab{b}})\citenamefont {Hayami}, \citenamefont {Yanagi},\
  and\ \citenamefont {Kusunose}}]{PhysRevB.101.220403}%
  \BibitemOpen
  \bibfield  {author} {\bibinfo {author} {\bibfnamefont {S.}~\bibnamefont
  {Hayami}}, \bibinfo {author} {\bibfnamefont {Y.}~\bibnamefont {Yanagi}},\
  and\ \bibinfo {author} {\bibfnamefont {H.}~\bibnamefont {Kusunose}},\
  }\bibfield  {title} {\bibinfo {title} {Spontaneous antisymmetric spin
  splitting in noncollinear antiferromagnets without spin-orbit coupling},\
  }\href {https://doi.org/10.1103/PhysRevB.101.220403} {\bibfield  {journal}
  {\bibinfo  {journal} {Phys. Rev. B}\ }\textbf {\bibinfo {volume} {101}},\
  \bibinfo {pages} {220403} (\bibinfo {year} {2020}{\natexlab{b}})}\BibitemShut
  {NoStop}%
\bibitem [{\citenamefont {\ifmmode~\check{S}\else \v{S}\fi{}mejkal}\ \emph
  {et~al.}(2022{\natexlab{b}})\citenamefont {\ifmmode~\check{S}\else
  \v{S}\fi{}mejkal}, \citenamefont {Hellenes}, \citenamefont
  {Gonz\'alez-Hern\'andez}, \citenamefont {Sinova},\ and\ \citenamefont
  {Jungwirth}}]{Libor011028}%
  \BibitemOpen
  \bibfield  {author} {\bibinfo {author} {\bibfnamefont {L.}~\bibnamefont
  {\ifmmode~\check{S}\else \v{S}\fi{}mejkal}}, \bibinfo {author} {\bibfnamefont
  {A.~B.}\ \bibnamefont {Hellenes}}, \bibinfo {author} {\bibfnamefont
  {R.}~\bibnamefont {Gonz\'alez-Hern\'andez}}, \bibinfo {author} {\bibfnamefont
  {J.}~\bibnamefont {Sinova}},\ and\ \bibinfo {author} {\bibfnamefont
  {T.}~\bibnamefont {Jungwirth}},\ }\bibfield  {title} {\bibinfo {title} {Giant
  and tunneling magnetoresistance in unconventional collinear antiferromagnets
  with nonrelativistic spin-momentum coupling},\ }\href
  {https://doi.org/10.1103/PhysRevX.12.011028} {\bibfield  {journal} {\bibinfo
  {journal} {Phys. Rev. X}\ }\textbf {\bibinfo {volume} {12}},\ \bibinfo
  {pages} {011028} (\bibinfo {year} {2022}{\natexlab{b}})}\BibitemShut
  {NoStop}%
\bibitem [{\citenamefont {{Gonzalez Betancourt}}\ \emph
  {et~al.}(2023)\citenamefont {{Gonzalez Betancourt}}, \citenamefont
  {Zub{\'{a}}{\v{c}}}, \citenamefont {Gonzalez-Hernandez}, \citenamefont
  {Geishendorf}, \citenamefont {{\v{S}}ob{\'{a}}ň}, \citenamefont
  {Springholz}, \citenamefont {Olejn{\'{i}}k}, \citenamefont {{\v{S}}mejkal},
  \citenamefont {Sinova}, \citenamefont {Jungwirth}, \citenamefont
  {Goennenwein}, \citenamefont {Thomas}, \citenamefont {Reichlov{\'{a}}},
  \citenamefont {{\v{Z}}elezn{\'{y}}},\ and\ \citenamefont
  {Kriegner}}]{GonzalezBetancourt2023}%
  \BibitemOpen
  \bibfield  {author} {\bibinfo {author} {\bibfnamefont {R.~D.}\ \bibnamefont
  {{Gonzalez Betancourt}}}, \bibinfo {author} {\bibfnamefont {J.}~\bibnamefont
  {Zub{\'{a}}{\v{c}}}}, \bibinfo {author} {\bibfnamefont {R.}~\bibnamefont
  {Gonzalez-Hernandez}}, \bibinfo {author} {\bibfnamefont {K.}~\bibnamefont
  {Geishendorf}}, \bibinfo {author} {\bibfnamefont {Z.}~\bibnamefont
  {{\v{S}}ob{\'{a}}ň}}, \bibinfo {author} {\bibfnamefont {G.}~\bibnamefont
  {Springholz}}, \bibinfo {author} {\bibfnamefont {K.}~\bibnamefont
  {Olejn{\'{i}}k}}, \bibinfo {author} {\bibfnamefont {L.}~\bibnamefont
  {{\v{S}}mejkal}}, \bibinfo {author} {\bibfnamefont {J.}~\bibnamefont
  {Sinova}}, \bibinfo {author} {\bibfnamefont {T.}~\bibnamefont {Jungwirth}},
  \bibinfo {author} {\bibfnamefont {S.~T.~B.}\ \bibnamefont {Goennenwein}},
  \bibinfo {author} {\bibfnamefont {A.}~\bibnamefont {Thomas}}, \bibinfo
  {author} {\bibfnamefont {H.}~\bibnamefont {Reichlov{\'{a}}}}, \bibinfo
  {author} {\bibfnamefont {J.}~\bibnamefont {{\v{Z}}elezn{\'{y}}}},\ and\
  \bibinfo {author} {\bibfnamefont {D.}~\bibnamefont {Kriegner}},\ }\bibfield
  {title} {\bibinfo {title} {{Spontaneous Anomalous Hall Effect Arising from an
  Unconventional Compensated Magnetic Phase in a Semiconductor}},\ }\href
  {https://doi.org/10.1103/PhysRevLett.130.036702} {\bibfield  {journal}
  {\bibinfo  {journal} {Phys. Rev. Lett.}\ }\textbf {\bibinfo {volume} {130}},\
  \bibinfo {pages} {036702} (\bibinfo {year} {2023})}\BibitemShut {NoStop}%
\bibitem [{\citenamefont {Tschirner}\ \emph {et~al.}(2023)\citenamefont
  {Tschirner}, \citenamefont {Keßler}, \citenamefont {Gonzalez~Betancourt},
  \citenamefont {Kotte}, \citenamefont {Kriegner}, \citenamefont {Büchner},
  \citenamefont {Dufouleur}, \citenamefont {Kamp}, \citenamefont {Jovic},
  \citenamefont {Smejkal}, \citenamefont {Sinova}, \citenamefont {Claessen},
  \citenamefont {Jungwirth}, \citenamefont {Moser}, \citenamefont {Reichlova},\
  and\ \citenamefont {Veyrat}}]{Tschirner2023Saturation}%
  \BibitemOpen
  \bibfield  {author} {\bibinfo {author} {\bibfnamefont {T.}~\bibnamefont
  {Tschirner}}, \bibinfo {author} {\bibfnamefont {P.}~\bibnamefont {Keßler}},
  \bibinfo {author} {\bibfnamefont {R.~D.}\ \bibnamefont
  {Gonzalez~Betancourt}}, \bibinfo {author} {\bibfnamefont {T.}~\bibnamefont
  {Kotte}}, \bibinfo {author} {\bibfnamefont {D.}~\bibnamefont {Kriegner}},
  \bibinfo {author} {\bibfnamefont {B.}~\bibnamefont {Büchner}}, \bibinfo
  {author} {\bibfnamefont {J.}~\bibnamefont {Dufouleur}}, \bibinfo {author}
  {\bibfnamefont {M.}~\bibnamefont {Kamp}}, \bibinfo {author} {\bibfnamefont
  {V.}~\bibnamefont {Jovic}}, \bibinfo {author} {\bibfnamefont
  {L.}~\bibnamefont {Smejkal}}, \bibinfo {author} {\bibfnamefont
  {J.}~\bibnamefont {Sinova}}, \bibinfo {author} {\bibfnamefont
  {R.}~\bibnamefont {Claessen}}, \bibinfo {author} {\bibfnamefont
  {T.}~\bibnamefont {Jungwirth}}, \bibinfo {author} {\bibfnamefont
  {S.}~\bibnamefont {Moser}}, \bibinfo {author} {\bibfnamefont
  {H.}~\bibnamefont {Reichlova}},\ and\ \bibinfo {author} {\bibfnamefont
  {L.}~\bibnamefont {Veyrat}},\ }\bibfield  {title} {\bibinfo {title}
  {Saturation of the anomalous {H}all effect at high magnetic fields in
  altermagnetic {RuO$_2$}},\ }\href {https://doi.org/10.1063/5.0160335}
  {\bibfield  {journal} {\bibinfo  {journal} {APL Materials}\ }\textbf
  {\bibinfo {volume} {11}},\ \bibinfo {pages} {101103} (\bibinfo {year}
  {2023})}\BibitemShut {NoStop}%
\bibitem [{\citenamefont {Reichlova}\ \emph {et~al.}(2024)\citenamefont
  {Reichlova}, \citenamefont {{Lopes Seeger}}, \citenamefont
  {Gonz{\'{a}}lez-Hern{\'{a}}ndez}, \citenamefont {Kounta}, \citenamefont
  {Schlitz}, \citenamefont {Kriegner}, \citenamefont {Ritzinger}, \citenamefont
  {Lammel}, \citenamefont {Leivisk{\"{a}}}, \citenamefont {{Birk Hellenes}},
  \citenamefont {Olejn{\'{i}}k}, \citenamefont {Petři{\v{c}}ek}, \citenamefont
  {Dole{\v{z}}al}, \citenamefont {Horak}, \citenamefont {Schmoranzerova},
  \citenamefont {Badura}, \citenamefont {Bertaina}, \citenamefont {Thomas},
  \citenamefont {Baltz}, \citenamefont {Michez}, \citenamefont {Sinova},
  \citenamefont {Goennenwein}, \citenamefont {Jungwirth},\ and\ \citenamefont
  {{\v{S}}mejkal}}]{Reichlova2024}%
  \BibitemOpen
  \bibfield  {author} {\bibinfo {author} {\bibfnamefont {H.}~\bibnamefont
  {Reichlova}}, \bibinfo {author} {\bibfnamefont {R.}~\bibnamefont {{Lopes
  Seeger}}}, \bibinfo {author} {\bibfnamefont {R.}~\bibnamefont
  {Gonz{\'{a}}lez-Hern{\'{a}}ndez}}, \bibinfo {author} {\bibfnamefont
  {I.}~\bibnamefont {Kounta}}, \bibinfo {author} {\bibfnamefont
  {R.}~\bibnamefont {Schlitz}}, \bibinfo {author} {\bibfnamefont
  {D.}~\bibnamefont {Kriegner}}, \bibinfo {author} {\bibfnamefont
  {P.}~\bibnamefont {Ritzinger}}, \bibinfo {author} {\bibfnamefont
  {M.}~\bibnamefont {Lammel}}, \bibinfo {author} {\bibfnamefont
  {M.}~\bibnamefont {Leivisk{\"{a}}}}, \bibinfo {author} {\bibfnamefont
  {A.}~\bibnamefont {{Birk Hellenes}}}, \bibinfo {author} {\bibfnamefont
  {K.}~\bibnamefont {Olejn{\'{i}}k}}, \bibinfo {author} {\bibfnamefont
  {V.}~\bibnamefont {Petři{\v{c}}ek}}, \bibinfo {author} {\bibfnamefont
  {P.}~\bibnamefont {Dole{\v{z}}al}}, \bibinfo {author} {\bibfnamefont
  {L.}~\bibnamefont {Horak}}, \bibinfo {author} {\bibfnamefont
  {E.}~\bibnamefont {Schmoranzerova}}, \bibinfo {author} {\bibfnamefont
  {A.}~\bibnamefont {Badura}}, \bibinfo {author} {\bibfnamefont
  {S.}~\bibnamefont {Bertaina}}, \bibinfo {author} {\bibfnamefont
  {A.}~\bibnamefont {Thomas}}, \bibinfo {author} {\bibfnamefont
  {V.}~\bibnamefont {Baltz}}, \bibinfo {author} {\bibfnamefont
  {L.}~\bibnamefont {Michez}}, \bibinfo {author} {\bibfnamefont
  {J.}~\bibnamefont {Sinova}}, \bibinfo {author} {\bibfnamefont {S.~T.~B.}\
  \bibnamefont {Goennenwein}}, \bibinfo {author} {\bibfnamefont
  {T.}~\bibnamefont {Jungwirth}},\ and\ \bibinfo {author} {\bibfnamefont
  {L.}~\bibnamefont {{\v{S}}mejkal}},\ }\bibfield  {title} {\bibinfo {title}
  {{Observation of a spontaneous anomalous Hall response in the Mn$_5$Si$_3$
  $d$-wave altermagnet candidate}},\ }\href
  {https://doi.org/10.1038/s41467-024-48493-w} {\bibfield  {journal} {\bibinfo
  {journal} {Nat. Commun.}\ }\textbf {\bibinfo {volume} {15}},\ \bibinfo
  {pages} {4961} (\bibinfo {year} {2024})}\BibitemShut {NoStop}%
\bibitem [{\citenamefont {Samanta}\ \emph {et~al.}(2024)\citenamefont
  {Samanta}, \citenamefont {Shao},\ and\ \citenamefont
  {Tsymbal}}]{Samanta2024Spin}%
  \BibitemOpen
  \bibfield  {author} {\bibinfo {author} {\bibfnamefont {K.}~\bibnamefont
  {Samanta}}, \bibinfo {author} {\bibfnamefont {D.-F.}\ \bibnamefont {Shao}},\
  and\ \bibinfo {author} {\bibfnamefont {E.~Y.}\ \bibnamefont {Tsymbal}},\
  }\bibfield  {title} {\bibinfo {title} {Spin filtering with insulating
  altermagnets},\ }\href {https://arxiv.org/abs/2409.00195} {\bibfield
  {journal} {\bibinfo  {journal} {arXiv:2409.00195}\ } (\bibinfo {year}
  {2024})}\BibitemShut {NoStop}%
\bibitem [{\citenamefont {Sun}\ and\ \citenamefont
  {Linder}(2023)}]{Sun2023Spin}%
  \BibitemOpen
  \bibfield  {author} {\bibinfo {author} {\bibfnamefont {C.}~\bibnamefont
  {Sun}}\ and\ \bibinfo {author} {\bibfnamefont {J.}~\bibnamefont {Linder}},\
  }\bibfield  {title} {\bibinfo {title} {Spin pumping from a ferromagnetic
  insulator into an altermagnet},\ }\href
  {https://doi.org/10.1103/PhysRevB.108.L140408} {\bibfield  {journal}
  {\bibinfo  {journal} {Phys. Rev. B}\ }\textbf {\bibinfo {volume} {108}},\
  \bibinfo {pages} {L140408} (\bibinfo {year} {2023})}\BibitemShut {NoStop}%
\bibitem [{\citenamefont {Reja}\ and\ \citenamefont
  {Narayan}(2024)}]{Reja2024}%
  \BibitemOpen
  \bibfield  {author} {\bibinfo {author} {\bibfnamefont {M.~A.}\ \bibnamefont
  {Reja}}\ and\ \bibinfo {author} {\bibfnamefont {A.}~\bibnamefont {Narayan}},\
  }\bibfield  {title} {\bibinfo {title} {{Emergence of tunable exceptional
  points in altermagnet-ferromagnet junctions}},\ }\href
  {https://doi.org/10.1103/PhysRevB.110.235401} {\bibfield  {journal} {\bibinfo
   {journal} {Phys. Rev. B}\ }\textbf {\bibinfo {volume} {110}},\ \bibinfo
  {pages} {235401} (\bibinfo {year} {2024})}\BibitemShut {NoStop}%
\bibitem [{\citenamefont {Werner}\ \emph {et~al.}(2024)\citenamefont {Werner},
  \citenamefont {Lysne},\ and\ \citenamefont {Murakami}}]{Werner2024High}%
  \BibitemOpen
  \bibfield  {author} {\bibinfo {author} {\bibfnamefont {P.}~\bibnamefont
  {Werner}}, \bibinfo {author} {\bibfnamefont {M.}~\bibnamefont {Lysne}},\ and\
  \bibinfo {author} {\bibfnamefont {Y.}~\bibnamefont {Murakami}},\ }\bibfield
  {title} {\bibinfo {title} {High harmonic generation in altermagnets},\ }\href
  {https://doi.org/10.1103/PhysRevB.110.235101} {\bibfield  {journal} {\bibinfo
   {journal} {Phys. Rev. B}\ }\textbf {\bibinfo {volume} {110}},\ \bibinfo
  {pages} {235101} (\bibinfo {year} {2024})}\BibitemShut {NoStop}%
\bibitem [{\citenamefont {Farajollahpour}\ \emph {et~al.}(2025)\citenamefont
  {Farajollahpour}, \citenamefont {Ganesh},\ and\ \citenamefont
  {Samokhin}}]{Farajollahpour2025Light}%
  \BibitemOpen
  \bibfield  {author} {\bibinfo {author} {\bibfnamefont {T.}~\bibnamefont
  {Farajollahpour}}, \bibinfo {author} {\bibfnamefont {R.}~\bibnamefont
  {Ganesh}},\ and\ \bibinfo {author} {\bibfnamefont {K.}~\bibnamefont
  {Samokhin}},\ }\bibfield  {title} {\bibinfo {title} {Light-induced charge and
  spin hall currents in materials with {$C_4K$} symmetry},\ }\href
  {https://doi.org/10.1038/s41535-025-00746-7} {\bibfield  {journal} {\bibinfo
  {journal} {npj Quantum Materials}\ }\textbf {\bibinfo {volume} {10}},\
  \bibinfo {pages} {29} (\bibinfo {year} {2025})}\BibitemShut {NoStop}%
\bibitem [{\citenamefont {Ezawa}(2025)}]{Ezawa2025Third}%
  \BibitemOpen
  \bibfield  {author} {\bibinfo {author} {\bibfnamefont {M.}~\bibnamefont
  {Ezawa}},\ }\bibfield  {title} {\bibinfo {title} {Third-order and fifth-order
  nonlinear spin-current generation in $g$-wave and $i$-wave altermagnets and
  perfectly nonreciprocal spin current in $f$-wave magnets},\ }\href
  {https://doi.org/10.1103/PhysRevB.111.125420} {\bibfield  {journal} {\bibinfo
   {journal} {Phys. Rev. B}\ }\textbf {\bibinfo {volume} {111}},\ \bibinfo
  {pages} {125420} (\bibinfo {year} {2025})}\BibitemShut {NoStop}%
\bibitem [{\citenamefont {Fu}\ \emph {et~al.}(2025{\natexlab{a}})\citenamefont
  {Fu}, \citenamefont {Lv}, \citenamefont {Xu}, \citenamefont {Cayao},
  \citenamefont {Liu},\ and\ \citenamefont {Yu}}]{Fu2025All}%
  \BibitemOpen
  \bibfield  {author} {\bibinfo {author} {\bibfnamefont {P.}~\bibnamefont
  {Fu}}, \bibinfo {author} {\bibfnamefont {Q.}~\bibnamefont {Lv}}, \bibinfo
  {author} {\bibfnamefont {Y.}~\bibnamefont {Xu}}, \bibinfo {author}
  {\bibfnamefont {J.}~\bibnamefont {Cayao}}, \bibinfo {author} {\bibfnamefont
  {J.}~\bibnamefont {Liu}},\ and\ \bibinfo {author} {\bibfnamefont
  {X.}~\bibnamefont {Yu}},\ }\bibfield  {title} {\bibinfo {title}
  {All‑electrically controlled spintronics in altermagnetic
  heterostructures},\ }\href {https://doi.org/10.1038/s41535-025-00827-7}
  {\bibfield  {journal} {\bibinfo  {journal} {npj Quantum Mater.}\ }\textbf
  {\bibinfo {volume} {10}},\ \bibinfo {pages} {111} (\bibinfo {year}
  {2025}{\natexlab{a}})}\BibitemShut {NoStop}%
\bibitem [{\citenamefont {Yang}\ \emph {et~al.}(2025)\citenamefont {Yang},
  \citenamefont {Yang}, \citenamefont {Wang}, \citenamefont {Li}, \citenamefont
  {Peng}, \citenamefont {Lee}, \citenamefont {Ang}, \citenamefont {Lu},
  \citenamefont {Ang},\ and\ \citenamefont {Fang}}]{Yang2025Unconventional}%
  \BibitemOpen
  \bibfield  {author} {\bibinfo {author} {\bibfnamefont {Z.}~\bibnamefont
  {Yang}}, \bibinfo {author} {\bibfnamefont {X.}~\bibnamefont {Yang}}, \bibinfo
  {author} {\bibfnamefont {J.}~\bibnamefont {Wang}}, \bibinfo {author}
  {\bibfnamefont {Q.}~\bibnamefont {Li}}, \bibinfo {author} {\bibfnamefont
  {R.}~\bibnamefont {Peng}}, \bibinfo {author} {\bibfnamefont {C.~H.}\
  \bibnamefont {Lee}}, \bibinfo {author} {\bibfnamefont {L.~K.}\ \bibnamefont
  {Ang}}, \bibinfo {author} {\bibfnamefont {J.}~\bibnamefont {Lu}}, \bibinfo
  {author} {\bibfnamefont {Y.~S.}\ \bibnamefont {Ang}},\ and\ \bibinfo {author}
  {\bibfnamefont {S.}~\bibnamefont {Fang}},\ }\bibfield  {title} {\bibinfo
  {title} {Unconventional thickness scaling of coherent tunnel
  magnetoresistance in altermagnets},\ }\href
  {https://doi.org/10.1103/2thy-fzzj} {\bibfield  {journal} {\bibinfo
  {journal} {Phys. Rev. B}\ }\textbf {\bibinfo {volume} {112}},\ \bibinfo
  {pages} {205202} (\bibinfo {year} {2025})}\BibitemShut {NoStop}%
\bibitem [{\citenamefont {Peng}\ \emph {et~al.}(2025)\citenamefont {Peng},
  \citenamefont {Fang}, \citenamefont {Ho}, \citenamefont {Liu}, \citenamefont
  {Zhou}, \citenamefont {Liu},\ and\ \citenamefont
  {Ang}}]{Peng2025Ferroelastic}%
  \BibitemOpen
  \bibfield  {author} {\bibinfo {author} {\bibfnamefont {R.}~\bibnamefont
  {Peng}}, \bibinfo {author} {\bibfnamefont {S.}~\bibnamefont {Fang}}, \bibinfo
  {author} {\bibfnamefont {P.}~\bibnamefont {Ho}}, \bibinfo {author}
  {\bibfnamefont {F.}~\bibnamefont {Liu}}, \bibinfo {author} {\bibfnamefont
  {T.}~\bibnamefont {Zhou}}, \bibinfo {author} {\bibfnamefont {J.}~\bibnamefont
  {Liu}},\ and\ \bibinfo {author} {\bibfnamefont {Y.~S.}\ \bibnamefont {Ang}},\
  }\bibfield  {title} {\bibinfo {title} {Ferroelastic altermagnetism},\ }\href
  {https://www.nature.com/articles/s41535-025-00835-7} {\bibfield  {journal}
  {\bibinfo  {journal} {npj Quantum Mater.}\ }\textbf {\bibinfo {volume}
  {11}},\ \bibinfo {pages} {5} (\bibinfo {year} {2025})}\BibitemShut {NoStop}%
\bibitem [{\citenamefont {Zhu}\ \emph {et~al.}(2025)\citenamefont {Zhu},
  \citenamefont {Duan}, \citenamefont {Zhang}, \citenamefont {Hao},
  \citenamefont {{\v{Z}}uti{\'c}},\ and\ \citenamefont
  {Zhou}}]{Zhu2025TwoDimensional}%
  \BibitemOpen
  \bibfield  {author} {\bibinfo {author} {\bibfnamefont {Z.}~\bibnamefont
  {Zhu}}, \bibinfo {author} {\bibfnamefont {X.}~\bibnamefont {Duan}}, \bibinfo
  {author} {\bibfnamefont {J.}~\bibnamefont {Zhang}}, \bibinfo {author}
  {\bibfnamefont {B.}~\bibnamefont {Hao}}, \bibinfo {author} {\bibfnamefont
  {I.}~\bibnamefont {{\v{Z}}uti{\'c}}},\ and\ \bibinfo {author} {\bibfnamefont
  {T.}~\bibnamefont {Zhou}},\ }\bibfield  {title} {\bibinfo {title}
  {Two‑dimensional ferroelectric altermagnets: From model to material
  realization},\ }\href {https://doi.org/10.1021/acs.nanolett.5c02121}
  {\bibfield  {journal} {\bibinfo  {journal} {Nano Lett.}\ }\textbf {\bibinfo
  {volume} {25}},\ \bibinfo {pages} {9456} (\bibinfo {year}
  {2025})}\BibitemShut {NoStop}%
\bibitem [{\citenamefont {Papaj}(2023)}]{PhysRevB.108.L060508}%
  \BibitemOpen
  \bibfield  {author} {\bibinfo {author} {\bibfnamefont {M.}~\bibnamefont
  {Papaj}},\ }\bibfield  {title} {\bibinfo {title} {Andreev reflection at the
  altermagnet-superconductor interface},\ }\href
  {https://doi.org/10.1103/PhysRevB.108.L060508} {\bibfield  {journal}
  {\bibinfo  {journal} {Phys. Rev. B}\ }\textbf {\bibinfo {volume} {108}},\
  \bibinfo {pages} {L060508} (\bibinfo {year} {2023})}\BibitemShut {NoStop}%
\bibitem [{\citenamefont {Sun}\ \emph {et~al.}(2023)\citenamefont {Sun},
  \citenamefont {Brataas},\ and\ \citenamefont {Linder}}]{PhysRevB.108.054511}%
  \BibitemOpen
  \bibfield  {author} {\bibinfo {author} {\bibfnamefont {C.}~\bibnamefont
  {Sun}}, \bibinfo {author} {\bibfnamefont {A.}~\bibnamefont {Brataas}},\ and\
  \bibinfo {author} {\bibfnamefont {J.}~\bibnamefont {Linder}},\ }\bibfield
  {title} {\bibinfo {title} {Andreev reflection in altermagnets},\ }\href
  {https://doi.org/10.1103/PhysRevB.108.054511} {\bibfield  {journal} {\bibinfo
   {journal} {Phys. Rev. B}\ }\textbf {\bibinfo {volume} {108}},\ \bibinfo
  {pages} {054511} (\bibinfo {year} {2023})}\BibitemShut {NoStop}%
\bibitem [{\citenamefont {Maeda}\ \emph {et~al.}(2024)\citenamefont {Maeda},
  \citenamefont {Lu}, \citenamefont {Yada},\ and\ \citenamefont
  {Tanaka}}]{Maeda_2024}%
  \BibitemOpen
  \bibfield  {author} {\bibinfo {author} {\bibfnamefont {K.}~\bibnamefont
  {Maeda}}, \bibinfo {author} {\bibfnamefont {B.}~\bibnamefont {Lu}}, \bibinfo
  {author} {\bibfnamefont {K.}~\bibnamefont {Yada}},\ and\ \bibinfo {author}
  {\bibfnamefont {Y.}~\bibnamefont {Tanaka}},\ }\bibfield  {title} {\bibinfo
  {title} {Theory of tunneling spectroscopy in unconventional p-wave
  magnet-superconductor hybrid structures},\ }\href
  {http://dx.doi.org/10.7566/JPSJ.93.114703} {\bibfield  {journal} {\bibinfo
  {journal} {J. Phys. Soc. Jpn.}\ }\textbf {\bibinfo {volume} {93}} (\bibinfo
  {year} {2024})}\BibitemShut {NoStop}%
\bibitem [{\citenamefont {Ouassou}\ \emph {et~al.}(2023)\citenamefont
  {Ouassou}, \citenamefont {Brataas},\ and\ \citenamefont
  {Linder}}]{PhysRevLett.131.076003}%
  \BibitemOpen
  \bibfield  {author} {\bibinfo {author} {\bibfnamefont {J.~A.}\ \bibnamefont
  {Ouassou}}, \bibinfo {author} {\bibfnamefont {A.}~\bibnamefont {Brataas}},\
  and\ \bibinfo {author} {\bibfnamefont {J.}~\bibnamefont {Linder}},\
  }\bibfield  {title} {\bibinfo {title} {dc {J}osephson effect in
  altermagnets},\ }\href {https://doi.org/10.1103/PhysRevLett.131.076003}
  {\bibfield  {journal} {\bibinfo  {journal} {Phys. Rev. Lett.}\ }\textbf
  {\bibinfo {volume} {131}},\ \bibinfo {pages} {076003} (\bibinfo {year}
  {2023})}\BibitemShut {NoStop}%
\bibitem [{\citenamefont {Lu}\ \emph {et~al.}(2024)\citenamefont {Lu},
  \citenamefont {Maeda}, \citenamefont {Ito}, \citenamefont {Yada},\ and\
  \citenamefont {Tanaka}}]{PhysRevLett.133.226002}%
  \BibitemOpen
  \bibfield  {author} {\bibinfo {author} {\bibfnamefont {B.}~\bibnamefont
  {Lu}}, \bibinfo {author} {\bibfnamefont {K.}~\bibnamefont {Maeda}}, \bibinfo
  {author} {\bibfnamefont {H.}~\bibnamefont {Ito}}, \bibinfo {author}
  {\bibfnamefont {K.}~\bibnamefont {Yada}},\ and\ \bibinfo {author}
  {\bibfnamefont {Y.}~\bibnamefont {Tanaka}},\ }\bibfield  {title} {\bibinfo
  {title} {$\ensuremath{\varphi}$ {J}osephson junction induced by
  altermagnetism},\ }\href {https://doi.org/10.1103/PhysRevLett.133.226002}
  {\bibfield  {journal} {\bibinfo  {journal} {Phys. Rev. Lett.}\ }\textbf
  {\bibinfo {volume} {133}},\ \bibinfo {pages} {226002} (\bibinfo {year}
  {2024})}\BibitemShut {NoStop}%
\bibitem [{\citenamefont {Brekke}\ \emph
  {et~al.}(2023{\natexlab{a}})\citenamefont {Brekke}, \citenamefont {Brataas},\
  and\ \citenamefont {Sudb\o{}}}]{PhysRevB.108.224421}%
  \BibitemOpen
  \bibfield  {author} {\bibinfo {author} {\bibfnamefont {B.}~\bibnamefont
  {Brekke}}, \bibinfo {author} {\bibfnamefont {A.}~\bibnamefont {Brataas}},\
  and\ \bibinfo {author} {\bibfnamefont {A.}~\bibnamefont {Sudb\o{}}},\
  }\bibfield  {title} {\bibinfo {title} {Two-dimensional altermagnets:
  Superconductivity in a minimal microscopic model},\ }\href
  {https://doi.org/10.1103/PhysRevB.108.224421} {\bibfield  {journal} {\bibinfo
   {journal} {Phys. Rev. B}\ }\textbf {\bibinfo {volume} {108}},\ \bibinfo
  {pages} {224421} (\bibinfo {year} {2023}{\natexlab{a}})}\BibitemShut
  {NoStop}%
\bibitem [{\citenamefont {Das}\ and\ \citenamefont
  {Soori}(2024)}]{PhysRevB.109.245424}%
  \BibitemOpen
  \bibfield  {author} {\bibinfo {author} {\bibfnamefont {S.}~\bibnamefont
  {Das}}\ and\ \bibinfo {author} {\bibfnamefont {A.}~\bibnamefont {Soori}},\
  }\bibfield  {title} {\bibinfo {title} {Crossed {A}ndreev reflection in
  altermagnets},\ }\href {https://doi.org/10.1103/PhysRevB.109.245424}
  {\bibfield  {journal} {\bibinfo  {journal} {Phys. Rev. B}\ }\textbf {\bibinfo
  {volume} {109}},\ \bibinfo {pages} {245424} (\bibinfo {year}
  {2024})}\BibitemShut {NoStop}%
\bibitem [{\citenamefont {Chakraborty}\ and\ \citenamefont
  {Black-Schaffer}(2024{\natexlab{a}})}]{PhysRevB.110.L060508}%
  \BibitemOpen
  \bibfield  {author} {\bibinfo {author} {\bibfnamefont {D.}~\bibnamefont
  {Chakraborty}}\ and\ \bibinfo {author} {\bibfnamefont {A.~M.}\ \bibnamefont
  {Black-Schaffer}},\ }\bibfield  {title} {\bibinfo {title} {Zero-field
  finite-momentum and field-induced superconductivity in altermagnets},\ }\href
  {https://doi.org/10.1103/PhysRevB.110.L060508} {\bibfield  {journal}
  {\bibinfo  {journal} {Phys. Rev. B}\ }\textbf {\bibinfo {volume} {110}},\
  \bibinfo {pages} {L060508} (\bibinfo {year}
  {2024}{\natexlab{a}})}\BibitemShut {NoStop}%
\bibitem [{\citenamefont {Nagae}\ \emph
  {et~al.}(2025{\natexlab{a}})\citenamefont {Nagae}, \citenamefont {Schnyder},\
  and\ \citenamefont {Ikegaya}}]{PhysRevB.111.L100507}%
  \BibitemOpen
  \bibfield  {author} {\bibinfo {author} {\bibfnamefont {Y.}~\bibnamefont
  {Nagae}}, \bibinfo {author} {\bibfnamefont {A.~P.}\ \bibnamefont
  {Schnyder}},\ and\ \bibinfo {author} {\bibfnamefont {S.}~\bibnamefont
  {Ikegaya}},\ }\bibfield  {title} {\bibinfo {title} {Spin-polarized specular
  {A}ndreev reflections in altermagnets},\ }\href
  {https://doi.org/10.1103/PhysRevB.111.L100507} {\bibfield  {journal}
  {\bibinfo  {journal} {Phys. Rev. B}\ }\textbf {\bibinfo {volume} {111}},\
  \bibinfo {pages} {L100507} (\bibinfo {year}
  {2025}{\natexlab{a}})}\BibitemShut {NoStop}%
\bibitem [{\citenamefont {Sukhachov}\ \emph {et~al.}(2025)\citenamefont
  {Sukhachov}, \citenamefont {Giil}, \citenamefont {Brekke},\ and\
  \citenamefont {Linder}}]{sukhachov2025}%
  \BibitemOpen
  \bibfield  {author} {\bibinfo {author} {\bibfnamefont {P.}~\bibnamefont
  {Sukhachov}}, \bibinfo {author} {\bibfnamefont {H.~G.}\ \bibnamefont {Giil}},
  \bibinfo {author} {\bibfnamefont {B.}~\bibnamefont {Brekke}},\ and\ \bibinfo
  {author} {\bibfnamefont {J.}~\bibnamefont {Linder}},\ }\bibfield  {title}
  {\bibinfo {title} {Coexistence of $p$-wave magnetism and superconductivity},\
  }\href {https://doi.org/10.1103/PhysRevB.111.L220403} {\bibfield  {journal}
  {\bibinfo  {journal} {Phys. Rev. B}\ }\textbf {\bibinfo {volume} {111}},\
  \bibinfo {pages} {L220403} (\bibinfo {year} {2025})}\BibitemShut {NoStop}%
\bibitem [{\citenamefont {Ghorashi}\ \emph {et~al.}(2024)\citenamefont
  {Ghorashi}, \citenamefont {Hughes},\ and\ \citenamefont
  {Cano}}]{PhysRevLett.133.106601}%
  \BibitemOpen
  \bibfield  {author} {\bibinfo {author} {\bibfnamefont {S.~A.~A.}\
  \bibnamefont {Ghorashi}}, \bibinfo {author} {\bibfnamefont {T.~L.}\
  \bibnamefont {Hughes}},\ and\ \bibinfo {author} {\bibfnamefont
  {J.}~\bibnamefont {Cano}},\ }\bibfield  {title} {\bibinfo {title}
  {Altermagnetic routes to {M}ajorana modes in zero net magnetization},\ }\href
  {https://doi.org/10.1103/PhysRevLett.133.106601} {\bibfield  {journal}
  {\bibinfo  {journal} {Phys. Rev. Lett.}\ }\textbf {\bibinfo {volume} {133}},\
  \bibinfo {pages} {106601} (\bibinfo {year} {2024})}\BibitemShut {NoStop}%
\bibitem [{\citenamefont {Mondal}\ \emph {et~al.}(2025)\citenamefont {Mondal},
  \citenamefont {Pal}, \citenamefont {Saha},\ and\ \citenamefont
  {Nag}}]{PhysRevB.111.L121401}%
  \BibitemOpen
  \bibfield  {author} {\bibinfo {author} {\bibfnamefont {D.}~\bibnamefont
  {Mondal}}, \bibinfo {author} {\bibfnamefont {A.}~\bibnamefont {Pal}},
  \bibinfo {author} {\bibfnamefont {A.}~\bibnamefont {Saha}},\ and\ \bibinfo
  {author} {\bibfnamefont {T.}~\bibnamefont {Nag}},\ }\bibfield  {title}
  {\bibinfo {title} {Distinguishing between topological {M}ajorana and trivial
  zero modes via transport and shot noise study in an altermagnet
  heterostructure},\ }\href {https://doi.org/10.1103/PhysRevB.111.L121401}
  {\bibfield  {journal} {\bibinfo  {journal} {Phys. Rev. B}\ }\textbf {\bibinfo
  {volume} {111}},\ \bibinfo {pages} {L121401} (\bibinfo {year}
  {2025})}\BibitemShut {NoStop}%
\bibitem [{\citenamefont {Li}\ and\ \citenamefont
  {Liu}(2023)}]{PhysRevB.108.205410}%
  \BibitemOpen
  \bibfield  {author} {\bibinfo {author} {\bibfnamefont {Y.-X.}\ \bibnamefont
  {Li}}\ and\ \bibinfo {author} {\bibfnamefont {C.-C.}\ \bibnamefont {Liu}},\
  }\bibfield  {title} {\bibinfo {title} {Majorana corner modes and tunable
  patterns in an altermagnet heterostructure},\ }\href
  {https://doi.org/10.1103/PhysRevB.108.205410} {\bibfield  {journal} {\bibinfo
   {journal} {Phys. Rev. B}\ }\textbf {\bibinfo {volume} {108}},\ \bibinfo
  {pages} {205410} (\bibinfo {year} {2023})}\BibitemShut {NoStop}%
\bibitem [{\citenamefont {Zhao}\ \emph {et~al.}(2025)\citenamefont {Zhao},
  \citenamefont {Fukaya}, \citenamefont {Burset}, \citenamefont {Cayao},
  \citenamefont {Tanaka},\ and\ \citenamefont {Lu}}]{PhysRevB.111.184515}%
  \BibitemOpen
  \bibfield  {author} {\bibinfo {author} {\bibfnamefont {W.}~\bibnamefont
  {Zhao}}, \bibinfo {author} {\bibfnamefont {Y.}~\bibnamefont {Fukaya}},
  \bibinfo {author} {\bibfnamefont {P.}~\bibnamefont {Burset}}, \bibinfo
  {author} {\bibfnamefont {J.}~\bibnamefont {Cayao}}, \bibinfo {author}
  {\bibfnamefont {Y.}~\bibnamefont {Tanaka}},\ and\ \bibinfo {author}
  {\bibfnamefont {B.}~\bibnamefont {Lu}},\ }\bibfield  {title} {\bibinfo
  {title} {Orientation-dependent transport in junctions formed by $d$-wave
  altermagnets and $d$-wave superconductors},\ }\href
  {https://doi.org/10.1103/PhysRevB.111.184515} {\bibfield  {journal} {\bibinfo
   {journal} {Phys. Rev. B}\ }\textbf {\bibinfo {volume} {111}},\ \bibinfo
  {pages} {184515} (\bibinfo {year} {2025})}\BibitemShut {NoStop}%
\bibitem [{\citenamefont {Fukaya}\ \emph
  {et~al.}(2025{\natexlab{b}})\citenamefont {Fukaya}, \citenamefont {Maeda},
  \citenamefont {Yada}, \citenamefont {Cayao}, \citenamefont {Tanaka},\ and\
  \citenamefont {Lu}}]{PhysRevB.111.064502}%
  \BibitemOpen
  \bibfield  {author} {\bibinfo {author} {\bibfnamefont {Y.}~\bibnamefont
  {Fukaya}}, \bibinfo {author} {\bibfnamefont {K.}~\bibnamefont {Maeda}},
  \bibinfo {author} {\bibfnamefont {K.}~\bibnamefont {Yada}}, \bibinfo {author}
  {\bibfnamefont {J.}~\bibnamefont {Cayao}}, \bibinfo {author} {\bibfnamefont
  {Y.}~\bibnamefont {Tanaka}},\ and\ \bibinfo {author} {\bibfnamefont
  {B.}~\bibnamefont {Lu}},\ }\bibfield  {title} {\bibinfo {title} {Josephson
  effect and odd-frequency pairing in superconducting junctions with
  unconventional magnets},\ }\href
  {https://doi.org/10.1103/PhysRevB.111.064502} {\bibfield  {journal} {\bibinfo
   {journal} {Phys. Rev. B}\ }\textbf {\bibinfo {volume} {111}},\ \bibinfo
  {pages} {064502} (\bibinfo {year} {2025}{\natexlab{b}})}\BibitemShut
  {NoStop}%
\bibitem [{\citenamefont {Sun}\ \emph {et~al.}(2025{\natexlab{a}})\citenamefont
  {Sun}, \citenamefont {Zhang}, \citenamefont {Li},\ and\ \citenamefont
  {Trauzettel}}]{PhysRevB.111.165406}%
  \BibitemOpen
  \bibfield  {author} {\bibinfo {author} {\bibfnamefont {H.-P.}\ \bibnamefont
  {Sun}}, \bibinfo {author} {\bibfnamefont {S.-B.}\ \bibnamefont {Zhang}},
  \bibinfo {author} {\bibfnamefont {C.-A.}\ \bibnamefont {Li}},\ and\ \bibinfo
  {author} {\bibfnamefont {B.}~\bibnamefont {Trauzettel}},\ }\bibfield  {title}
  {\bibinfo {title} {Tunable second harmonic in altermagnetic {J}osephson
  junctions},\ }\href {https://doi.org/10.1103/PhysRevB.111.165406} {\bibfield
  {journal} {\bibinfo  {journal} {Phys. Rev. B}\ }\textbf {\bibinfo {volume}
  {111}},\ \bibinfo {pages} {165406} (\bibinfo {year}
  {2025}{\natexlab{a}})}\BibitemShut {NoStop}%
\bibitem [{\citenamefont {Hu}\ \emph {et~al.}(2025)\citenamefont {Hu},
  \citenamefont {Matsyshyn},\ and\ \citenamefont
  {Song}}]{PhysRevLett.134.026001}%
  \BibitemOpen
  \bibfield  {author} {\bibinfo {author} {\bibfnamefont {J.-X.}\ \bibnamefont
  {Hu}}, \bibinfo {author} {\bibfnamefont {O.}~\bibnamefont {Matsyshyn}},\ and\
  \bibinfo {author} {\bibfnamefont {J.~C.~W.}\ \bibnamefont {Song}},\
  }\bibfield  {title} {\bibinfo {title} {Nonlinear superconducting
  magnetoelectric effect},\ }\href
  {https://doi.org/10.1103/PhysRevLett.134.026001} {\bibfield  {journal}
  {\bibinfo  {journal} {Phys. Rev. Lett.}\ }\textbf {\bibinfo {volume} {134}},\
  \bibinfo {pages} {026001} (\bibinfo {year} {2025})}\BibitemShut {NoStop}%
\bibitem [{\citenamefont {Chakraborty}\ and\ \citenamefont
  {Black-Schaffer}(2024{\natexlab{b}})}]{chakraborty2024constr}%
  \BibitemOpen
  \bibfield  {author} {\bibinfo {author} {\bibfnamefont {D.}~\bibnamefont
  {Chakraborty}}\ and\ \bibinfo {author} {\bibfnamefont {A.~M.}\ \bibnamefont
  {Black-Schaffer}},\ }\bibfield  {title} {\bibinfo {title} {Constraints on
  superconducting pairing in altermagnets},\ }\href
  {https://arxiv.org/abs/2408.03999} {\bibfield  {journal} {\bibinfo  {journal}
  {arXiv:2408.03999}\ } (\bibinfo {year} {2024}{\natexlab{b}})}\BibitemShut
  {NoStop}%
\bibitem [{\citenamefont {Maeda}\ \emph {et~al.}(2025)\citenamefont {Maeda},
  \citenamefont {Fukaya}, \citenamefont {Yada}, \citenamefont {Lu},
  \citenamefont {Tanaka},\ and\ \citenamefont {Cayao}}]{maeda2025pair}%
  \BibitemOpen
  \bibfield  {author} {\bibinfo {author} {\bibfnamefont {K.}~\bibnamefont
  {Maeda}}, \bibinfo {author} {\bibfnamefont {Y.}~\bibnamefont {Fukaya}},
  \bibinfo {author} {\bibfnamefont {K.}~\bibnamefont {Yada}}, \bibinfo {author}
  {\bibfnamefont {B.}~\bibnamefont {Lu}}, \bibinfo {author} {\bibfnamefont
  {Y.}~\bibnamefont {Tanaka}},\ and\ \bibinfo {author} {\bibfnamefont
  {J.}~\bibnamefont {Cayao}},\ }\bibfield  {title} {\bibinfo {title}
  {Classification of pair symmetries in superconductors with unconventional
  magnetism},\ }\href {https://doi.org/10.1103/PhysRevB.111.144508} {\bibfield
  {journal} {\bibinfo  {journal} {Phys. Rev. B}\ }\textbf {\bibinfo {volume}
  {111}},\ \bibinfo {pages} {144508} (\bibinfo {year} {2025})}\BibitemShut
  {NoStop}%
\bibitem [{\citenamefont {Zyuzin}(2024)}]{PhysRevB.109.L220505}%
  \BibitemOpen
  \bibfield  {author} {\bibinfo {author} {\bibfnamefont {A.~A.}\ \bibnamefont
  {Zyuzin}},\ }\bibfield  {title} {\bibinfo {title} {Magnetoelectric effect in
  superconductors with $d$-wave magnetization},\ }\href
  {https://doi.org/10.1103/PhysRevB.109.L220505} {\bibfield  {journal}
  {\bibinfo  {journal} {Phys. Rev. B}\ }\textbf {\bibinfo {volume} {109}},\
  \bibinfo {pages} {L220505} (\bibinfo {year} {2024})}\BibitemShut {NoStop}%
\bibitem [{\citenamefont {Chatterjee}\ and\ \citenamefont {Juri\ifmmode
  \check{c}\else \v{c}\fi{}i\ifmmode~\acute{c}\else
  \'{c}\fi{}}(2025)}]{chatterjee2025inter}%
  \BibitemOpen
  \bibfield  {author} {\bibinfo {author} {\bibfnamefont {P.}~\bibnamefont
  {Chatterjee}}\ and\ \bibinfo {author} {\bibfnamefont {V.}~\bibnamefont
  {Juri\ifmmode \check{c}\else \v{c}\fi{}i\ifmmode~\acute{c}\else
  \'{c}\fi{}}},\ }\bibfield  {title} {\bibinfo {title} {Interplay between
  altermagnetism and topological superconductivity on an unconventional
  superconducting platform},\ }\href {https://doi.org/10.1103/4318-ttvf}
  {\bibfield  {journal} {\bibinfo  {journal} {Phys. Rev. B}\ }\textbf {\bibinfo
  {volume} {112}},\ \bibinfo {pages} {054503} (\bibinfo {year}
  {2025})}\BibitemShut {NoStop}%
\bibitem [{\citenamefont {Beenakker}\ and\ \citenamefont
  {Vakhtel}(2023)}]{PhysRevB.108.075425}%
  \BibitemOpen
  \bibfield  {author} {\bibinfo {author} {\bibfnamefont {C.~W.~J.}\
  \bibnamefont {Beenakker}}\ and\ \bibinfo {author} {\bibfnamefont
  {T.}~\bibnamefont {Vakhtel}},\ }\bibfield  {title} {\bibinfo {title}
  {Phase-shifted {A}ndreev levels in an altermagnet {J}osephson junction},\
  }\href {https://doi.org/10.1103/PhysRevB.108.075425} {\bibfield  {journal}
  {\bibinfo  {journal} {Phys. Rev. B}\ }\textbf {\bibinfo {volume} {108}},\
  \bibinfo {pages} {075425} (\bibinfo {year} {2023})}\BibitemShut {NoStop}%
\bibitem [{\citenamefont {Sun}\ \emph {et~al.}(2025{\natexlab{b}})\citenamefont
  {Sun}, \citenamefont {Feng}, \citenamefont {Xie}, \citenamefont {Zhou},
  \citenamefont {Hu},\ and\ \citenamefont {Law}}]{sun2025pseudopwave}%
  \BibitemOpen
  \bibfield  {author} {\bibinfo {author} {\bibfnamefont {Z.-T.}\ \bibnamefont
  {Sun}}, \bibinfo {author} {\bibfnamefont {X.}~\bibnamefont {Feng}}, \bibinfo
  {author} {\bibfnamefont {Y.-M.}\ \bibnamefont {Xie}}, \bibinfo {author}
  {\bibfnamefont {B.~T.}\ \bibnamefont {Zhou}}, \bibinfo {author}
  {\bibfnamefont {J.-X.}\ \bibnamefont {Hu}},\ and\ \bibinfo {author}
  {\bibfnamefont {K.~T.}\ \bibnamefont {Law}},\ }\bibfield  {title} {\bibinfo
  {title} {Pseudo-{I}sing superconductivity induced by $p$-wave magnetism},\
  }\href {https://doi.org/10.1103/cm1l-1rsh} {\bibfield  {journal} {\bibinfo
  {journal} {Phys. Rev. B}\ }\textbf {\bibinfo {volume} {112}},\ \bibinfo
  {pages} {214504} (\bibinfo {year} {2025}{\natexlab{b}})}\BibitemShut
  {NoStop}%
\bibitem [{\citenamefont {Banerjee}\ and\ \citenamefont
  {Scheurer}(2024)}]{PhysRevB.110.024503}%
  \BibitemOpen
  \bibfield  {author} {\bibinfo {author} {\bibfnamefont {S.}~\bibnamefont
  {Banerjee}}\ and\ \bibinfo {author} {\bibfnamefont {M.~S.}\ \bibnamefont
  {Scheurer}},\ }\bibfield  {title} {\bibinfo {title} {Altermagnetic
  superconducting diode effect},\ }\href
  {https://doi.org/10.1103/PhysRevB.110.024503} {\bibfield  {journal} {\bibinfo
   {journal} {Phys. Rev. B}\ }\textbf {\bibinfo {volume} {110}},\ \bibinfo
  {pages} {024503} (\bibinfo {year} {2024})}\BibitemShut {NoStop}%
\bibitem [{\citenamefont {Nagae}\ \emph
  {et~al.}(2025{\natexlab{b}})\citenamefont {Nagae}, \citenamefont {Katayama},\
  and\ \citenamefont {Ikegaya}}]{Nagae2025Flat}%
  \BibitemOpen
  \bibfield  {author} {\bibinfo {author} {\bibfnamefont {Y.}~\bibnamefont
  {Nagae}}, \bibinfo {author} {\bibfnamefont {L.}~\bibnamefont {Katayama}},\
  and\ \bibinfo {author} {\bibfnamefont {S.}~\bibnamefont {Ikegaya}},\
  }\bibfield  {title} {\bibinfo {title} {Flat-band zero-energy states and
  anomalous proximity effects in $p$-wave magnet--superconductor hybrid
  systems},\ }\href {https://doi.org/10.1103/PhysRevB.111.174519} {\bibfield
  {journal} {\bibinfo  {journal} {Phys. Rev. B}\ }\textbf {\bibinfo {volume}
  {111}},\ \bibinfo {pages} {174519} (\bibinfo {year}
  {2025}{\natexlab{b}})}\BibitemShut {NoStop}%
\bibitem [{\citenamefont {Fukaya}\ \emph
  {et~al.}(2025{\natexlab{c}})\citenamefont {Fukaya}, \citenamefont {Yada},\
  and\ \citenamefont {Tanaka}}]{Fukaya2025Tunneling}%
  \BibitemOpen
  \bibfield  {author} {\bibinfo {author} {\bibfnamefont {Y.}~\bibnamefont
  {Fukaya}}, \bibinfo {author} {\bibfnamefont {K.}~\bibnamefont {Yada}},\ and\
  \bibinfo {author} {\bibfnamefont {Y.}~\bibnamefont {Tanaka}},\ }\bibfield
  {title} {\bibinfo {title} {Tunneling conductance in superconducting junctions
  with $p$‑wave unconventional magnets breaking time‑reversal symmetry},\
  }\href {https://doi.org/10.1007/s10948-025-07064-w} {\bibfield  {journal}
  {\bibinfo  {journal} {J. Supercond. Nov. Magn.}\ }\textbf {\bibinfo {volume}
  {38}},\ \bibinfo {pages} {228} (\bibinfo {year}
  {2025}{\natexlab{c}})}\BibitemShut {NoStop}%
\bibitem [{\citenamefont {Lu}\ \emph {et~al.}(2025)\citenamefont {Lu},
  \citenamefont {Mercebach}, \citenamefont {Burset}, \citenamefont {Yada},
  \citenamefont {Cayao}, \citenamefont {Tanaka},\ and\ \citenamefont
  {Fukaya}}]{lu2025engineeringsubgapstatessuperconductors}%
  \BibitemOpen
  \bibfield  {author} {\bibinfo {author} {\bibfnamefont {B.}~\bibnamefont
  {Lu}}, \bibinfo {author} {\bibfnamefont {P.}~\bibnamefont {Mercebach}},
  \bibinfo {author} {\bibfnamefont {P.}~\bibnamefont {Burset}}, \bibinfo
  {author} {\bibfnamefont {K.}~\bibnamefont {Yada}}, \bibinfo {author}
  {\bibfnamefont {J.}~\bibnamefont {Cayao}}, \bibinfo {author} {\bibfnamefont
  {Y.}~\bibnamefont {Tanaka}},\ and\ \bibinfo {author} {\bibfnamefont
  {Y.}~\bibnamefont {Fukaya}},\ }\bibfield  {title} {\bibinfo {title}
  {Engineering subgap states in superconductors by altermagnetism},\ }\href
  {https://arxiv.org/abs/2508.03364} {\bibfield  {journal} {\bibinfo  {journal}
  {arXiv:2508.03364}\ } (\bibinfo {year} {2025})}\BibitemShut {NoStop}%
\bibitem [{\citenamefont {Fukaya}\ \emph
  {et~al.}(2025{\natexlab{d}})\citenamefont {Fukaya}, \citenamefont {Lu},
  \citenamefont {Yada}, \citenamefont {Tanaka},\ and\ \citenamefont
  {Cayao}}]{fukaya2025crossedsurfaceflatbands}%
  \BibitemOpen
  \bibfield  {author} {\bibinfo {author} {\bibfnamefont {Y.}~\bibnamefont
  {Fukaya}}, \bibinfo {author} {\bibfnamefont {B.}~\bibnamefont {Lu}}, \bibinfo
  {author} {\bibfnamefont {K.}~\bibnamefont {Yada}}, \bibinfo {author}
  {\bibfnamefont {Y.}~\bibnamefont {Tanaka}},\ and\ \bibinfo {author}
  {\bibfnamefont {J.}~\bibnamefont {Cayao}},\ }\bibfield  {title} {\bibinfo
  {title} {Crossed surface flat bands in three-dimensional superconducting
  altermagnets},\ }\href {https://arxiv.org/abs/2510.14724} {\bibfield
  {journal} {\bibinfo  {journal} {arXiv:2510.14724}\ } (\bibinfo {year}
  {2025}{\natexlab{d}})}\BibitemShut {NoStop}%
\bibitem [{\citenamefont {Fu}\ \emph {et~al.}(2025{\natexlab{b}})\citenamefont
  {Fu}, \citenamefont {Mondal}, \citenamefont {Liu}, \citenamefont {Tanaka},\
  and\ \citenamefont {Cayao}}]{Fu2025Floquet}%
  \BibitemOpen
  \bibfield  {author} {\bibinfo {author} {\bibfnamefont {P.-H.}\ \bibnamefont
  {Fu}}, \bibinfo {author} {\bibfnamefont {S.}~\bibnamefont {Mondal}}, \bibinfo
  {author} {\bibfnamefont {J.-F.}\ \bibnamefont {Liu}}, \bibinfo {author}
  {\bibfnamefont {Y.}~\bibnamefont {Tanaka}},\ and\ \bibinfo {author}
  {\bibfnamefont {J.}~\bibnamefont {Cayao}},\ }\bibfield  {title} {\bibinfo
  {title} {Floquet engineering spin triplet states in unconventional magnets},\
  }\href {https://arxiv.org/abs/2505.20205} {\bibfield  {journal} {\bibinfo
  {journal} {arXiv:2505.20205}\ } (\bibinfo {year}
  {2025}{\natexlab{b}})}\BibitemShut {NoStop}%
\bibitem [{\citenamefont {Roig}\ \emph {et~al.}(2024)\citenamefont {Roig},
  \citenamefont {Kreisel}, \citenamefont {Yu}, \citenamefont {Andersen},\ and\
  \citenamefont {Agterberg}}]{PhysRevB.110.144412}%
  \BibitemOpen
  \bibfield  {author} {\bibinfo {author} {\bibfnamefont {M.}~\bibnamefont
  {Roig}}, \bibinfo {author} {\bibfnamefont {A.}~\bibnamefont {Kreisel}},
  \bibinfo {author} {\bibfnamefont {Y.}~\bibnamefont {Yu}}, \bibinfo {author}
  {\bibfnamefont {B.~M.}\ \bibnamefont {Andersen}},\ and\ \bibinfo {author}
  {\bibfnamefont {D.~F.}\ \bibnamefont {Agterberg}},\ }\bibfield  {title}
  {\bibinfo {title} {Minimal models for altermagnetism},\ }\href
  {https://doi.org/10.1103/PhysRevB.110.144412} {\bibfield  {journal} {\bibinfo
   {journal} {Phys. Rev. B}\ }\textbf {\bibinfo {volume} {110}},\ \bibinfo
  {pages} {144412} (\bibinfo {year} {2024})}\BibitemShut {NoStop}%
\bibitem [{\citenamefont {Brekke}\ \emph
  {et~al.}(2023{\natexlab{b}})\citenamefont {Brekke}, \citenamefont {Brataas},\
  and\ \citenamefont {Sudb\o{}}}]{Brekke23}%
  \BibitemOpen
  \bibfield  {author} {\bibinfo {author} {\bibfnamefont {B.}~\bibnamefont
  {Brekke}}, \bibinfo {author} {\bibfnamefont {A.}~\bibnamefont {Brataas}},\
  and\ \bibinfo {author} {\bibfnamefont {A.}~\bibnamefont {Sudb\o{}}},\
  }\bibfield  {title} {\bibinfo {title} {Two-dimensional altermagnets:
  Superconductivity in a minimal microscopic model},\ }\href
  {https://doi.org/10.1103/PhysRevB.108.224421} {\bibfield  {journal} {\bibinfo
   {journal} {Phys. Rev. B}\ }\textbf {\bibinfo {volume} {108}},\ \bibinfo
  {pages} {224421} (\bibinfo {year} {2023}{\natexlab{b}})}\BibitemShut
  {NoStop}%
\bibitem [{sm()}]{sm}%
  \BibitemOpen
  \href@noop {} {}\bibinfo {note} {See Supplemental Material at
  URL-will-be-inserted-by-publisher for additional details of Floquet
  engineering spin density and spin-triplet correlations in unconventional
  magnets with $s$-wave superconductivity. We also present details of driven
  unconventional magnets with higher parities and assess the feasibility of our
  findings under realistic conditions in available experiments. The
  Supplemental Material also includes
  Refs.\,\cite{Esin2020Floquet,mciver2020light,Merboldt2025Observation,Matsuda2020Room,Zhang2023Light,Shan2021Giant,Shan2021Giant,Jungwirth2025Symmetry,Dehghani2021Light,Kennes2019Light,Kitamura2022Floquet,Ashcroft1976Solid,Jackson1999Classical,wang2013observation,aeschlimann2021survival,FukayaCayaoReviewUMs,Ezawa2025Third,Chen2025Quasicrystalline,zagoskin,Prada2020From,Machida1980Theory,Golubov2025Physics,Vagov2023Intertype,Zareapour2012Proximity,Lofwander2004Proximity,Dehghani2021Light,Kennes2019Light,Kitamura2022Floquet,yokoyama2025FloquetSC,PhysRevB.109.134517,Mizushima2025Detecting,Luo2021LightInduced}.}\BibitemShut
  {Stop}%
\bibitem [{\citenamefont {Fu}\ \emph {et~al.}(2019)\citenamefont {Fu},
  \citenamefont {Wang}, \citenamefont {Liu},\ and\ \citenamefont
  {Wang}}]{Fu2019Josephson}%
  \BibitemOpen
  \bibfield  {author} {\bibinfo {author} {\bibfnamefont {P.-H.}\ \bibnamefont
  {Fu}}, \bibinfo {author} {\bibfnamefont {J.}~\bibnamefont {Wang}}, \bibinfo
  {author} {\bibfnamefont {J.-F.}\ \bibnamefont {Liu}},\ and\ \bibinfo {author}
  {\bibfnamefont {R.-Q.}\ \bibnamefont {Wang}},\ }\bibfield  {title} {\bibinfo
  {title} {Josephson signatures of {W}eyl node creation and annihilation in
  irradiated {D}irac semimetals},\ }\href
  {https://doi.org/10.1103/PhysRevB.100.115414} {\bibfield  {journal} {\bibinfo
   {journal} {Phys. Rev. B}\ }\textbf {\bibinfo {volume} {100}},\ \bibinfo
  {pages} {115414} (\bibinfo {year} {2019})}\BibitemShut {NoStop}%
\bibitem [{\citenamefont {Fu}\ \emph {et~al.}(2022{\natexlab{a}})\citenamefont
  {Fu}, \citenamefont {Xu}, \citenamefont {Yu}, \citenamefont {Liu},\ and\
  \citenamefont {Wu}}]{Fu2022Electrically}%
  \BibitemOpen
  \bibfield  {author} {\bibinfo {author} {\bibfnamefont {P.-H.}\ \bibnamefont
  {Fu}}, \bibinfo {author} {\bibfnamefont {Y.}~\bibnamefont {Xu}}, \bibinfo
  {author} {\bibfnamefont {X.-L.}\ \bibnamefont {Yu}}, \bibinfo {author}
  {\bibfnamefont {J.-F.}\ \bibnamefont {Liu}},\ and\ \bibinfo {author}
  {\bibfnamefont {J.}~\bibnamefont {Wu}},\ }\bibfield  {title} {\bibinfo
  {title} {Electrically modulated {J}osephson junction of light-dressed
  topological insulators},\ }\href
  {https://doi.org/10.1103/PhysRevB.105.064503} {\bibfield  {journal} {\bibinfo
   {journal} {Phys. Rev. B}\ }\textbf {\bibinfo {volume} {105}},\ \bibinfo
  {pages} {064503} (\bibinfo {year} {2022}{\natexlab{a}})}\BibitemShut
  {NoStop}%
\bibitem [{\citenamefont {Lee}\ \emph {et~al.}(2025)\citenamefont {Lee},
  \citenamefont {Calderon}, \citenamefont {Yu}, \citenamefont {Lee},
  \citenamefont {Ang},\ and\ \citenamefont {Fu}}]{Lee2025Floquet}%
  \BibitemOpen
  \bibfield  {author} {\bibinfo {author} {\bibfnamefont {K.~W.}\ \bibnamefont
  {Lee}}, \bibinfo {author} {\bibfnamefont {M.~J.~A.}\ \bibnamefont
  {Calderon}}, \bibinfo {author} {\bibfnamefont {X.-L.}\ \bibnamefont {Yu}},
  \bibinfo {author} {\bibfnamefont {C.~H.}\ \bibnamefont {Lee}}, \bibinfo
  {author} {\bibfnamefont {Y.~S.}\ \bibnamefont {Ang}},\ and\ \bibinfo {author}
  {\bibfnamefont {P.-H.}\ \bibnamefont {Fu}},\ }\bibfield  {title} {\bibinfo
  {title} {Floquet engineering of topological phase transitions in a quantum
  spin {H}all $\alpha$-${T}_{3}$ system},\ }\href
  {https://doi.org/10.1103/PhysRevB.111.045406} {\bibfield  {journal} {\bibinfo
   {journal} {Phys. Rev. B}\ }\textbf {\bibinfo {volume} {111}},\ \bibinfo
  {pages} {045406} (\bibinfo {year} {2025})}\BibitemShut {NoStop}%
\bibitem [{\citenamefont {Fu}\ \emph {et~al.}(2022{\natexlab{b}})\citenamefont
  {Fu}, \citenamefont {Lv}, \citenamefont {Yu}, \citenamefont {Liu},\ and\
  \citenamefont {Wu}}]{Fu2022The}%
  \BibitemOpen
  \bibfield  {author} {\bibinfo {author} {\bibfnamefont {P.-H.}\ \bibnamefont
  {Fu}}, \bibinfo {author} {\bibfnamefont {Q.}~\bibnamefont {Lv}}, \bibinfo
  {author} {\bibfnamefont {X.-L.}\ \bibnamefont {Yu}}, \bibinfo {author}
  {\bibfnamefont {J.-F.}\ \bibnamefont {Liu}},\ and\ \bibinfo {author}
  {\bibfnamefont {J.}~\bibnamefont {Wu}},\ }\bibfield  {title} {\bibinfo
  {title} {The generation of switchable polarized currents in nodal ring
  semimetals using high-frequency periodic driving},\ }\href
  {https://doi.org/10.1088/1361-648X/ac37db} {\bibfield  {journal} {\bibinfo
  {journal} {J. Phys.: Condens. Matter}\ }\textbf {\bibinfo {volume} {34}},\
  \bibinfo {pages} {075401} (\bibinfo {year} {2022}{\natexlab{b}})}\BibitemShut
  {NoStop}%
\bibitem [{\citenamefont {Li}\ \emph {et~al.}(2019)\citenamefont {Li},
  \citenamefont {Wang}, \citenamefont {Deng}, \citenamefont {Duan},
  \citenamefont {Fu}, \citenamefont {Wang}, \citenamefont {Sheng},\ and\
  \citenamefont {Xing}}]{Li2019Photon}%
  \BibitemOpen
  \bibfield  {author} {\bibinfo {author} {\bibfnamefont {X.-S.}\ \bibnamefont
  {Li}}, \bibinfo {author} {\bibfnamefont {C.}~\bibnamefont {Wang}}, \bibinfo
  {author} {\bibfnamefont {M.-X.}\ \bibnamefont {Deng}}, \bibinfo {author}
  {\bibfnamefont {H.-J.}\ \bibnamefont {Duan}}, \bibinfo {author}
  {\bibfnamefont {P.-H.}\ \bibnamefont {Fu}}, \bibinfo {author} {\bibfnamefont
  {R.-Q.}\ \bibnamefont {Wang}}, \bibinfo {author} {\bibfnamefont
  {L.}~\bibnamefont {Sheng}},\ and\ \bibinfo {author} {\bibfnamefont {D.~Y.}\
  \bibnamefont {Xing}},\ }\bibfield  {title} {\bibinfo {title} {Photon-induced
  {W}eyl half-metal phase and spin filter effect from topological {D}irac
  semimetals},\ }\href {https://doi.org/10.1103/PhysRevLett.123.206601}
  {\bibfield  {journal} {\bibinfo  {journal} {Phys. Rev. Lett.}\ }\textbf
  {\bibinfo {volume} {123}},\ \bibinfo {pages} {206601} (\bibinfo {year}
  {2019})}\BibitemShut {NoStop}%
\bibitem [{\citenamefont {Wang}\ \emph {et~al.}(2023)\citenamefont {Wang},
  \citenamefont {Balembois}, \citenamefont {Ran{\v{c}}i{\'c}}, \citenamefont
  {Billaud}, \citenamefont {Le~Dantec}, \citenamefont {Ferrier}, \citenamefont
  {Goldner}, \citenamefont {Bertaina}, \citenamefont {Chaneli{\`e}re},
  \citenamefont {Est{\`e}ve} \emph {et~al.}}]{wang2023single}%
  \BibitemOpen
  \bibfield  {author} {\bibinfo {author} {\bibfnamefont {Z.}~\bibnamefont
  {Wang}}, \bibinfo {author} {\bibfnamefont {L.}~\bibnamefont {Balembois}},
  \bibinfo {author} {\bibfnamefont {M.}~\bibnamefont {Ran{\v{c}}i{\'c}}},
  \bibinfo {author} {\bibfnamefont {E.}~\bibnamefont {Billaud}}, \bibinfo
  {author} {\bibfnamefont {M.}~\bibnamefont {Le~Dantec}}, \bibinfo {author}
  {\bibfnamefont {A.}~\bibnamefont {Ferrier}}, \bibinfo {author} {\bibfnamefont
  {P.}~\bibnamefont {Goldner}}, \bibinfo {author} {\bibfnamefont
  {S.}~\bibnamefont {Bertaina}}, \bibinfo {author} {\bibfnamefont
  {T.}~\bibnamefont {Chaneli{\`e}re}}, \bibinfo {author} {\bibfnamefont
  {D.}~\bibnamefont {Est{\`e}ve}}, \emph {et~al.},\ }\bibfield  {title}
  {\bibinfo {title} {Single-electron spin resonance detection by microwave
  photon counting},\ }\href {https://doi.org/10.1038/s41586-023-06097-2}
  {\bibfield  {journal} {\bibinfo  {journal} {Nature}\ }\textbf {\bibinfo
  {volume} {619}},\ \bibinfo {pages} {276} (\bibinfo {year}
  {2023})}\BibitemShut {NoStop}%
\bibitem [{Note1()}]{Note1}%
  \BibitemOpen
  \bibinfo {note} {We note that low frequency CPL and LPL drives can also
  induce a spin-triplet density in both $d$- and $p$-wave UMs; here, the
  Floquet bands play a role.}\BibitemShut {Stop}%
\bibitem [{\citenamefont {Zagoskin}(2014)}]{zagoskin}%
  \BibitemOpen
  \bibfield  {author} {\bibinfo {author} {\bibfnamefont {A.}~\bibnamefont
  {Zagoskin}},\ }\href@noop {} {\emph {\bibinfo {title} {Quantum Theory of
  Many-Body Systems: Techniques and Applications}}}\ (\bibinfo  {publisher}
  {Springer},\ \bibinfo {year} {2014})\BibitemShut {NoStop}%
\bibitem [{\citenamefont {Mahan}(2013)}]{mahan2013many}%
  \BibitemOpen
  \bibfield  {author} {\bibinfo {author} {\bibfnamefont {G.~D.}\ \bibnamefont
  {Mahan}},\ }\href@noop {} {\emph {\bibinfo {title} {Many-particle physics}}}\
  (\bibinfo  {publisher} {Springer Science \& Business Media},\ \bibinfo {year}
  {2013})\BibitemShut {NoStop}%
\bibitem [{\citenamefont {Bergeret}\ \emph {et~al.}(2005)\citenamefont
  {Bergeret}, \citenamefont {Volkov},\ and\ \citenamefont
  {Efetov}}]{RevModPhys.77.1321}%
  \BibitemOpen
  \bibfield  {author} {\bibinfo {author} {\bibfnamefont {F.~S.}\ \bibnamefont
  {Bergeret}}, \bibinfo {author} {\bibfnamefont {A.~F.}\ \bibnamefont
  {Volkov}},\ and\ \bibinfo {author} {\bibfnamefont {K.~B.}\ \bibnamefont
  {Efetov}},\ }\bibfield  {title} {\bibinfo {title} {Odd triplet
  superconductivity and related phenomena in superconductor-ferromagnet
  structures},\ }\href {https://doi.org/10.1103/RevModPhys.77.1321} {\bibfield
  {journal} {\bibinfo  {journal} {Rev. Mod. Phys.}\ }\textbf {\bibinfo {volume}
  {77}},\ \bibinfo {pages} {1321} (\bibinfo {year} {2005})}\BibitemShut
  {NoStop}%
\bibitem [{\citenamefont {Tanaka}\ \emph {et~al.}(2012)\citenamefont {Tanaka},
  \citenamefont {Sato},\ and\ \citenamefont {Nagaosa}}]{tanaka2012symmetry}%
  \BibitemOpen
  \bibfield  {author} {\bibinfo {author} {\bibfnamefont {Y.}~\bibnamefont
  {Tanaka}}, \bibinfo {author} {\bibfnamefont {M.}~\bibnamefont {Sato}},\ and\
  \bibinfo {author} {\bibfnamefont {N.}~\bibnamefont {Nagaosa}},\ }\bibfield
  {title} {\bibinfo {title} {Symmetry and topology in
  superconductors--odd-frequency pairing and edge states--},\ }\href
  {https://doi.org/10.1143/JPSJ.81.011013} {\bibfield  {journal} {\bibinfo
  {journal} {J. Phys. Soc. Jpn.}\ }\textbf {\bibinfo {volume} {81}},\ \bibinfo
  {pages} {011013} (\bibinfo {year} {2012})}\BibitemShut {NoStop}%
\bibitem [{\citenamefont {Cayao}\ \emph {et~al.}(2020)\citenamefont {Cayao},
  \citenamefont {Triola},\ and\ \citenamefont {Black-Schaffer}}]{cayao2019odd}%
  \BibitemOpen
  \bibfield  {author} {\bibinfo {author} {\bibfnamefont {J.}~\bibnamefont
  {Cayao}}, \bibinfo {author} {\bibfnamefont {C.}~\bibnamefont {Triola}},\ and\
  \bibinfo {author} {\bibfnamefont {A.~M.}\ \bibnamefont {Black-Schaffer}},\
  }\bibfield  {title} {\bibinfo {title} {Odd-frequency superconducting pairing
  in one-dimensional systems},\ }\href
  {https://link.springer.com/article/10.1140/epjst/e2019-900168-0} {\bibfield
  {journal} {\bibinfo  {journal} {Eur. Phys. J. Spec. Top.}\ }\textbf {\bibinfo
  {volume} {229}},\ \bibinfo {pages} {545} (\bibinfo {year}
  {2020})}\BibitemShut {NoStop}%
\bibitem [{\citenamefont {Linder}\ and\ \citenamefont
  {Balatsky}(2019)}]{RevModPhys.91.045005}%
  \BibitemOpen
  \bibfield  {author} {\bibinfo {author} {\bibfnamefont {J.}~\bibnamefont
  {Linder}}\ and\ \bibinfo {author} {\bibfnamefont {A.~V.}\ \bibnamefont
  {Balatsky}},\ }\bibfield  {title} {\bibinfo {title} {Odd-frequency
  superconductivity},\ }\href {https://doi.org/10.1103/RevModPhys.91.045005}
  {\bibfield  {journal} {\bibinfo  {journal} {Rev. Mod. Phys.}\ }\textbf
  {\bibinfo {volume} {91}},\ \bibinfo {pages} {045005} (\bibinfo {year}
  {2019})}\BibitemShut {NoStop}%
\bibitem [{\citenamefont {Triola}\ \emph {et~al.}(2020)\citenamefont {Triola},
  \citenamefont {Cayao},\ and\ \citenamefont
  {Black-Schaffer}}]{triola2020role}%
  \BibitemOpen
  \bibfield  {author} {\bibinfo {author} {\bibfnamefont {C.}~\bibnamefont
  {Triola}}, \bibinfo {author} {\bibfnamefont {J.}~\bibnamefont {Cayao}},\ and\
  \bibinfo {author} {\bibfnamefont {A.~M.}\ \bibnamefont {Black-Schaffer}},\
  }\bibfield  {title} {\bibinfo {title} {The role of odd-frequency pairing in
  multiband superconductors},\ }\href
  {https://onlinelibrary.wiley.com/doi/10.1002/andp.201900298} {\bibfield
  {journal} {\bibinfo  {journal} {Ann. Phys.}\ }\textbf {\bibinfo {volume}
  {532}},\ \bibinfo {pages} {1900298} (\bibinfo {year} {2020})}\BibitemShut
  {NoStop}%
\bibitem [{\citenamefont {Tanaka}\ \emph {et~al.}(2024)\citenamefont {Tanaka},
  \citenamefont {Tamura},\ and\ \citenamefont {Cayao}}]{tanaka2024theory}%
  \BibitemOpen
  \bibfield  {author} {\bibinfo {author} {\bibfnamefont {Y.}~\bibnamefont
  {Tanaka}}, \bibinfo {author} {\bibfnamefont {S.}~\bibnamefont {Tamura}},\
  and\ \bibinfo {author} {\bibfnamefont {J.}~\bibnamefont {Cayao}},\ }\bibfield
   {title} {\bibinfo {title} {Theory of {M}ajorana zero modes in unconventional
  superconductors},\ }\href {https://doi.org/10.1093/ptep/ptae065} {\bibfield
  {journal} {\bibinfo  {journal} {Prog. Theor. Exp. Phys.}\ }\textbf {\bibinfo
  {volume} {2024}},\ \bibinfo {pages} {08C105} (\bibinfo {year}
  {2024})}\BibitemShut {NoStop}%
\bibitem [{\citenamefont {Tanaka}\ and\ \citenamefont
  {Golubov}(2007)}]{Tanaka2007Theory}%
  \BibitemOpen
  \bibfield  {author} {\bibinfo {author} {\bibfnamefont {Y.}~\bibnamefont
  {Tanaka}}\ and\ \bibinfo {author} {\bibfnamefont {A.~A.}\ \bibnamefont
  {Golubov}},\ }\bibfield  {title} {\bibinfo {title} {Theory of the proximity
  effect in junctions with unconventional superconductors},\ }\href
  {https://doi.org/10.1103/PhysRevLett.98.037003} {\bibfield  {journal}
  {\bibinfo  {journal} {Phys. Rev. Lett.}\ }\textbf {\bibinfo {volume} {98}},\
  \bibinfo {pages} {037003} (\bibinfo {year} {2007})}\BibitemShut {NoStop}%
\bibitem [{\citenamefont {Berezinskii}(1974)}]{Berezinskii1974New}%
  \BibitemOpen
  \bibfield  {author} {\bibinfo {author} {\bibfnamefont {V.~L.}\ \bibnamefont
  {Berezinskii}},\ }\bibfield  {title} {\bibinfo {title} {New model of the
  anisotropic phase of superfluid {He}$^3$},\ }\href@noop {} {\bibfield
  {journal} {\bibinfo  {journal} {Zh. Eksp. Teor. Fiz. Pis’ma Red.}\ }\textbf
  {\bibinfo {volume} {20}},\ \bibinfo {pages} {628} (\bibinfo {year} {1974})},\
  \bibinfo {note} {[JETP Lett. {\bf 20}, 287 (1974)]}\BibitemShut {NoStop}%
\bibitem [{Note2()}]{Note2}%
  \BibitemOpen
  \bibinfo {note} {We have verified that light-driven unconventional magnets
  with unconventional superconductivity allow to engineer even more exotic
  Cooper pairs \cite {phjc}.}\BibitemShut {Stop}%
\bibitem [{\citenamefont {Li}\ \emph {et~al.}(1993)\citenamefont {Li},
  \citenamefont {Koltenbah},\ and\ \citenamefont {Joynt}}]{PhysRevB.48.437}%
  \BibitemOpen
  \bibfield  {author} {\bibinfo {author} {\bibfnamefont {Q.~P.}\ \bibnamefont
  {Li}}, \bibinfo {author} {\bibfnamefont {B.~E.~C.}\ \bibnamefont
  {Koltenbah}},\ and\ \bibinfo {author} {\bibfnamefont {R.}~\bibnamefont
  {Joynt}},\ }\bibfield  {title} {\bibinfo {title} {Mixed s-wave and d-wave
  superconductivity in high-{$T_{c}$} systems},\ }\href
  {https://doi.org/10.1103/PhysRevB.48.437} {\bibfield  {journal} {\bibinfo
  {journal} {Phys. Rev. B}\ }\textbf {\bibinfo {volume} {48}},\ \bibinfo
  {pages} {437} (\bibinfo {year} {1993})}\BibitemShut {NoStop}%
\bibitem [{\citenamefont {O'Donovan}\ and\ \citenamefont
  {Carbotte}(1995)}]{PhysRevB.52.16208}%
  \BibitemOpen
  \bibfield  {author} {\bibinfo {author} {\bibfnamefont {C.}~\bibnamefont
  {O'Donovan}}\ and\ \bibinfo {author} {\bibfnamefont {J.~P.}\ \bibnamefont
  {Carbotte}},\ }\bibfield  {title} {\bibinfo {title} {s- and d-wave mixing in
  high-${T}_{c}$ superconductors},\ }\href
  {https://doi.org/10.1103/PhysRevB.52.16208} {\bibfield  {journal} {\bibinfo
  {journal} {Phys. Rev. B}\ }\textbf {\bibinfo {volume} {52}},\ \bibinfo
  {pages} {16208} (\bibinfo {year} {1995})}\BibitemShut {NoStop}%
\bibitem [{\citenamefont {Hanaguri}\ \emph {et~al.}(2007)\citenamefont
  {Hanaguri}, \citenamefont {Kohsaka}, \citenamefont {Hoffman}, \citenamefont
  {Kim}, \citenamefont {Lang}, \citenamefont {Davis} \emph
  {et~al.}}]{hanaguri2007quasiparticle}%
  \BibitemOpen
  \bibfield  {author} {\bibinfo {author} {\bibfnamefont {T.}~\bibnamefont
  {Hanaguri}}, \bibinfo {author} {\bibfnamefont {Y.}~\bibnamefont {Kohsaka}},
  \bibinfo {author} {\bibfnamefont {J.~E.}\ \bibnamefont {Hoffman}}, \bibinfo
  {author} {\bibfnamefont {E.-A.}\ \bibnamefont {Kim}}, \bibinfo {author}
  {\bibfnamefont {K.~M.}\ \bibnamefont {Lang}}, \bibinfo {author}
  {\bibfnamefont {A.}~\bibnamefont {Davis}}, \emph {et~al.},\ }\bibfield
  {title} {\bibinfo {title} {Quasiparticle interference and superconducting gap
  in {Ca$_{2-x}$Na$_x$CuO$_2$Cl$_2$}},\ }\href
  {https://doi.org/10.1038/nphys753} {\bibfield  {journal} {\bibinfo  {journal}
  {Nat. Phys.}\ }\textbf {\bibinfo {volume} {3}},\ \bibinfo {pages} {865}
  (\bibinfo {year} {2007})}\BibitemShut {NoStop}%
\bibitem [{\citenamefont {Sharma}\ \emph {et~al.}(2020)\citenamefont {Sharma},
  \citenamefont {Edkins}, \citenamefont {Wang}, \citenamefont {Kostin},
  \citenamefont {Sow}, \citenamefont {Maeno}, \citenamefont {Mackenzie},
  \citenamefont {Davis},\ and\ \citenamefont {Madhavan}}]{sharma2020momentum}%
  \BibitemOpen
  \bibfield  {author} {\bibinfo {author} {\bibfnamefont {R.}~\bibnamefont
  {Sharma}}, \bibinfo {author} {\bibfnamefont {S.~D.}\ \bibnamefont {Edkins}},
  \bibinfo {author} {\bibfnamefont {Z.}~\bibnamefont {Wang}}, \bibinfo {author}
  {\bibfnamefont {A.}~\bibnamefont {Kostin}}, \bibinfo {author} {\bibfnamefont
  {C.}~\bibnamefont {Sow}}, \bibinfo {author} {\bibfnamefont {Y.}~\bibnamefont
  {Maeno}}, \bibinfo {author} {\bibfnamefont {A.~P.}\ \bibnamefont
  {Mackenzie}}, \bibinfo {author} {\bibfnamefont {J.~C.~S.}\ \bibnamefont
  {Davis}},\ and\ \bibinfo {author} {\bibfnamefont {V.}~\bibnamefont
  {Madhavan}},\ }\bibfield  {title} {\bibinfo {title} {Momentum-resolved
  superconducting energy gaps of {Sr$_2$RuO$_4$} from quasiparticle
  interference imaging},\ }\href {https://doi.org/10.1073/pnas.1916463117}
  {\bibfield  {journal} {\bibinfo  {journal} {Proc. Natl. Acad. Sci. U.S.A.}\
  }\textbf {\bibinfo {volume} {117}},\ \bibinfo {pages} {5222} (\bibinfo {year}
  {2020})}\BibitemShut {NoStop}%
\bibitem [{\citenamefont {Allan}\ \emph {et~al.}(2012)\citenamefont {Allan},
  \citenamefont {Rost}, \citenamefont {Mackenzie}, \citenamefont {Xie},
  \citenamefont {Davis}, \citenamefont {Kihou}, \citenamefont {Lee},
  \citenamefont {Iyo}, \citenamefont {Eisaki},\ and\ \citenamefont
  {Chuang}}]{allan2012anisotropic}%
  \BibitemOpen
  \bibfield  {author} {\bibinfo {author} {\bibfnamefont {M.~P.}\ \bibnamefont
  {Allan}}, \bibinfo {author} {\bibfnamefont {A.~W.}\ \bibnamefont {Rost}},
  \bibinfo {author} {\bibfnamefont {A.~P.}\ \bibnamefont {Mackenzie}}, \bibinfo
  {author} {\bibfnamefont {Y.}~\bibnamefont {Xie}}, \bibinfo {author}
  {\bibfnamefont {J.~C.}\ \bibnamefont {Davis}}, \bibinfo {author}
  {\bibfnamefont {K.}~\bibnamefont {Kihou}}, \bibinfo {author} {\bibfnamefont
  {C.~H.}\ \bibnamefont {Lee}}, \bibinfo {author} {\bibfnamefont
  {A.}~\bibnamefont {Iyo}}, \bibinfo {author} {\bibfnamefont {H.}~\bibnamefont
  {Eisaki}},\ and\ \bibinfo {author} {\bibfnamefont {T.}~\bibnamefont
  {Chuang}},\ }\bibfield  {title} {\bibinfo {title} {Anisotropic energy gaps of
  iron-based superconductivity from intraband quasiparticle interference in
  {LiFeAs}},\ }\href {https://doi.org/10.1126/science.1218726} {\bibfield
  {journal} {\bibinfo  {journal} {Science}\ }\textbf {\bibinfo {volume}
  {336}},\ \bibinfo {pages} {563} (\bibinfo {year} {2012})}\BibitemShut
  {NoStop}%
\bibitem [{Note3()}]{Note3}%
  \BibitemOpen
  \bibinfo {note} {We note that the same conclusions can be obtained from the
  imaginary part ${\protect \rm Im}\protect \bar {F}_{t}^{\protect \rm eff}$
  demonstrating peak-dip structures and competition between ${\protect \rm
  Im}\protect \bar {F}_{t,\protect \rm M}^{\protect \rm eff}$ and ${\protect
  \rm Im}\protect \bar {F}_{t,\Omega }^{\protect \rm eff}$.}\BibitemShut
  {Stop}%
\bibitem [{\citenamefont {Kashiwaya}\ and\ \citenamefont
  {Tanaka}(2000)}]{kashiwaya2000}%
  \BibitemOpen
  \bibfield  {author} {\bibinfo {author} {\bibfnamefont {S.}~\bibnamefont
  {Kashiwaya}}\ and\ \bibinfo {author} {\bibfnamefont {Y.}~\bibnamefont
  {Tanaka}},\ }\bibfield  {title} {\bibinfo {title} {Tunnelling effects on
  surface bound states in unconventional superconductors},\ }\href
  {https://doi.org/10.1088/0034-4885/63/10/202} {\bibfield  {journal} {\bibinfo
   {journal} {Rep. Prog. Phys.}\ }\textbf {\bibinfo {volume} {63}},\ \bibinfo
  {pages} {1641} (\bibinfo {year} {2000})}\BibitemShut {NoStop}%
\bibitem [{\citenamefont {Cuevas}\ \emph {et~al.}(1996)\citenamefont {Cuevas},
  \citenamefont {Mart\'{\i}n-Rodero},\ and\ \citenamefont
  {Yeyati}}]{PhysRevB.54.7366}%
  \BibitemOpen
  \bibfield  {author} {\bibinfo {author} {\bibfnamefont {J.~C.}\ \bibnamefont
  {Cuevas}}, \bibinfo {author} {\bibfnamefont {A.}~\bibnamefont
  {Mart\'{\i}n-Rodero}},\ and\ \bibinfo {author} {\bibfnamefont {A.~L.}\
  \bibnamefont {Yeyati}},\ }\bibfield  {title} {\bibinfo {title} {Hamiltonian
  approach to the transport properties of superconducting quantum point
  contacts},\ }\href {https://doi.org/10.1103/PhysRevB.54.7366} {\bibfield
  {journal} {\bibinfo  {journal} {Phys. Rev. B}\ }\textbf {\bibinfo {volume}
  {54}},\ \bibinfo {pages} {7366} (\bibinfo {year} {1996})}\BibitemShut
  {NoStop}%
\bibitem [{\citenamefont {Burset}\ \emph {et~al.}(2016)\citenamefont {Burset},
  \citenamefont {Lu}, \citenamefont {Ebisu}, \citenamefont {Asano},\ and\
  \citenamefont {Tanaka}}]{PhysRevB.93.201402}%
  \BibitemOpen
  \bibfield  {author} {\bibinfo {author} {\bibfnamefont {P.}~\bibnamefont
  {Burset}}, \bibinfo {author} {\bibfnamefont {B.}~\bibnamefont {Lu}}, \bibinfo
  {author} {\bibfnamefont {H.}~\bibnamefont {Ebisu}}, \bibinfo {author}
  {\bibfnamefont {Y.}~\bibnamefont {Asano}},\ and\ \bibinfo {author}
  {\bibfnamefont {Y.}~\bibnamefont {Tanaka}},\ }\bibfield  {title} {\bibinfo
  {title} {All-electrical generation and control of odd-frequency $s$-wave
  cooper pairs in double quantum dots},\ }\href
  {https://doi.org/10.1103/PhysRevB.93.201402} {\bibfield  {journal} {\bibinfo
  {journal} {Phys. Rev. B}\ }\textbf {\bibinfo {volume} {93}},\ \bibinfo
  {pages} {201402} (\bibinfo {year} {2016})}\BibitemShut {NoStop}%
\bibitem [{\citenamefont {Ahmed}\ \emph {et~al.}(2025)\citenamefont {Ahmed},
  \citenamefont {Tamura}, \citenamefont {Tanaka},\ and\ \citenamefont
  {Cayao}}]{PhysRevB.111.024507}%
  \BibitemOpen
  \bibfield  {author} {\bibinfo {author} {\bibfnamefont {E.}~\bibnamefont
  {Ahmed}}, \bibinfo {author} {\bibfnamefont {S.}~\bibnamefont {Tamura}},
  \bibinfo {author} {\bibfnamefont {Y.}~\bibnamefont {Tanaka}},\ and\ \bibinfo
  {author} {\bibfnamefont {J.}~\bibnamefont {Cayao}},\ }\bibfield  {title}
  {\bibinfo {title} {Odd-frequency superconducting pairing due to multiple
  {M}ajorana edge modes in driven topological superconductors},\ }\href
  {https://doi.org/10.1103/PhysRevB.111.024507} {\bibfield  {journal} {\bibinfo
   {journal} {Phys. Rev. B}\ }\textbf {\bibinfo {volume} {111}},\ \bibinfo
  {pages} {024507} (\bibinfo {year} {2025})}\BibitemShut {NoStop}%
\bibitem [{\citenamefont {Cayao}\ \emph {et~al.}(2024)\citenamefont {Cayao},
  \citenamefont {Burset},\ and\ \citenamefont {Tanaka}}]{PhysRevB.109.205406}%
  \BibitemOpen
  \bibfield  {author} {\bibinfo {author} {\bibfnamefont {J.}~\bibnamefont
  {Cayao}}, \bibinfo {author} {\bibfnamefont {P.}~\bibnamefont {Burset}},\ and\
  \bibinfo {author} {\bibfnamefont {Y.}~\bibnamefont {Tanaka}},\ }\bibfield
  {title} {\bibinfo {title} {Controllable odd-frequency {C}ooper pairs in
  multisuperconductor {J}osephson junctions},\ }\href
  {https://doi.org/10.1103/PhysRevB.109.205406} {\bibfield  {journal} {\bibinfo
   {journal} {Phys. Rev. B}\ }\textbf {\bibinfo {volume} {109}},\ \bibinfo
  {pages} {205406} (\bibinfo {year} {2024})}\BibitemShut {NoStop}%
\bibitem [{\citenamefont {Matsuda}\ \emph {et~al.}(2020)\citenamefont
  {Matsuda}, \citenamefont {Kanda}, \citenamefont {Higo} \emph
  {et~al.}}]{Matsuda2020Room}%
  \BibitemOpen
  \bibfield  {author} {\bibinfo {author} {\bibfnamefont {T.}~\bibnamefont
  {Matsuda}}, \bibinfo {author} {\bibfnamefont {N.}~\bibnamefont {Kanda}},
  \bibinfo {author} {\bibfnamefont {T.}~\bibnamefont {Higo}}, \emph {et~al.},\
  }\bibfield  {title} {\bibinfo {title} {Room‑temperature terahertz anomalous
  {H}all effect in {W}eyl antiferromagnet {Mn$_3$Sn} thin films},\ }\href
  {https://doi.org/10.1038/s41467-020-14690-6} {\bibfield  {journal} {\bibinfo
  {journal} {Nat. Commun.}\ }\textbf {\bibinfo {volume} {11}},\ \bibinfo
  {pages} {909} (\bibinfo {year} {2020})}\BibitemShut {NoStop}%
\bibitem [{\citenamefont {Zhang}\ \emph {et~al.}(2024)\citenamefont {Zhang},
  \citenamefont {Carbin}, \citenamefont {Culver}, \citenamefont {Du},
  \citenamefont {Wang}, \citenamefont {Cheong}, \citenamefont {Roy},\ and\
  \citenamefont {Kogar}}]{Zhang2023Light}%
  \BibitemOpen
  \bibfield  {author} {\bibinfo {author} {\bibfnamefont {X.}~\bibnamefont
  {Zhang}}, \bibinfo {author} {\bibfnamefont {T.}~\bibnamefont {Carbin}},
  \bibinfo {author} {\bibfnamefont {A.~B.}\ \bibnamefont {Culver}}, \bibinfo
  {author} {\bibfnamefont {K.}~\bibnamefont {Du}}, \bibinfo {author}
  {\bibfnamefont {K.}~\bibnamefont {Wang}}, \bibinfo {author} {\bibfnamefont
  {S.}~\bibnamefont {Cheong}}, \bibinfo {author} {\bibfnamefont
  {R.}~\bibnamefont {Roy}},\ and\ \bibinfo {author} {\bibfnamefont
  {A.}~\bibnamefont {Kogar}},\ }\bibfield  {title} {\bibinfo {title}
  {{Light‑induced electronic polarization in antiferromagnetic
  Cr$_{2}$O$_{3}$}},\ }\href {https://doi.org/10.1038/s41563-024-01852-w}
  {\bibfield  {journal} {\bibinfo  {journal} {Nat. Mater.}\ }\textbf {\bibinfo
  {volume} {23}},\ \bibinfo {pages} {790} (\bibinfo {year} {2024})}\BibitemShut
  {NoStop}%
\bibitem [{\citenamefont {Shan}\ \emph {et~al.}(2021)\citenamefont {Shan},
  \citenamefont {Ye}, \citenamefont {Chu}, \citenamefont {Lee}, \citenamefont
  {Park}, \citenamefont {Balents},\ and\ \citenamefont
  {Hsieh}}]{Shan2021Giant}%
  \BibitemOpen
  \bibfield  {author} {\bibinfo {author} {\bibfnamefont {J.}~\bibnamefont
  {Shan}}, \bibinfo {author} {\bibfnamefont {M.}~\bibnamefont {Ye}}, \bibinfo
  {author} {\bibfnamefont {H.}~\bibnamefont {Chu}}, \bibinfo {author}
  {\bibfnamefont {S.}~\bibnamefont {Lee}}, \bibinfo {author} {\bibfnamefont
  {J.}~\bibnamefont {Park}}, \bibinfo {author} {\bibfnamefont {L.}~\bibnamefont
  {Balents}},\ and\ \bibinfo {author} {\bibfnamefont {D.}~\bibnamefont
  {Hsieh}},\ }\bibfield  {title} {\bibinfo {title} {Giant modulation of optical
  nonlinearity by {F}loquet engineering},\ }\href
  {https://doi.org/10.1038/s41586-021-04051-8} {\bibfield  {journal} {\bibinfo
  {journal} {Nature}\ }\textbf {\bibinfo {volume} {600}},\ \bibinfo {pages}
  {235} (\bibinfo {year} {2021})}\BibitemShut {NoStop}%
\bibitem [{\citenamefont {Luo}\ \emph {et~al.}(2021)\citenamefont {Luo},
  \citenamefont {Cheng}, \citenamefont {Song}, \citenamefont {Wang},
  \citenamefont {Vaswani}, \citenamefont {Lozano}, \citenamefont {Gu},
  \citenamefont {Huang}, \citenamefont {Kim}, \citenamefont {Liu},
  \citenamefont {Park}, \citenamefont {Yao}, \citenamefont {Ho}, \citenamefont
  {Perakis}, \citenamefont {Li},\ and\ \citenamefont
  {Wang}}]{Luo2021LightInduced}%
  \BibitemOpen
  \bibfield  {author} {\bibinfo {author} {\bibfnamefont {L.}~\bibnamefont
  {Luo}}, \bibinfo {author} {\bibfnamefont {D.}~\bibnamefont {Cheng}}, \bibinfo
  {author} {\bibfnamefont {B.}~\bibnamefont {Song}}, \bibinfo {author}
  {\bibfnamefont {L.}~\bibnamefont {Wang}}, \bibinfo {author} {\bibfnamefont
  {C.}~\bibnamefont {Vaswani}}, \bibinfo {author} {\bibfnamefont {P.~M.}\
  \bibnamefont {Lozano}}, \bibinfo {author} {\bibfnamefont {G.}~\bibnamefont
  {Gu}}, \bibinfo {author} {\bibfnamefont {C.}~\bibnamefont {Huang}}, \bibinfo
  {author} {\bibfnamefont {R.~H.~J.}\ \bibnamefont {Kim}}, \bibinfo {author}
  {\bibfnamefont {Z.}~\bibnamefont {Liu}}, \bibinfo {author} {\bibfnamefont
  {J.}~\bibnamefont {Park}}, \bibinfo {author} {\bibfnamefont {Y.}~\bibnamefont
  {Yao}}, \bibinfo {author} {\bibfnamefont {K.}~\bibnamefont {Ho}}, \bibinfo
  {author} {\bibfnamefont {I.~E.}\ \bibnamefont {Perakis}}, \bibinfo {author}
  {\bibfnamefont {Q.}~\bibnamefont {Li}},\ and\ \bibinfo {author}
  {\bibfnamefont {J.}~\bibnamefont {Wang}},\ }\bibfield  {title} {\bibinfo
  {title} {A light‑induced phononic symmetry switch and giant dissipationless
  topological photocurrent in {ZrTe$_5$}},\ }\href
  {https://doi.org/10.1038/s41563-020-00882-4} {\bibfield  {journal} {\bibinfo
  {journal} {Nat. Mater.}\ }\textbf {\bibinfo {volume} {20}},\ \bibinfo {pages}
  {329} (\bibinfo {year} {2021})}\BibitemShut {NoStop}%
\bibitem [{\citenamefont {Merboldt}\ \emph {et~al.}(2025)\citenamefont
  {Merboldt}, \citenamefont {Sch\"uler}, \citenamefont {Schmitt}, \citenamefont
  {Bange}, \citenamefont {Bennecke}, \citenamefont {Gadge}, \citenamefont
  {Pierz}, \citenamefont {Schumacher}, \citenamefont {Momeni}, \citenamefont
  {Steil}, \citenamefont {Manmana}, \citenamefont {Sentef}, \citenamefont
  {Reutzel},\ and\ \citenamefont {Mathias}}]{Merboldt2025Observation}%
  \BibitemOpen
  \bibfield  {author} {\bibinfo {author} {\bibfnamefont {M.}~\bibnamefont
  {Merboldt}}, \bibinfo {author} {\bibfnamefont {M.}~\bibnamefont {Sch\"uler}},
  \bibinfo {author} {\bibfnamefont {D.}~\bibnamefont {Schmitt}}, \bibinfo
  {author} {\bibfnamefont {J.~P.}\ \bibnamefont {Bange}}, \bibinfo {author}
  {\bibfnamefont {W.}~\bibnamefont {Bennecke}}, \bibinfo {author}
  {\bibfnamefont {K.}~\bibnamefont {Gadge}}, \bibinfo {author} {\bibfnamefont
  {K.}~\bibnamefont {Pierz}}, \bibinfo {author} {\bibfnamefont {H.~W.}\
  \bibnamefont {Schumacher}}, \bibinfo {author} {\bibfnamefont
  {D.}~\bibnamefont {Momeni}}, \bibinfo {author} {\bibfnamefont
  {D.}~\bibnamefont {Steil}}, \bibinfo {author} {\bibfnamefont {S.~R.}\
  \bibnamefont {Manmana}}, \bibinfo {author} {\bibfnamefont {M.~A.}\
  \bibnamefont {Sentef}}, \bibinfo {author} {\bibfnamefont {M.}~\bibnamefont
  {Reutzel}},\ and\ \bibinfo {author} {\bibfnamefont {S.}~\bibnamefont
  {Mathias}},\ }\bibfield  {title} {\bibinfo {title} {Observation of {F}loquet
  states in graphene},\ }\href {https://doi.org/10.1038/s41567-025-02889-7}
  {\bibfield  {journal} {\bibinfo  {journal} {Nat. Phys.}\ }\textbf {\bibinfo
  {volume} {21}},\ \bibinfo {pages} {1093} (\bibinfo {year}
  {2025})}\BibitemShut {NoStop}%
\bibitem [{\citenamefont {Ghorashi}\ and\ \citenamefont
  {Li}(2025)}]{Ghorashi2025Dynamical}%
  \BibitemOpen
  \bibfield  {author} {\bibinfo {author} {\bibfnamefont {S.~A.~A.}\
  \bibnamefont {Ghorashi}}\ and\ \bibinfo {author} {\bibfnamefont
  {Q.}~\bibnamefont {Li}},\ }\bibfield  {title} {\bibinfo {title} {Dynamical
  generation of higher-order spin-orbit coupling, topology, and persistent spin
  texture in light-irradiated altermagnets},\ }\href
  {https://doi.org/10.1103/tm58-lbdl} {\bibfield  {journal} {\bibinfo
  {journal} {Phys. Rev. Lett.}\ }\textbf {\bibinfo {volume} {135}},\ \bibinfo
  {pages} {236702} (\bibinfo {year} {2025})}\BibitemShut {NoStop}%
\bibitem [{\citenamefont {Yarmohammadi}\ \emph {et~al.}(2025)\citenamefont
  {Yarmohammadi}, \citenamefont {Z\"ulicke}, \citenamefont {Berakdar},
  \citenamefont {Linder},\ and\ \citenamefont
  {Freericks}}]{Yarmohammadi2025Anisotropic}%
  \BibitemOpen
  \bibfield  {author} {\bibinfo {author} {\bibfnamefont {M.}~\bibnamefont
  {Yarmohammadi}}, \bibinfo {author} {\bibfnamefont {U.}~\bibnamefont
  {Z\"ulicke}}, \bibinfo {author} {\bibfnamefont {J.}~\bibnamefont {Berakdar}},
  \bibinfo {author} {\bibfnamefont {J.}~\bibnamefont {Linder}},\ and\ \bibinfo
  {author} {\bibfnamefont {J.~K.}\ \bibnamefont {Freericks}},\ }\bibfield
  {title} {\bibinfo {title} {Anisotropic light‑tailored {RKKY} interaction in
  two‑dimensional \(d\)-wave altermagnets},\ }\href
  {https://doi.org/10.1103/k3xb-8pts} {\bibfield  {journal} {\bibinfo
  {journal} {Phys. Rev. B}\ }\textbf {\bibinfo {volume} {111}},\ \bibinfo
  {pages} {224412} (\bibinfo {year} {2025})}\BibitemShut {NoStop}%
\bibitem [{\citenamefont {Yokoyama}(2025)}]{yokoyama2025FloquetSC}%
  \BibitemOpen
  \bibfield  {author} {\bibinfo {author} {\bibfnamefont {T.}~\bibnamefont
  {Yokoyama}},\ }\bibfield  {title} {\bibinfo {title} {Floquet engineering
  triplet superconductivity in superconductors with spin-orbit coupling or
  altermagnetism},\ }\href {https://doi.org/10.1103/4tng-rhc4} {\bibfield
  {journal} {\bibinfo  {journal} {Phys. Rev. B}\ }\textbf {\bibinfo {volume}
  {112}},\ \bibinfo {pages} {024512} (\bibinfo {year} {2025})}\BibitemShut
  {NoStop}%
\bibitem [{\citenamefont {Song}\ \emph {et~al.}(2025)\citenamefont {Song},
  \citenamefont {Stavri{\'c}}, \citenamefont {Barone}, \citenamefont
  {Droghetti}, \citenamefont {Antonenko}, \citenamefont {Venderbos},
  \citenamefont {Occhialini}, \citenamefont {Ilyas}, \citenamefont {Ergeçen},
  \citenamefont {Gedik}, \citenamefont {Cheong}, \citenamefont {Fernandes},
  \citenamefont {Picozzi},\ and\ \citenamefont {Comin}}]{Song2025Electrical}%
  \BibitemOpen
  \bibfield  {author} {\bibinfo {author} {\bibfnamefont {Q.}~\bibnamefont
  {Song}}, \bibinfo {author} {\bibfnamefont {S.}~\bibnamefont {Stavri{\'c}}},
  \bibinfo {author} {\bibfnamefont {P.}~\bibnamefont {Barone}}, \bibinfo
  {author} {\bibfnamefont {A.}~\bibnamefont {Droghetti}}, \bibinfo {author}
  {\bibfnamefont {D.~S.}\ \bibnamefont {Antonenko}}, \bibinfo {author}
  {\bibfnamefont {J.~W.~F.}\ \bibnamefont {Venderbos}}, \bibinfo {author}
  {\bibfnamefont {C.~A.}\ \bibnamefont {Occhialini}}, \bibinfo {author}
  {\bibfnamefont {B.}~\bibnamefont {Ilyas}}, \bibinfo {author} {\bibfnamefont
  {E.}~\bibnamefont {Ergeçen}}, \bibinfo {author} {\bibfnamefont
  {N.}~\bibnamefont {Gedik}}, \bibinfo {author} {\bibfnamefont {S.-W.}\
  \bibnamefont {Cheong}}, \bibinfo {author} {\bibfnamefont {R.~M.}\
  \bibnamefont {Fernandes}}, \bibinfo {author} {\bibfnamefont {S.}~\bibnamefont
  {Picozzi}},\ and\ \bibinfo {author} {\bibfnamefont {R.}~\bibnamefont
  {Comin}},\ }\bibfield  {title} {\bibinfo {title} {Electrical switching of a
  $p$‑wave magnet},\ }\href {https://doi.org/10.1038/s41586-025-09034-7}
  {\bibfield  {journal} {\bibinfo  {journal} {Nature}\ }\textbf {\bibinfo
  {volume} {642}},\ \bibinfo {pages} {64} (\bibinfo {year} {2025})}\BibitemShut
  {NoStop}%
\bibitem [{\citenamefont {Yamada}\ \emph {et~al.}(2025)\citenamefont {Yamada},
  \citenamefont {Birch}, \citenamefont {Baral}, \citenamefont {Okumura},
  \citenamefont {Nakano}, \citenamefont {Gao}, \citenamefont {Ezawa},
  \citenamefont {Nomoto}, \citenamefont {Masell}, \citenamefont {Ishihara},
  \citenamefont {Kolincio}, \citenamefont {Belopolski}, \citenamefont
  {Sagayama}, \citenamefont {Nakao}, \citenamefont {Ohishi}, \citenamefont
  {Ohhara}, \citenamefont {Kiyanagi}, \citenamefont {Nakajima}, \citenamefont
  {Tokura}, \citenamefont {Arima}, \citenamefont {Motome}, \citenamefont
  {Hirschmann},\ and\ \citenamefont {Hirschberger}}]{Yamada2025Metallic}%
  \BibitemOpen
  \bibfield  {author} {\bibinfo {author} {\bibfnamefont {R.}~\bibnamefont
  {Yamada}}, \bibinfo {author} {\bibfnamefont {M.~T.}\ \bibnamefont {Birch}},
  \bibinfo {author} {\bibfnamefont {P.~R.}\ \bibnamefont {Baral}}, \bibinfo
  {author} {\bibfnamefont {S.}~\bibnamefont {Okumura}}, \bibinfo {author}
  {\bibfnamefont {R.}~\bibnamefont {Nakano}}, \bibinfo {author} {\bibfnamefont
  {S.}~\bibnamefont {Gao}}, \bibinfo {author} {\bibfnamefont {M.}~\bibnamefont
  {Ezawa}}, \bibinfo {author} {\bibfnamefont {T.}~\bibnamefont {Nomoto}},
  \bibinfo {author} {\bibfnamefont {J.}~\bibnamefont {Masell}}, \bibinfo
  {author} {\bibfnamefont {Y.}~\bibnamefont {Ishihara}}, \bibinfo {author}
  {\bibfnamefont {K.~K.}\ \bibnamefont {Kolincio}}, \bibinfo {author}
  {\bibfnamefont {I.}~\bibnamefont {Belopolski}}, \bibinfo {author}
  {\bibfnamefont {H.}~\bibnamefont {Sagayama}}, \bibinfo {author}
  {\bibfnamefont {H.}~\bibnamefont {Nakao}}, \bibinfo {author} {\bibfnamefont
  {K.}~\bibnamefont {Ohishi}}, \bibinfo {author} {\bibfnamefont
  {T.}~\bibnamefont {Ohhara}}, \bibinfo {author} {\bibfnamefont
  {R.}~\bibnamefont {Kiyanagi}}, \bibinfo {author} {\bibfnamefont
  {T.}~\bibnamefont {Nakajima}}, \bibinfo {author} {\bibfnamefont
  {Y.}~\bibnamefont {Tokura}}, \bibinfo {author} {\bibfnamefont
  {T.}~\bibnamefont {Arima}}, \bibinfo {author} {\bibfnamefont
  {Y.}~\bibnamefont {Motome}}, \bibinfo {author} {\bibfnamefont {M.~M.}\
  \bibnamefont {Hirschmann}},\ and\ \bibinfo {author} {\bibfnamefont
  {M.}~\bibnamefont {Hirschberger}},\ }\bibfield  {title} {\bibinfo {title} {A
  metallic $p$‑wave magnet with commensurate spin helix},\ }\href
  {https://doi.org/10.1038/s41586-025-09633-4} {\bibfield  {journal} {\bibinfo
  {journal} {Nature}\ }\textbf {\bibinfo {volume} {646}},\ \bibinfo {pages}
  {837} (\bibinfo {year} {2025})}\BibitemShut {NoStop}%
\bibitem [{\citenamefont {Esin}\ \emph {et~al.}(2020)\citenamefont {Esin},
  \citenamefont {Rudner},\ and\ \citenamefont {Lindner}}]{Esin2020Floquet}%
  \BibitemOpen
  \bibfield  {author} {\bibinfo {author} {\bibfnamefont {I.}~\bibnamefont
  {Esin}}, \bibinfo {author} {\bibfnamefont {M.~S.}\ \bibnamefont {Rudner}},\
  and\ \bibinfo {author} {\bibfnamefont {N.~H.}\ \bibnamefont {Lindner}},\
  }\bibfield  {title} {\bibinfo {title} {Floquet metal‑to‑insulator phase
  transitions in semiconductor nanowires},\ }\href
  {https://doi.org/10.1126/sciadv.aay4922} {\bibfield  {journal} {\bibinfo
  {journal} {Sci. Adv.}\ }\textbf {\bibinfo {volume} {6}},\ \bibinfo {pages}
  {eaay4922} (\bibinfo {year} {2020})}\BibitemShut {NoStop}%
\bibitem [{\citenamefont {Jungwirth}\ \emph {et~al.}(2025)\citenamefont
  {Jungwirth}, \citenamefont {Sinova}, \citenamefont {Fernandes}, \citenamefont
  {Liu}, \citenamefont {Watanabe}, \citenamefont {Murakami}, \citenamefont
  {Nakatsuji},\ and\ \citenamefont {Šmejkal}}]{Jungwirth2025Symmetry}%
  \BibitemOpen
  \bibfield  {author} {\bibinfo {author} {\bibfnamefont {T.}~\bibnamefont
  {Jungwirth}}, \bibinfo {author} {\bibfnamefont {J.}~\bibnamefont {Sinova}},
  \bibinfo {author} {\bibfnamefont {R.~M.}\ \bibnamefont {Fernandes}}, \bibinfo
  {author} {\bibfnamefont {Q.}~\bibnamefont {Liu}}, \bibinfo {author}
  {\bibfnamefont {H.}~\bibnamefont {Watanabe}}, \bibinfo {author}
  {\bibfnamefont {S.}~\bibnamefont {Murakami}}, \bibinfo {author}
  {\bibfnamefont {S.}~\bibnamefont {Nakatsuji}},\ and\ \bibinfo {author}
  {\bibfnamefont {L.}~\bibnamefont {Šmejkal}},\ }\bibfield  {title} {\bibinfo
  {title} {Symmetry, microscopy and spectroscopy signatures of
  altermagnetism},\ }\href {10.48550/arXiv.2506.22860} {\bibfield  {journal}
  {\bibinfo  {journal} {arXiv:2506.22860}\ } (\bibinfo {year}
  {2025})}\BibitemShut {NoStop}%
\bibitem [{\citenamefont {Dehghani}\ \emph {et~al.}(2021)\citenamefont
  {Dehghani}, \citenamefont {Hafezi},\ and\ \citenamefont
  {Ghaemi}}]{Dehghani2021Light}%
  \BibitemOpen
  \bibfield  {author} {\bibinfo {author} {\bibfnamefont {H.}~\bibnamefont
  {Dehghani}}, \bibinfo {author} {\bibfnamefont {M.}~\bibnamefont {Hafezi}},\
  and\ \bibinfo {author} {\bibfnamefont {P.}~\bibnamefont {Ghaemi}},\
  }\bibfield  {title} {\bibinfo {title} {Light‑induced topological
  superconductivity via {F}loquet interaction engineering},\ }\href
  {https://doi.org/10.1103/PhysRevResearch.3.023039} {\bibfield  {journal}
  {\bibinfo  {journal} {Phys. Rev. Research}\ }\textbf {\bibinfo {volume}
  {3}},\ \bibinfo {pages} {023039} (\bibinfo {year} {2021})}\BibitemShut
  {NoStop}%
\bibitem [{\citenamefont {Kennes}\ \emph {et~al.}(2019)\citenamefont {Kennes},
  \citenamefont {Claassen}, \citenamefont {Sentef},\ and\ \citenamefont
  {Karrasch}}]{Kennes2019Light}%
  \BibitemOpen
  \bibfield  {author} {\bibinfo {author} {\bibfnamefont {D.~M.}\ \bibnamefont
  {Kennes}}, \bibinfo {author} {\bibfnamefont {M.}~\bibnamefont {Claassen}},
  \bibinfo {author} {\bibfnamefont {M.~A.}\ \bibnamefont {Sentef}},\ and\
  \bibinfo {author} {\bibfnamefont {C.}~\bibnamefont {Karrasch}},\ }\bibfield
  {title} {\bibinfo {title} {Light‑induced d‑wave superconductivity through
  {F}loquet‑engineered {F}ermi surfaces in cuprates},\ }\href
  {https://doi.org/10.1103/PhysRevB.100.075115} {\bibfield  {journal} {\bibinfo
   {journal} {Phys. Rev. B}\ }\textbf {\bibinfo {volume} {100}},\ \bibinfo
  {pages} {075115} (\bibinfo {year} {2019})}\BibitemShut {NoStop}%
\bibitem [{\citenamefont {Kitamura}\ and\ \citenamefont
  {Aoki}(2022)}]{Kitamura2022Floquet}%
  \BibitemOpen
  \bibfield  {author} {\bibinfo {author} {\bibfnamefont {S.}~\bibnamefont
  {Kitamura}}\ and\ \bibinfo {author} {\bibfnamefont {H.}~\bibnamefont
  {Aoki}},\ }\bibfield  {title} {\bibinfo {title} {Floquet topological
  superconductivity induced by chiral many-body interaction},\ }\href
  {https://doi.org/10.1038/s42005-022-00936-w} {\bibfield  {journal} {\bibinfo
  {journal} {Commun. Phys.}\ }\textbf {\bibinfo {volume} {5}},\ \bibinfo
  {pages} {174} (\bibinfo {year} {2022})}\BibitemShut {NoStop}%
\bibitem [{\citenamefont {Ashcroft}\ and\ \citenamefont
  {Mermin}(1976)}]{Ashcroft1976Solid}%
  \BibitemOpen
  \bibfield  {author} {\bibinfo {author} {\bibfnamefont {N.~W.}\ \bibnamefont
  {Ashcroft}}\ and\ \bibinfo {author} {\bibfnamefont {N.~D.}\ \bibnamefont
  {Mermin}},\ }\href@noop {} {\emph {\bibinfo {title} {Solid State Physics}}}\
  (\bibinfo  {publisher} {Holt, Rinehart and Winston},\ \bibinfo {address} {New
  York, NY},\ \bibinfo {year} {1976})\BibitemShut {NoStop}%
\bibitem [{\citenamefont {Jackson}(1999)}]{Jackson1999Classical}%
  \BibitemOpen
  \bibfield  {author} {\bibinfo {author} {\bibfnamefont {J.~D.}\ \bibnamefont
  {Jackson}},\ }\href@noop {} {\emph {\bibinfo {title} {Classical
  Electrodynamics}}},\ \bibinfo {edition} {3rd}\ ed.\ (\bibinfo  {publisher}
  {John Wiley \& Sons},\ \bibinfo {address} {Hoboken, NJ},\ \bibinfo {year}
  {1999})\BibitemShut {NoStop}%
\bibitem [{\citenamefont {Chen}\ \emph {et~al.}(2025)\citenamefont {Chen},
  \citenamefont {Zhou},\ and\ \citenamefont {Xu}}]{Chen2025Quasicrystalline}%
  \BibitemOpen
  \bibfield  {author} {\bibinfo {author} {\bibfnamefont {R.}~\bibnamefont
  {Chen}}, \bibinfo {author} {\bibfnamefont {B.}~\bibnamefont {Zhou}},\ and\
  \bibinfo {author} {\bibfnamefont {D.-H.}\ \bibnamefont {Xu}},\ }\bibfield
  {title} {\bibinfo {title} {Quasicrystalline altermagnetism},\ }\href
  {10.48550/arXiv.2507.18408} {\bibfield  {journal} {\bibinfo  {journal}
  {arXiv:2507.18408}\ } (\bibinfo {year} {2025})}\BibitemShut {NoStop}%
\bibitem [{\citenamefont {Prada}\ \emph {et~al.}(2020)\citenamefont {Prada},
  \citenamefont {San‑Jose}, \citenamefont {de~Moor}, \citenamefont {Geresdi},
  \citenamefont {Lee}, \citenamefont {Klinovaja}, \citenamefont {Loss},
  \citenamefont {Nygård}, \citenamefont {Aguado},\ and\ \citenamefont
  {Kouwenhoven}}]{Prada2020From}%
  \BibitemOpen
  \bibfield  {author} {\bibinfo {author} {\bibfnamefont {E.}~\bibnamefont
  {Prada}}, \bibinfo {author} {\bibfnamefont {P.}~\bibnamefont {San‑Jose}},
  \bibinfo {author} {\bibfnamefont {M.~W.~A.}\ \bibnamefont {de~Moor}},
  \bibinfo {author} {\bibfnamefont {A.}~\bibnamefont {Geresdi}}, \bibinfo
  {author} {\bibfnamefont {E.~J.~H.}\ \bibnamefont {Lee}}, \bibinfo {author}
  {\bibfnamefont {J.}~\bibnamefont {Klinovaja}}, \bibinfo {author}
  {\bibfnamefont {D.}~\bibnamefont {Loss}}, \bibinfo {author} {\bibfnamefont
  {J.}~\bibnamefont {Nygård}}, \bibinfo {author} {\bibfnamefont
  {R.}~\bibnamefont {Aguado}},\ and\ \bibinfo {author} {\bibfnamefont {L.~P.}\
  \bibnamefont {Kouwenhoven}},\ }\bibfield  {title} {\bibinfo {title} {From
  {A}ndreev to {M}ajorana bound states in hybrid superconductor–semiconductor
  nanowires},\ }\href {https://doi.org/10.1038/s42254-020-0228-y} {\bibfield
  {journal} {\bibinfo  {journal} {Nat. Rev. Phys.}\ }\textbf {\bibinfo {volume}
  {2}},\ \bibinfo {pages} {575} (\bibinfo {year} {2020})}\BibitemShut {NoStop}%
\bibitem [{\citenamefont {Machida}\ \emph {et~al.}(1980)\citenamefont
  {Machida}, \citenamefont {Nokura},\ and\ \citenamefont
  {Matsubara}}]{Machida1980Theory}%
  \BibitemOpen
  \bibfield  {author} {\bibinfo {author} {\bibfnamefont {K.}~\bibnamefont
  {Machida}}, \bibinfo {author} {\bibfnamefont {K.}~\bibnamefont {Nokura}},\
  and\ \bibinfo {author} {\bibfnamefont {T.}~\bibnamefont {Matsubara}},\
  }\bibfield  {title} {\bibinfo {title} {Theory of antiferromagnetic
  superconductors},\ }\href {https://doi.org/10.1103/PhysRevB.22.2307}
  {\bibfield  {journal} {\bibinfo  {journal} {Phys. Rev. B}\ }\textbf {\bibinfo
  {volume} {22}},\ \bibinfo {pages} {2307} (\bibinfo {year}
  {1980})}\BibitemShut {NoStop}%
\bibitem [{\citenamefont {Golubov}\ \emph {et~al.}(2025)\citenamefont
  {Golubov}, \citenamefont {Bakurskiy}, \citenamefont {Kupriyanov},
  \citenamefont {Karabassov}, \citenamefont {Vasenko},\ and\ \citenamefont
  {Sidorenko}}]{Golubov2025Physics}%
  \BibitemOpen
  \bibfield  {author} {\bibinfo {author} {\bibfnamefont {A.~A.}\ \bibnamefont
  {Golubov}}, \bibinfo {author} {\bibfnamefont {S.~V.}\ \bibnamefont
  {Bakurskiy}}, \bibinfo {author} {\bibfnamefont {M.~Y.}\ \bibnamefont
  {Kupriyanov}}, \bibinfo {author} {\bibfnamefont {T.}~\bibnamefont
  {Karabassov}}, \bibinfo {author} {\bibfnamefont {A.~S.}\ \bibnamefont
  {Vasenko}},\ and\ \bibinfo {author} {\bibfnamefont {A.~S.}\ \bibnamefont
  {Sidorenko}},\ }\bibfield  {title} {\bibinfo {title} {The physics of
  superconductor–ferromagnet hybrid structures},\ }\href
  {10.48550/arXiv.2509.16387} {\bibfield  {journal} {\bibinfo  {journal}
  {arXiv:2509.16387}\ } (\bibinfo {year} {2025})}\BibitemShut {NoStop}%
\bibitem [{\citenamefont {Vagov}\ \emph {et~al.}(2023)\citenamefont {Vagov},
  \citenamefont {Saraiva}, \citenamefont {Shanenko}, \citenamefont {Vasenko},
  \citenamefont {Aguiar}, \citenamefont {Stolyarov},\ and\ \citenamefont
  {Roditchev}}]{Vagov2023Intertype}%
  \BibitemOpen
  \bibfield  {author} {\bibinfo {author} {\bibfnamefont {A.}~\bibnamefont
  {Vagov}}, \bibinfo {author} {\bibfnamefont {T.~T.}\ \bibnamefont {Saraiva}},
  \bibinfo {author} {\bibfnamefont {A.~A.}\ \bibnamefont {Shanenko}}, \bibinfo
  {author} {\bibfnamefont {A.~S.}\ \bibnamefont {Vasenko}}, \bibinfo {author}
  {\bibfnamefont {J.~A.}\ \bibnamefont {Aguiar}}, \bibinfo {author}
  {\bibfnamefont {V.~S.}\ \bibnamefont {Stolyarov}},\ and\ \bibinfo {author}
  {\bibfnamefont {D.}~\bibnamefont {Roditchev}},\ }\bibfield  {title} {\bibinfo
  {title} {Intertype superconductivity in ferromagnetic superconductors},\
  }\href {https://www.nature.com/articles/s42005-023-01395-7} {\bibfield
  {journal} {\bibinfo  {journal} {Commun. Phys.}\ }\textbf {\bibinfo {volume}
  {6}},\ \bibinfo {pages} {284} (\bibinfo {year} {2023})}\BibitemShut {NoStop}%
\bibitem [{\citenamefont {Zareapour}\ \emph {et~al.}(2012)\citenamefont
  {Zareapour}, \citenamefont {Hayat}, \citenamefont {Zhao}, \citenamefont
  {Kreshchuk}, \citenamefont {Jain}, \citenamefont {Kwok}, \citenamefont {Lee},
  \citenamefont {Cheong}, \citenamefont {Xu}, \citenamefont {Yang},
  \citenamefont {Gu}, \citenamefont {Jia}, \citenamefont {Cava},\ and\
  \citenamefont {Burch}}]{Zareapour2012Proximity}%
  \BibitemOpen
  \bibfield  {author} {\bibinfo {author} {\bibfnamefont {P.}~\bibnamefont
  {Zareapour}}, \bibinfo {author} {\bibfnamefont {A.}~\bibnamefont {Hayat}},
  \bibinfo {author} {\bibfnamefont {S.~Y.~F.}\ \bibnamefont {Zhao}}, \bibinfo
  {author} {\bibfnamefont {M.}~\bibnamefont {Kreshchuk}}, \bibinfo {author}
  {\bibfnamefont {A.}~\bibnamefont {Jain}}, \bibinfo {author} {\bibfnamefont
  {D.~C.}\ \bibnamefont {Kwok}}, \bibinfo {author} {\bibfnamefont
  {N.}~\bibnamefont {Lee}}, \bibinfo {author} {\bibfnamefont {S.-W.}\
  \bibnamefont {Cheong}}, \bibinfo {author} {\bibfnamefont {Z.}~\bibnamefont
  {Xu}}, \bibinfo {author} {\bibfnamefont {A.}~\bibnamefont {Yang}}, \bibinfo
  {author} {\bibfnamefont {G.}~\bibnamefont {Gu}}, \bibinfo {author}
  {\bibfnamefont {S.}~\bibnamefont {Jia}}, \bibinfo {author} {\bibfnamefont
  {R.~J.}\ \bibnamefont {Cava}},\ and\ \bibinfo {author} {\bibfnamefont
  {K.~S.}\ \bibnamefont {Burch}},\ }\bibfield  {title} {\bibinfo {title}
  {Proximity-induced high-temperature superconductivity in the topological
  insulators {Bi$_2$Se$_3$} and {Bi$_2$Te$_3$}},\ }\href
  {https://doi.org/10.1038/ncomms2042} {\bibfield  {journal} {\bibinfo
  {journal} {Nat. Commun.}\ }\textbf {\bibinfo {volume} {3}},\ \bibinfo {pages}
  {1056} (\bibinfo {year} {2012})}\BibitemShut {NoStop}%
\bibitem [{\citenamefont {L\"ofwander}(2004)}]{Lofwander2004Proximity}%
  \BibitemOpen
  \bibfield  {author} {\bibinfo {author} {\bibfnamefont {T.}~\bibnamefont
  {L\"ofwander}},\ }\bibfield  {title} {\bibinfo {title} {Proximity effect in
  normal metal–high-{$T_c$} superconductor contacts},\ }\href
  {https://doi.org/10.1103/PhysRevB.70.094518} {\bibfield  {journal} {\bibinfo
  {journal} {Phys. Rev. B}\ }\textbf {\bibinfo {volume} {70}},\ \bibinfo
  {pages} {094518} (\bibinfo {year} {2004})}\BibitemShut {NoStop}%
\bibitem [{\citenamefont {Kuhn}\ \emph {et~al.}(2024)\citenamefont {Kuhn},
  \citenamefont {Sothmann},\ and\ \citenamefont {Cayao}}]{PhysRevB.109.134517}%
  \BibitemOpen
  \bibfield  {author} {\bibinfo {author} {\bibfnamefont {T.}~\bibnamefont
  {Kuhn}}, \bibinfo {author} {\bibfnamefont {B.}~\bibnamefont {Sothmann}},\
  and\ \bibinfo {author} {\bibfnamefont {J.}~\bibnamefont {Cayao}},\ }\bibfield
   {title} {\bibinfo {title} {Floquet engineering {H}iggs dynamics in
  time-periodic superconductors},\ }\href
  {https://doi.org/10.1103/PhysRevB.109.134517} {\bibfield  {journal} {\bibinfo
   {journal} {Phys. Rev. B}\ }\textbf {\bibinfo {volume} {109}},\ \bibinfo
  {pages} {134517} (\bibinfo {year} {2024})}\BibitemShut {NoStop}%
\bibitem [{\citenamefont {Mizushima}\ \emph {et~al.}(2025)\citenamefont
  {Mizushima}, \citenamefont {Tanaka},\ and\ \citenamefont
  {Cayao}}]{Mizushima2025Detecting}%
  \BibitemOpen
  \bibfield  {author} {\bibinfo {author} {\bibfnamefont {T.}~\bibnamefont
  {Mizushima}}, \bibinfo {author} {\bibfnamefont {Y.}~\bibnamefont {Tanaka}},\
  and\ \bibinfo {author} {\bibfnamefont {J.}~\bibnamefont {Cayao}},\ }\bibfield
   {title} {\bibinfo {title} {Detecting the topological phase transition in
  superconductor-semiconductor hybrids by electronic {R}aman spectroscopy},\
  }\href {https://doi.org/10.1103/c325-kgbf} {\bibfield  {journal} {\bibinfo
  {journal} {Phys. Rev. B}\ }\textbf {\bibinfo {volume} {112}},\ \bibinfo
  {pages} {174504} (\bibinfo {year} {2025})}\BibitemShut {NoStop}%
\bibitem [{\citenamefont {Fu}\ and\ \citenamefont {Cayao}()}]{phjc}%
  \BibitemOpen
  \bibfield  {author} {\bibinfo {author} {\bibfnamefont {P.-H.}\ \bibnamefont
  {Fu}}\ and\ \bibinfo {author} {\bibfnamefont {J.}~\bibnamefont {Cayao}},\
  }\href@noop {} {}\bibinfo {note} {To be published somewhere else}\BibitemShut
  {NoStop}%
\bibitem [{\citenamefont {Cayao}\ \emph {et~al.}(2021)\citenamefont {Cayao},
  \citenamefont {Triola},\ and\ \citenamefont
  {Black-Schaffer}}]{PhysRevB.103.104505}%
  \BibitemOpen
  \bibfield  {author} {\bibinfo {author} {\bibfnamefont {J.}~\bibnamefont
  {Cayao}}, \bibinfo {author} {\bibfnamefont {C.}~\bibnamefont {Triola}},\ and\
  \bibinfo {author} {\bibfnamefont {A.~M.}\ \bibnamefont {Black-Schaffer}},\
  }\bibfield  {title} {\bibinfo {title} {Floquet engineering bulk odd-frequency
  superconducting pairs},\ }\href {https://doi.org/10.1103/PhysRevB.103.104505}
  {\bibfield  {journal} {\bibinfo  {journal} {Phys. Rev. B}\ }\textbf {\bibinfo
  {volume} {103}},\ \bibinfo {pages} {104505} (\bibinfo {year}
  {2021})}\BibitemShut {NoStop}%
\end{thebibliography}
\end{document}

\title{Supplemental Material for "Floquet engineering spin triplet states in unconventional magnets"}

\author{Pei-Hao Fu}
\email{phy.phfu@gmail.com}
%\affiliation{Department of Physics and Astronomy, Uppsala University, Box 516, S-751 20 Uppsala, Sweden}
\affiliation{School of Physics and Materials Science, Guangzhou University, Guangzhou 510006, China}

\author{Sayan Mondal}
%\email{sayan.mondal@physics.uu.se}
\affiliation{Department of Physics and Astronomy, Uppsala University, Box 516, S-751 20 Uppsala, Sweden}

\author{Jun-Feng Liu}
%\email{phjfliu@gzhu.edu.cn}
\affiliation{School of Physics and Materials Science, Guangzhou University, Guangzhou 510006, China}

 \author{Yukio Tanaka}
 %\email{ytanaka@nuap.nagoya-u.ac.jp}
 \affiliation{Department of Applied Physics,
 Nagoya University, 464-8603 Nagoya, Japan}
 \affiliation{Research Center for Crystalline Materials Engineering, Nagoya University, 464-8603  Nagoya Japan}

\author{Jorge Cayao}
\email{jorge.cayao@physics.uu.se}
\affiliation{Department of Physics and Astronomy, Uppsala University, Box 516, S-751 20 Uppsala, Sweden}

\date{\today}
\maketitle
\tableofcontents
\newpage
%%%%%%%%%%%%%%%%%%%%%%%%%%%%%%%%%%%%%%%%%%%%%%%%%%%%%%%%%%%%%
\section{Floquet engineering spin density in unconventional magnets} \label{SecI}

In the Floquet engineering spin density studied in the main text, we focus on two representative unconventional magnets, i.e., the $d$-wave and $p$-wave magnets, which are the simplest cases for unconventional magnets with even- and odd-parity momentum, respectively.
In this section, for completeness and universality, we generalize our discussion to all unconventional magnets with higher-order parities, including $f$-, $g$-, and $i$-wave magnets. 
We begin with introducing a general Hamiltonian for unconventional magnets in the static regime in Sec. \ref{SecIA}, and then we demonstrate the high-frequency effective Hamiltonian in Sec. \ref{SecIB} under both circularly and linearly polarized driving fields.
In Sec. \ref{SecIC}, the light-induced spin density and its peak-dip structures are studied.
A summary is given in Sec. \ref{SecID}, followed by some calculation details in Sec. \ref{SecIE}.

By generalizing the discussion to unconventional magnets with higher-order parities, we find that the central conclusion in the main text, based on $d$-wave and $p$-wave magnets, can be generalized:
For even-parity unconventional magnets (i.e., $d-$, $g$- and $i$-wave magnets), the peak–dip structure in the spin density provides a direct probe of the altermagnetic strength and orientation, whereas for their odd-parity counterparts (i.e., $p$- and $f$-wave magnets), this feature is absent. 
The details are elaborated below.

\subsection{General Hamiltonian of unconventional magnets in the static regime} \label{SecIA}
We begin by introducing a general Hamiltonian of unconventional magnets in the static regime.
For an unconventional magnet with $j$-wave parity, the Hamiltonian is given by
\begin{equation}
H^{j}(\bm{k}) = \xi_{\bm{k}}\,\sigma_{0} + M_{\bm{k}}^{j}\,\sigma_{z},
\label{eq_hjk}
\end{equation}
where
\begin{equation}
\xi_{\bm{k}} = \frac{\hbar^{2}}{2m}(k_{x}^{2} + k_{y}^{2}) - \mu,
\label{eq_xik}
\end{equation}
is the kinetic energy measured from the chemical potential $\mu$, and $M_{\bm{k}}^{j}$ describes the unconventional magnetic order,
\begin{equation}
M_{\bm{k}}^{j} = \bar{M}^{j}\,k^{j}\cos\!\big[j(\theta_{k} - \theta_{j})\big],
\label{eq_mjk}
\end{equation}
with $j=\{0,1,2,3,4,6\}$ corresponding to $s$-, $p$-, $d$-, $f$-, $g$-, and $i$-wave parities of the unconventional magnets, respectively, reflecting the underlying Fermi surface symmetry. 
Expanding Eq.\,(\ref{eq_mjk}) yields explicit forms \cite{FukayaCayaoReviewUMs}:
\begin{eqnarray}
M_{\bm{k}}^{s} &=& \bar{M}^{s}, 
\label{eq_msk} \\
M_{\bm{k}}^{p} &=& \bar{M}^{p}\big(k_{x}\cos\theta_{p} + k_{y}\sin\theta_{p}\big),
\label{eq_mpk} \\
M_{\bm{k}}^{d} &=& \bar{M}^{d}\big[\,2k_{x}k_{y}\sin2\theta_{d} + (k_{x}^{2} - k_{y}^{2})\cos2\theta_{d}\big],
\label{eq_mdk} \\
M_{\bm{k}}^{f} &=& \bar{M}^{f}\big[\,k_{x}(k_{x}^{2}-3k_{y}^{2})\cos3\theta_{f}
+ k_{y}(k_{y}^{2}-3k_{x}^{2})\sin3\theta_{f}\big],
\label{eq_mfk} \\
M_{\bm{k}}^{g} &=& \bar{M}^{g}\big[\, (k_{x}^{4}-6k_{x}^{2}k_{y}^{2}+k_{y}^{4})\cos4\theta_{g}
+ 4k_{x}k_{y}(k_{x}^{2}-k_{y}^{2})\sin4\theta_{g}\big],
\label{eq_mgk} \\
M_{\bm{k}}^{i} &=& \bar{M}^{i}\big[\, (k_{x}^{6}-15k_{x}^{4}k_{y}^{2}+15k_{x}^{2}k_{y}^{4}-k_{y}^{6})\cos6\theta_{i}
+ 2k_{x}k_{y}(3k_{x}^{4}-10k_{x}^{2}k_{y}^{2}+3k_{y}^{4})\sin6\theta_{i}\big],
\label{eq_mik}
\end{eqnarray}
where $\bar{M}^{j}=M^{j}/k_{F}^{j}$ is the normalized magnetic strength. 
The case $j=0$ corresponds to the conventional ferromagnet Zeeman interaction, while $j=5$ is excluded because fivefold rotational symmetry is incompatible with crystalline periodicity \cite{Ezawa2025Third,Chen2025Quasicrystalline}. 
The $d$- and $p$-wave forms coincide with Eq.\,(2) in the main text.

Further classification within each unconventional magnet depends on the orientation of
the magnetic lobes, parameterized by the angle $\theta _{j}$ . For example,
in the $f$-wave magnet, the term $M_{\bm{k}}^{f}$ corresponds to the $%
f_{x(x^{2}-3y^{2})}$-wave configuration when $\theta _{f}=0$, and to the $%
f_{y(y^{2}-3x^{2})}$-wave configuration when $\theta _{f}=\pi /6$. One can
verify that along the momentum directions defined by $\theta _{k}=\theta
_{j}+n\pi /j$ ($n\in 
\mathbb{Z}
$), the unconventional magnetic effect is maximized, resulting in
spin-splitting that depends solely on the radial momentum magnitude \cite{FukayaCayaoReviewUMs}. In contrast, along directions $\theta _{k}=\theta _{j}+(2n+1)\pi
/(2j)$, the magnetic term vanishes, leading to spin-degenerate
band structures. Thus, these spin-degenerate nodal structures are directly
inherited from the symmetry of the underlying magnet.

\subsection{High-frequency effective Hamiltonian in the driven regime} \label{SecIB}
When the unconventional magnet is driven by light fields, 
the static Hamiltonian in Eq.\,(\ref{eq_hjk}) becomes time-dependent through minimal coupling, 
\begin{equation}
H^{j}(\bm{k}) \;\rightarrow\; H^{j}(\bm{k},t)
= H^{j}\!\left[\bm{k} + \frac{e}{\hbar}\bm{A}(t)\right].
\end{equation}
The time periodicity of $\bm{A}(t)$ allows a Floquet representation after Fourier transformation,
\begin{equation}
H_{+n}^{j}(\bm{k}) = \frac{1}{T}\!\int_{0}^{T}\! dt\, H^{j}(\bm{k},t)\,e^{in\Omega t},
\end{equation}
where $\Omega$ is the driving frequency.
As discussed in the manuscript, we consider two types of driving involving linearly and circularly polarized light.
The effects of both driving are described by the time-dependent vector potential as
\begin{equation}
\bm{A}_{\text{C}}\left( t\right) =\frac{E_{0}}{\Omega }\left( \cos \Omega t,\beta
\sin \Omega t\right) \text{,}  \label{eq_cpl}
\end{equation}%
for circularly polarized light and 
\begin{equation}
\bm{A}_{\text{L}}\left( t\right) =\frac{E_{0}}{\Omega }\cos \Omega t\left( \cos
\phi,\sin \phi\right) \text{,}  \label{eq_lpl}
\end{equation}%
for linearly polarized light.
Here, $\beta =+1$ ($-1$) denotes right-handed (left-handed) circular polarization, $\phi$ represents the real-space polarization direction as denoted in Fig. 1 in the main text, and $E_{0}$ is the magnitude of the electric component of light with a frequency $\Omega $. 
The definition the driving fields given Eqs.\,(\ref{eq_cpl}) and (\ref{eq_lpl}) are shown in the paragraph before Eq. (3) of the main text.

For circularly polarized light, the Floquet components of the Hamiltonian are given as
\begin{eqnarray}
H_{0}^{j}(\bm{k}) &=& H^{j}(\bm{k}) + \frac{\hbar^{2}}{2m}(eA_{0})^{2},
\label{eq_hq0} \\
H_{+n}^{j}(\bm{k}) &=&
\frac{e\hbar^{2}}{2m}A_{0}k\,e^{i\beta\theta_{k}}\,\delta_{n,1}
+ \sigma_{z}\,\bar{M}^{j}\,e^{\beta i j\theta_{j}}
\frac{j!}{2(j-n)!n!}\,
\big(k\,e^{-i\beta\theta_{k}}\big)^{j-n}(eA_{0})^{n},
\label{eq_hqn}
\end{eqnarray}
where $\beta=+/-1$ denotes the right/left handedness of circularly polarized light and $H_{-n}^{j}(\bm{k})=[H_{+n}^{j}(\bm{k})]^{\dagger}$.
By comparing $f$-, $g$-, and $i$-wave magnets with the $d$- and $p$-wave cases shown in Eq. (4) in the main text, higher-order Floquet components $H_{n>2}^{j}(\bm{k})$ may appear for $f$-, $g$-, and $i$-wave magnets due to their higher-order momentum dependence. 
However, since all $H_{\pm n}^{j}(\bm{k})$ are proportional to $\sigma_{0}$ and $\sigma_{z}$, their commutator vanishes,
\begin{equation}
\big[H_{+n}^{j}(\bm{k}), H_{-n}^{j}(\bm{k})\big] = 0.
\end{equation}
This then leads to the high-frequency effective Hamiltonian given by
\begin{equation}
H_{\text{eff}}^{j}(\bm{k})
= H_{0}^{j}(\bm{k})
+ \sum_{n}\frac{[H_{+n}^{j}(\bm{k}),H_{-n}^{j}(\bm{k})]}{n\hbar\Omega}
+ O(\Omega^{-2})
= H_{0}^{j}(\bm{k}),
\label{eq_hcpleff}
\end{equation}
solely determined by $H_{0}^{j}(\bm{k})$ [Eq.\,(\ref{eq_hq0})] and identical to the static Hamiltonian [Eq.\,(\ref{eq_hjk})] except for a renormalized chemical potential $(\hbar e A_{0})^{2}/(2m)$ due to the self-doping effect. 
Consequently, {no} light-induced effect is developed by a circularly polarized drive, since the system’s properties remain equivalent to those of the static state.
This result is consistent with Eq.\,(4) and the following discussion carried in the main text for the $d$- and $p$-wave cases, where we have verified that spin density or spin-triplet pair amplitude remains zero due to the vanishing net magnetization nature of unconventional magnets.

In the case of linearly polarized driving, the Floquet components become more complicated. 
Nevertheless, $H_{+n}^{j}(\bm{k})$ remains proportional to $\sigma_{0}$ and $\sigma_{z}$, 
so the commutator $\big[H_{+n}^{j}(\bm{k}),H_{-n}^{j}(\bm{k})\big]$ still vanishes, which is confirmed in Sec. \ref{SecIE}. 
Consequently, the effective Hamiltonian in the high-frequency limit is fully determined by $H_{0}^{j}(\bm{k})$,  
\begin{equation}
H_{\text{eff}}^{j}(\bm{k})
= H_{0}^{j}(\bm{k})
= H^{j}(\bm{k})
+ \frac{(e\hbar A_{0})^{2}}{4m}
+ \sigma_{z}M_{\Omega}^{j}
+ \sigma_{z}B_{\bm{k}\Omega}^{j},
\label{eq_hjeff}
\end{equation}
where $H^{j}(\bm{k})$ is the static Hamiltonian [Eq.\,(\ref{eq_hjk})] and $(e\hbar A_{0})^{2}/(4m)$ is a correction to the chemical potential and can be gauged away.
Moreover, $M_{\Omega}^{j}$ and $B_{\bm{k}\Omega}^{j}$ represent two light-induced fields that can be regarded as a light-induced Zeeman field and a light-induced unconventional magnetic field. 
Below, we present the explicit expressions of these two light-induced fields.

(i) Light-induced Zeeman field.
The term $M_{\Omega}^{j}$ in Eq.\,(\ref{eq_hjeff}) denotes the light-induced momentum-\textit{independent} spin splitting that can be seen as a light-induced Zeeman field. 
It appears in \textit{all} even-parity unconventional magnets and acquires the following form:
\begin{equation}
M_{\Omega}^{j}
= \frac{j-1}{2^{\,j-1}}\,
\bar{M}^{j}\,(eA_{0})^{j}\cos\!\big(j\Theta_{j}\big)\,\delta_{j,j^{\prime}},
\label{eq_mjO}
\end{equation}
where $\Theta_{j}=\theta_{j}-\phi$ and $j^{\prime}=\{2,4,6\}$ correspond to $d$-, $g$-, and $i$-wave symmetries, respectively. 
This light-induced Zeeman field originates from the coupling between the static unconventional magnetic field $\bar{M}^{j}$, 
its orientation $\Theta_{j}$, and the driving amplitude $(eA_{0})^{j}$, 
with the order $j$ determined by the symmetry of the underlying static magnets. 
The $d$-wave case ($j=2$) is discussed in detail in the main text.

(ii) Light-induced unconventional magnetic field.
The term $B_{\bm{k}\Omega}^{j}$ in Eq.\,(\ref{eq_hjeff}) describes a momentum-\textit{dependent} spin splitting, which can be regarded as a light-induced unconventional magnetic field. 
Its explicit form is given by
\begin{eqnarray} 
B_{\bm{k}\Omega}^{j}
&=&
\frac{3}{2}\,\bar{M}^{j}(eA_{0})^{2}k
\cos\!\big(3\theta_{f}-2\phi-\theta_{k}\big)\,\delta_{j,f}
\nonumber\\
&&+\,3\,\bar{M}^{j}(eA_{0})^{2}k^{2}
\cos\!\big(4\theta_{g}-2\phi-2\theta_{k}\big)\,\delta_{j,g}
\label{eq_BjkO}\\
&&+\frac{15}{8}\,\bar{M}^{j}(eA_{0})^{2}k^{2}
\!\left[4k^{2}\cos\!\big(6\theta_{i}-2\phi-4\theta_{k}\big)
+3(eA_{0})^{2}\cos\!\big(6\theta_{i}-4\phi-2\theta_{k}\big)\right]\!
\delta_{j,i}.
\nonumber
\end{eqnarray}
which is finite for $f$-, $g$-, and $i$-wave unconventional magnets with parity order $j \geq 3$ and therefore absent in the main text, which focuses on the simplest $p$- and $d$-wave cases.  
The crucial consequence of Eq.\,(\ref{eq_BjkO}) is that the light drive induces effective lower-order parity unconventional magnetism:
an effective $p$-wave magnet emerges from the $f$-wave state [first line of Eq.\,(\ref{eq_BjkO})], 
an effective $d$-wave magnet from the $g$-wave state [second line of Eq.\,(\ref{eq_BjkO})], 
and effective $d$- and $g$-wave components within the $i$-wave state [third line of Eq.\,(\ref{eq_BjkO})].

Eqs.\,(\ref{eq_mjO}) and (\ref{eq_BjkO}), therefore, demonstrate that applying linearly polarized light to unconventional magnets with $j$-wave symmetry can generate effective magnetic components with reduced parities $j' = j-2$, $j-4$, and $j-6$, provided that $j,\,j' > 0$. 
Consequently, for the $p$-wave case ($j=1$), the induced magnetic order would correspond to $j'=-1$, which is naturally forbidden. 
This explains why linearly polarized light does not affect $p$-wave magnets. 
Most importantly, for all even-parity unconventional magnets, a light-induced Zeeman term emerges [Eq.\,(\ref{eq_mjO})],  which lifts the spin degeneracy in the quasiparticle dispersion and generates a finite spin density, as we demonstrate below and aligned with the $d$-wave altermagnet studied in the main text.

\subsection{Light-indcued spin density} \label{SecIC} 
To demonstrate the effects of the light-induced Zeeman term and higher-order momentum contributions in higher-symmetry unconventional magnets, we evaluate the spin density following the same procedure described in the main text. 
The resulting spin density of states can be expressed as
\begin{equation}
\rho_{z}^{\text{eff}}(\bm{k},\omega)
= \rho_{z,\text{M}}^{\text{eff}}(\bm{k},\omega)
+ \rho_{z,\Omega}^{\text{eff}}(\bm{k},\omega)
+ \rho_{z,B}^{\text{eff}}(\bm{k},\omega),
\label{eq_rhoz}
\end{equation}
where
\begin{eqnarray}
\rho_{z,\text{M}}^{\text{eff}}(\bm{k},\omega)
&=& \frac{2}{\pi}\,\text{Im}\!\left[\frac{M_{\bm{k}}^{j}}{D_{\Omega}^{j}}\right],
\label{eq_rhozM} \\
\rho_{z,\Omega}^{\text{eff}}(\bm{k},\omega)
&=& \frac{2}{\pi}\,\text{Im}\!\left[\frac{M_{\Omega}^{j}}{D_{\Omega}^{j}}\right],
\label{eq_rhozO} \\
\rho_{z,B}^{\text{eff}}(\bm{k},\omega)
&=& \frac{2}{\pi}\,\text{Im}\!\left[\frac{B_{\bm{k}\Omega}^{j}}{D_{\Omega}^{j}}\right],
\label{eq_rhozF}
\end{eqnarray}
and the denominator is given by
\begin{equation}
D_{\Omega}^{j}
= \big(M_{\bm{k}}^{j} + M_{\Omega}^{j} + B_{\bm{k}\Omega}^{j}\big)^{2}
- \big(\omega + i\eta - \xi_{\bm{k}}\big)^{2}.
\end{equation}
where we have gauged out the chemical potential shift $(e\hbar A_{0})^{2}/(4m)$.
In Eq.\,(\ref{eq_rhoz})--(\ref{eq_rhozF}), $\rho_{z,\text{M}}^{\text{eff}}(\bm{k},\omega)$ arises from the static magnetic contribution and
$\rho_{z,\Omega}^{\text{eff}}(\bm{k},\omega)$ represents the spin density of states induced by the light-induced Zeeman field, which has been identified in the main text; while $\rho_{z,B}^{\text{eff}}(\bm{k},\omega)$ is caused by the light-induced unconventional magnet and occurs in the static unconventional magnet with parity order $j \geq 3$, which is therefore absent in the main text focusing on the simplest $p$- and $d$-wave cases.
To visualize the influence of the driving field, we perform momentum integration to obtain the total spin density of states,
\begin{equation}
\bar{\rho}_{z}^{\text{eff}}(\omega)
= \sum_{\alpha}\bar{\rho}_{z,\alpha}^{\text{eff}}(\omega).
\label{eq_sdosz}
\end{equation} 
where 
\begin{equation}
\bar{\rho}_{z,\alpha}^{\text{eff}}(\omega)
= \frac{1}{(2\pi)^{2}} \int d\bm{k}\,
\rho_{z,\alpha}^{\text{eff}}(\bm{k},\omega),
\end{equation}
with $\alpha = \{\text{M},\,\Omega,\,B\}$ corresponding to the components defined in Eqs.\,(\ref{eq_rhozM})–(\ref{eq_rhozF}).

\begin{figure}
    \centering
    \includegraphics[width=0.8\linewidth]{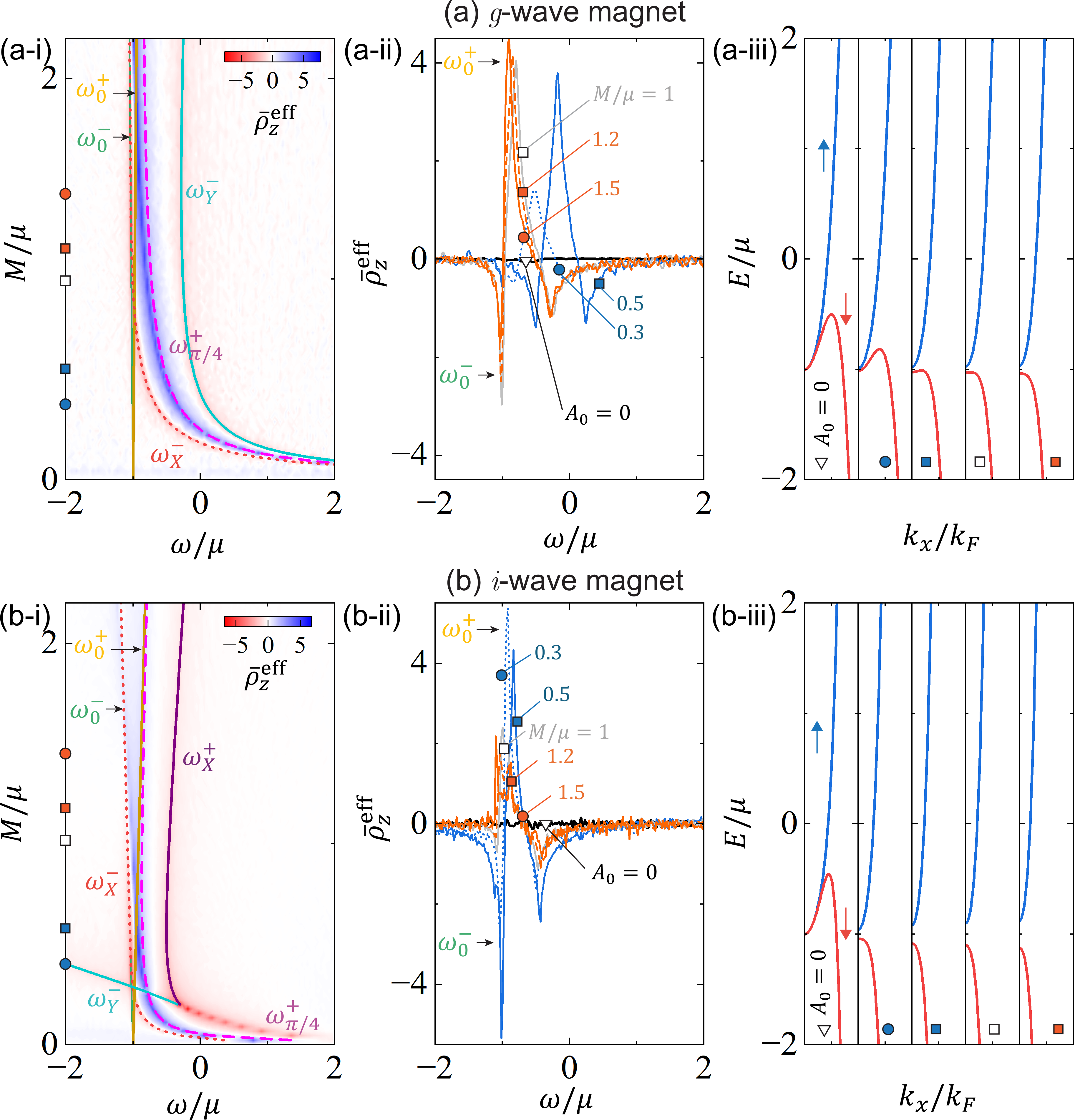}
    \caption{(a) Spin density and dispersion of driven $g$-wave magnet by linearly polarized light with $\phi=0$. 
    (a-i) Spin density $\bar{\rho}_z^\text{{eff}}$ [Eq.\,(\ref{eq_sdosz})] as a function of $M$ and $\omega$ at $eA_{0}/(\hbar k_{\rm F})=0.5$.
    (a-ii) shows line cuts at distinct values of $M$ marked in (a-i) and depicted by blue, gray, and orange curves for $M/\mu<1$, $M/\mu=1$, and $M/\mu>1$, respectively.
    (a-iii) Energy versus $k_{x}$ at $k_{y}=0$ for the distinct values corresponding to the distinct values of $M$ in (a-ii). 
    The spin-up (spin-down) bands are denoted in blue (red). Parameters: $\mu=1$, $\hbar^2/(2m)=1$ and $k_{\rm F}= 1$, $\theta_d=0$ and $\phi=0$, $eA_0/(\hbar k_{\rm F})=0.5$.  
    (b) same as (a) but for an $i$-wave magnet driven by linearly polarized light. 
    The driving amplitude is $eA_{0}/(\hbar k_{\rm F})=0.8$ for clarity.}
    \label{figureR1}
\end{figure}

Fig.\,\ref{figureR1}(a-i) present the driven $g$-wave case with $\theta_{g}=0$,  showing $\bar{\rho}_{z}^{\text{eff}}(\bm{k},\omega)$ as a function of frequency $\omega$ and the unconventional magnetic strength $M$. 
A pronounced peak–dip structure appears in $\bar{\rho}_{z}^{\text{eff}}$ around $\omega/\mu=-1$ when $M/\mu>1$, exhibiting behavior similar to that of the $d$-wave case discussed in the main text, see Fig.\,\ref{figureR1}(a–ii). 
In addition, multiple peaks and dips emerge due to the higher-order momentum dependence characteristic of the $g$-wave magnet. 
After some calculations, we confirm that the peak (dip) structures in Figs.\,\ref{figureR1}(a–i) and (a–ii) correspond to the local maxima (minima) of the spin-band dispersion,
\begin{equation}
E_{\sigma}^{j}
= \xi_{\bm{k}} + \sigma\big(M_{\bm{k}}^{j} + M_{\Omega}^{j} + B_{\bm{k}\Omega}^{j}\big),
\label{eq_ejs}
\end{equation}
where the chemical potential shift $(e\hbar A_{0})^{2}/(4m)$ has been gauged out, $M_{\bm{k}}^{j}$ is the static unconventional magnetic field [Eq.\,(\ref{eq_mjk})], $M_{\Omega}^{j}$ and $B_{\bm{k}\Omega}^{j}$ are the light-induced Zeeman field and unconventional magnetic field given by Eqs.\,(\ref{eq_mjO}) and (\ref{eq_BjkO}), respectively.
For the $g$-wave magnet, at $k=0$, a light-induced Zeeman splitting appears, producing an energy gap between the two spin subbands and a local minimum in the dispersion, which is shown in Fig.\,\ref{figureR1}(a–iii) with $A_0\neq0$. 
The local minimum in the dispersion is given by
\begin{equation}
\omega_{0}^{\sigma}
= -\mu + \sigma\,\frac{3}{8}\bar{M}^{g}(eA_{0})^{4}\cos(4\Theta_{g}),
\label{eq_omega0}
\end{equation}
while for $eA_{0}=0$, the bands remain spin-degenerate [see Fig.\,\ref{figureR1}(a–iii) with $A_0=0$].
The curvature of the dispersion is governed by the competition between the $k^{2}$ and $k^{4}$ terms in Eq.\,(\ref{eq_ejs}): 
for small $k<1$, the $k^{2}$ term dominates and $E_{\sigma}^{g}(k)$ increases with $k$, whereas the $k^{4}$ term dominates for large $k>1$, leading to a decrease in $E_{-}^{g}(k) \!\sim\! C_{1}k^{2} - C_{2}k^{4}$ 
($C_{1}, C_{2} > 0$) [see Fig.\,\ref{figureR1}(a–iii)]. 
This behavior results in a local maximum \footnote{The extrema can be obtained mathematically from 
$\big.\partial_{k}E_{\sigma}(k,\theta_{k})\big|_{k=k_{0}^{\sigma}}=0$.} 
around $k=k_{F}=\sqrt{2m\mu}/\hbar$, with the corresponding energy
\begin{equation}
\omega_{X}^{-}
= \omega_{0}^{-}
+ \frac{1}{4\bar{M}^{g}\cos(4\theta_{g})}
\Big[\frac{\hbar^{2}}{2m} + 3\bar{M}^{g}(eA_{0})^{2}\cos(4\theta_{g}-2\phi)\Big]^{2},
\label{eq_omegaX}
\end{equation}
along $k_x$-axis with $\theta_{k}=0$.
A similar energy-momentum dependence occurs along $k_y$-axis with $\theta_{k}=\pi/2$, yielding
\begin{equation}
\omega_{Y}^{-}
= \omega_{0}^{-}
+ \frac{1}{4\bar{M}^{g}\cos(4\theta_{g})}
\Big[\frac{\hbar^{2}}{2m} - 3\bar{M}^{g}(eA_{0})^{2}\cos(4\theta_{g}-2\phi)\Big]^{2},
\label{eq_omegaY}
\end{equation}
and for $\theta_{k}=\pi/4$,
\begin{equation}
\omega_{\pi/4}^{+}
= \omega_{0}^{+}
+ \frac{1}{4\bar{M}^{g}\cos(4\theta_{g})}
\Big[\frac{\hbar^{2}}{2m} + 3\bar{M}^{g}(eA_{0})^{2}\sin(4\theta_{g}-2\phi)\Big]^{2},
\label{eq_omegapi4}
\end{equation}
where the light-induced $d$-wave magnetic component vanishes, see Eq.\,(\ref{eq_BjkO}).  
These critical energy values correspond to the peak–dip structures observed in Fig.\,\ref{figureR1}(a–i).

The peaks and dips in the spin density correspond to local extrema in the spin-band dispersion, 
allowing one to extract the altermagnetic strength from the energy splitting between the peaks and dips. 
Consistent with the main text, the Zeeman splitting [Eq.\,(\ref{eq_omega0})] and the local extrema [Eqs.\,(\ref{eq_omegaX})–(\ref{eq_omegapi4})] are induced by the light drive and encode information about the strength and orientation of altermagnetism. 
Particularly, the energy splitting between the two spin subbands due to the light-induced Zeeman splitting is
\begin{equation}
\delta\omega_{1}
= \big|\omega_{0}^{+} - \omega_{0}^{-}\big|
= \frac{3}{4}\bar{M}^{g}(eA_{0})^{4}\cos[4(\theta_{g}-\phi)],
\end{equation}
which is directly related to the altermagnetic strength $M$, 
its orientation $\theta_{g}$, and the driving polarization angle $\phi$. 
Hence, for a fixed driving polarization direction $\phi$, the $g$-wave strength and orientation can be determined from
\begin{equation}
M\cos(4\Theta_{g})
= \frac{4}{3}\left(\frac{\hbar k_{F}}{eA_{0}}\right)^{4}\delta\omega_{1},
\label{eq_mgomega}
\end{equation}
with $\Theta_{g}=\theta_g-\phi$  providing an experimentally accessible method to probe altermagnetism, similar to the $d$-wave magnets discussed in the main text.

A similar analysis relating to the peak-dip structure of the spin density and the altermagnetism applies to the $i$-wave magnet, as shown in Fig.\,\ref{figureR1}(b). 
The spin density in Fig.\,\ref{figureR1}(b–i) exhibits a richer peak–dip structure, corresponding to multiple extrema in the dispersion. 
Although the explicit expressions of these values are more complex owing to the higher-order momentum dependence, the light-induced Zeeman splitting persists and corresponds to the peak–dip structures around $\omega = -\mu$, given by 
\begin{equation}
\omega_{0}^{\sigma}
= -\mu + \sigma\,\frac{5}{32}\bar{M}^{i}(eA_{0})^{6}\cos(6\Theta_{g}).
\label{eq_omega0i}
\end{equation}
Therefore, the $i$-wave altermagnetic strength and orientation can be determined from the energy difference between the peak and dip via
\begin{equation}
M\cos(6\Theta_{i})
= \frac{8}{5}\left(\frac{\hbar k_{F}}{eA_{0}}\right)^{6}\delta\omega_{1},
\label{eq_miomega}
\end{equation}
where $\Theta_{i}=\theta_i-\phi$ and $\delta\omega_{1} = \omega_{0}^{+} - \omega_{0}^{-}$ denotes the energy difference between the peak and dip in Fig.\,\ref{figureR1}(b–ii).

\subsection{Summary} \label{SecID}
For both $g$- and $i$-wave magnets, the main conclusion of this work remains: 
the strength and orientation of the altermagnetic field can be identified from the spin density. 
Although the increasing number of spin-degenerate nodes complicates the detailed peak–dip structure, 
the light-induced Zeeman field [Eq.\,(\ref{eq_mjO})] in even-parity unconventional magnets 
enables direct extraction of the magnitude and orientation of altermagnetism from Eqs.\,(\ref{eq_mgomega}) and (\ref{eq_miomega}). 
In contrast, for the $f$-wave (odd-parity) case, the spin density vanishes due to the absence of the light-induced Zeeman term, 
as in the $p$-wave magnet. 
Therefore, the central conclusion of the manuscript can be generalized as follows: 
for even-parity states, the light-induced Zeeman field produces a characteristic peak–dip structure in the spin density, 
providing a direct probe of the altermagnetic strength and orientation, 
whereas for odd-parity states, this feature is absent.

\subsection{Appendix: Floquet components of unconventional
magnets under linear polarized light} \label{SecIE}
For completeness, we list below the Floquet components of the unconventional magnet under linearly polarized light. 
These components satisfy $\big[ H_{+n}^{j}(\bm{k}), H_{-n}^{j}(\bm{k}) \big]=0$, 
leading to the effective time-averaged Hamiltonian given in Eq.\,(\ref{eq_hjeff}). 
The first harmonic component reads
\begin{equation}
H_{+1}^{j} = \frac{\hbar^{2}}{2m} eA_{0}k \cos(\theta_{k}-\phi)
+ \sigma_{z} eA_{0} \bar{M}^{j} m_{1}^{j},
\end{equation}
with
\begin{eqnarray}
m_{1}^{s} &=& 0, \\
m_{1}^{p} &=& \tfrac{1}{2}\cos(\theta_{J}-\phi), \\
m_{1}^{d} &=& k \cos(2\theta_{J}-\phi-\theta_{k}), \\
m_{1}^{f} &=& \tfrac{3}{8}\!\left[(eA_{0})^{2}\!\cos(3\theta_{J}-3\phi)
+4k^{2}\!\cos(3\theta_{J}-\phi-2\theta_{k})\right], \\
m_{1}^{g} &=& \tfrac{1}{2}k\!\left[4k^{2}\!\cos(4\theta_{J}-\phi-3\theta_{k})
+3(eA_{0})^{2}\!\cos(4\theta_{J}-3\phi-\theta_{k})\right], \\
m_{1}^{i} &=& \tfrac{3}{8}k\!\left[8k^{4}\!\cos(6\theta_{J}-\phi-5\theta_{k})
+20(eA_{0})^{2}k^{2}\!\cos(6\theta_{J}-3\phi-3\theta_{k})
+5(eA_{0})^{4}\!\cos(6\theta_{J}-5\phi-\theta_{k})\right].
\end{eqnarray}

The second harmonic component is
\begin{equation}
H_{+2}^{j,\sigma} = \frac{1}{4}\frac{\hbar^{2}}{2m}(eA_{0})^{2}
+ \sigma_{z}(eA_{0})^{2}\bar{M}^{j}m_{2}^{j},
\end{equation}
where
\begin{eqnarray}
m_{2}^{s} &=& m_{2}^{p}=0, \\
m_{2}^{d} &=& \tfrac{1}{4}\cos(2\Theta_{d}), \\
m_{2}^{f} &=& \tfrac{3}{4}k\cos(3\theta_{f}-2\phi-\theta_{k}), \\
m_{2}^{g} &=& \tfrac{1}{4}\!\left[k_{A}^{2}\!\cos(4\Theta_{g})
+6k^{2}\!\cos(4\theta_{g}-2\phi-2\theta_{k})\right], \\
m_{2}^{i} &=& \tfrac{15}{64}\!\left[k_{A}^{2}\!\cos(6\Theta_{i})
+16k^{2}\!\cos(6\theta_{i}-2\phi-4\theta_{k})
+k_{A}^{2}\!\cos(6\theta_{i}-4\phi-2\theta_{k})\right].
\end{eqnarray}

Higher-order harmonics are given by
\begin{eqnarray}
H_{+3}^{j} &=& \sigma_{z}\frac{\bar{M}^{j}(eA_{0})^{3}}{16}
\!\left[2\cos(3\Theta_{f})\delta_{j,f}
+8k\cos(4\theta_{g}-3\phi-\theta_{k})\delta_{j,g}
+5\!\left(8k^{2}\!\cos(6\theta_{i}-3\phi-3\theta_{k})
+3k_{A}^{2}\!\cos(6\theta_{i}-5\phi-\theta_{k})\right)\!\delta_{j,i}\right], \\
H_{+4}^{j} &=& \sigma_{z}\frac{\bar{M}^{j}(eA_{0})^{4}}{32}
\!\left[2\cos(4\Theta_{g})\delta_{j,g}
+3\!\left(k_{A}^{2}\!\cos(6\Theta_{i})
+10k^{2}\!\cos(6\theta_{i}-4\phi-2\theta_{k})\right)\!\delta_{j,i}\right], \\
H_{+5}^{j} &=& \sigma_{z}\frac{3}{16}\bar{M}^{j}(eA_{0})^{5}
\cos(6\theta_{J}-5\phi-\theta_{k})\delta_{j,i}, \\
H_{+6}^{j} &=& \sigma_{z}\frac{1}{64}\bar{M}^{j}(eA_{0})^{6}
\cos(6\Theta_{i})\delta_{j,i}.
\end{eqnarray}
These harmonic components collectively determine the symmetry and magnitude of the light-induced effective magnetic fields entering the high-frequency Hamiltonian, and thus govern the Floquet-engineered spin-triplet correlations discussed in the main text.

\section{Floquet engineering spin-triplet Cooper pairs in unconventional magnets with $s$-wave superconductivity}

In the main text, after demonstrating the Floquet engineering spin density in $p$- and $d$-wave unconventional magnet, we address the Floquet engineering spin-triplet Cooper pair.
This is because the spin-triplet pair amplitude is intrinsically related to the spin density via Eq.\,(18) in the main text.
Based on the extension from $p$- and $d$-wave magnet to higher-order unconventional magnets illustrated in Sec. \ref{SecI}, a question naturally raised is whether the universality of the distinction between unconventional magnets with odd and even parities also manifests in the spin-triplet pair amplitude in unconventional magnets with conventional $s$-wave spin-singlet superconductivity. 

To address the above-mentioned question, in this section, we first briefly review the parity of the pair amplitude in unconventional magnets with conventional $s$-wave superconductivity in Sec.,\,\ref{SecIIA}, which is consistent with the results reported by some previous works, see the review \cite{FukayaCayaoReviewUMs}.
Then, in Sec.\,\ref{SecIIB}, the high-frequency Hamiltonian in the driven regime is investigated, with particular focus on the role of the parity in unconventional magnets on the BdG spectrum and spin-triplet amplitude.
Furthermore, pairing correlations are investigated in the low-frequency regime.
As summarized in Sec.\,\ref{SecIIC}, with conventional $s$-wave spin-singlet superconductivity, the emergent spin-triplet pair amplitude is universally expected in all unconventional magnets with odd and even parities and under both high- and low-frequency driving fields.

Before we dive into the details, we can provide a short answer to the question above:
The conclusion of the main text, based on $d$-wave and $p$-wave magnets, is universal and can be applied to unconventional magnets with higher-order parities:
The strength and orientation of even-parity unconventional magnets (the $d$-, $g$-, and $i$-wave altermagnets) can be directly extracted from the peak–dip structures of the light-induced spin-triplet pair amplitude;
In contrast, for odd-parity magnets, such as $p$- and $f$-wave systems, the vanishing spin-triplet pair amplitude remains.
In the following, we ellaborate on this universal distinction between unconventional magnets with odd and even parities in detail.

\subsection{Hamiltonian and pair amplitude in the static regime}  \label{SecIIA}
When superconductivity is induced, in the basis $\{ c_{\bm{k},\uparrow}, c_{\bm{k},\downarrow}, c_{-\bm{k},\uparrow}^{\dagger}, c_{-\bm{k},\downarrow}^{\dagger} \}^{\text{T}}$, the resulting BdG Hamiltonian is
\begin{equation}
\mathcal{H}^{j,l}(\bm{k})
=
\begin{pmatrix}
H^{j}(\bm{k}) & \hat{\Delta}^{l}(\bm{k}) \\[4pt]
\big[\hat{\Delta}^{l}(\bm{k})\big]^{\dagger} & -\big[H^{j}(-\bm{k})\big]^{\ast}
\end{pmatrix},
\label{eq_hjl0}
\end{equation}
which describes a $j$-wave unconventional magnet with $l$-wave superconductivity. 
Here, $H^{j}(\bm{k})$ is the normal electron Hamiltonian defined in Eq.\,(\ref{eq_hjk}), with the index $j=\{0,1,2,3,4,6\}$ corresponding to magnets with $s$-, $p$-, $d$-, $f$-, $g$-, and $i$-wave parities, respectively, reflecting the Fermi surface anisotropy.
Also, $-\big[H^{j}(-\bm{k})\big]^{\ast}$ is the hole Hamiltonian obtained as 
\begin{equation}
-\big[H^{j}(-\bm{k})\big]^{\ast}
= -\xi_{\bm{k}}\sigma_{0} + (-1)^{j+1}M_{\bm{k}}^{j}\sigma_{z},
\label{eq_hhole}
\end{equation}
where an odd/even value of $j$ leaves a minus/plus sign in the second term. Hence, Eq.\,(\ref{eq_hhole}) is parity-dependent and has opposite momentum parity between even- and odd-$j$ unconventional magnets.
Moreover, $\hat{\Delta}^{l}(\bm{k})$ in Eq.\,(\ref{eq_hjl0}) is the pair potential representing spin-singlet superconductors with $l$-wave parity, where $l=\{s,d\}$ and $\hat{\Delta}^{l}(\bm{k})=\hat{\Delta}^{l}(-\bm{k})$ is an even function of momentum $\bm{k}$.

Consequently, the BdG Hamiltonian for a $j$-wave magnet with conventional spin-singlet $s$-wave superconductivity ($l=s$) can be rewritten as
\begin{equation}
\mathcal{H}^{j,s}(\bm{k})
= \xi_{\bm{k}}\tau_{z}\sigma_{0}
- \Delta\,\tau_{y}\sigma_{y}
+ 
\begin{cases}
M_{\bm{k}}^{j}\,\tau_{z}\sigma_{z}, & j~\text{even}, \\[4pt]
M_{\bm{k}}^{j}\,\tau_{0}\sigma_{z}, & j~\text{odd},
\end{cases}
\label{eq_hjl}
\end{equation}
where $\hat{\Delta}^{s}(\bm{k}) = i\Delta\sigma_{y}$ is chosen in Eq.\,(\ref{eq_hjl0}) for the spin-singlet $s$-wave pairing potential, 
and $\tau_{i}$ ($\tau_{0}$) are the Pauli (unit) matrices in Nambu space. 
Eq.\,(\ref{eq_hjl}) highlights the essential role of momentum parity of the unconventional magnets in determining the structure of the BdG Hamiltonian and, consequently, the induced pairing correlations, as we discuss next.

\begin{figure}
    \centering
    \includegraphics[width=1\linewidth]{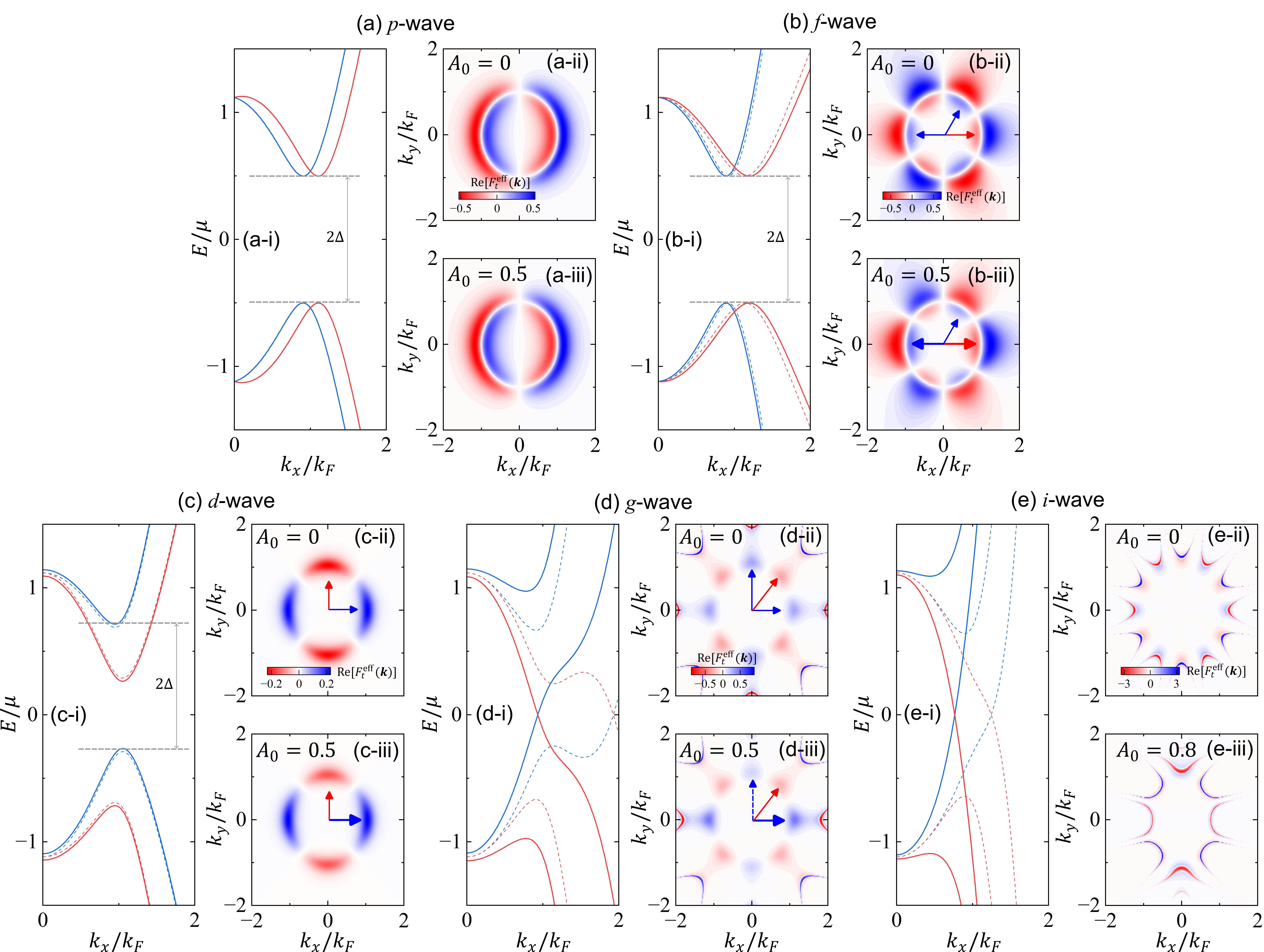}
    \caption{BdG spectrum [Eqs.\,(\ref{eq_ejng}) and (\ref{eq_ebdgeff})] along $k_x$ with $k_y=0$ and the real part of momentum-resolved spin-triplet amplitude [Eqs.\,(\ref{eq_ft}) and (\ref{eq_fteff})] for odd-parity (upper panel) and even-parity (lower panel) unconventional magnets with $s$-wave superconductivity in static ($A_0=0$) and driven regimes ($A_0\neq0$) with linearly polarized light.
    (a-i) the BdG spectrum for driven $p$-wave unconventional magnets with $s$-wave superconductivity in static (dashed lines) and driven (solid lines) states.
    (a-ii) and (a-iii): The static and driven regime spin-triplet amplitude, respectively.
    The arrangements from (b-e) are the same as (a) but for differet type of unconventional magnets.
    %The BdG spectrum of the static (driven) states is denoted by dashed (solid) lines.
    The blue (red) lines represent the superconducting states formed by pin-up (spin-down) electrons and spin-down (spin-up) holes.
    Parameters: $M=0.2$, $\theta_j=0$, $\mu=1$, $k_F=1$, $\Delta=0.5\mu$, $\phi=0$ and $z=0.1\Delta+0.005i$ is set in the spin-triplet amplitude. 
    No qualitative change occurs for a different set of parameters.
    }
    \label{figureR2}
\end{figure}

The parity dependence of the Hamiltonian manifests directly in the BdG spectrum. 
From Eq.\,(\ref{eq_hjl}), the quasiparticle energies are given by
\begin{equation}
E_{\nu}^{j} =
\begin{cases}
\nu\,M_{\bm{k}}^{j} \pm \sqrt{\xi_{\bm{k}}^{2} + \Delta^{2}}, & j~\text{even}, \\[4pt]
\pm \sqrt{(\xi_{\bm{k}} + \nu\,M_{\bm{k}}^{j})^{2} + \Delta^{2}}, & j~\text{odd},
\end{cases}
\label{eq_ejng}
\end{equation}
where $\nu = +1~(-1)$ labels the quasiparticle formed by spin-up (spin-down) electrons and spin-down (spin-up) holes, respectively. 
Fig.\,\ref{figureR2} presents the BdG spectra of various unconventional magnets in their static states (dashed lines), exhibiting an even–odd parity distinction: 
For odd-parity unconventional magnets, the spin splitting appears along the momentum axis [Figs.\,\ref{figureR2}(a–b)], and the BdG spectrum of Eq.\,(\ref{eq_ejng}) displays a superconducting gap of $2|\Delta|$ symmetric with respect to $E=0$, regardless of the strength and orientation of the unconventional magnetism.
In contrast, for even-parity unconventional magnets, the spin splitting occurs along the energy axis [Figs.\,\ref{figureR2}(c–e)], resulting in asymmetric BdG spectra for individual $\nu$ branches, while the overall spectrum remains symmetric with respect to zero energy. 
Specifically, in the $d$-wave altermagnet, each $\nu$ branch exhibits a superconducting gap of magnitude $2|\Delta|$, whereas in the $g$- and $i$-wave altermagnets, the BdG spectrum becomes gapless, because the magnetic effect becomes dominant as momentum increases
[Figs.\,\ref{figureR2}(d–e)]. 
This even–odd dichotomy underlies the fundamental difference in quasiparticle response and pairing symmetry 
across the family of unconventional magnets.

The parity-dependent properties also manifest in the emergent Cooper pair amplitudes, which are encoded in the anomalous (off-diagonal) components of the Green’s function \cite{zagoskin}. 
Associated with Eq.\,(\ref{eq_hjl}), the Gor’kov Green’s function is defined as $G(z,\bm{k}) = [\,z - \mathcal{H}^{j}(\bm{k})\,]^{-1}$ with complex frequency $z = \omega + i0^{+}$. 
After some algebra, the spin-singlet pairing amplitude is obtained as
\begin{equation}
F_{s}^{j}(z,\bm{k})
= \frac{\Delta}{D^{j}(z)}
\big[z^{2} - \xi_{\bm{k}}^{2} - \Delta^{2} + (-1)^{j}\big(M_{\bm{k}}^{j}\big)^{2}\big],
\label{eq_fs}
\end{equation}
and the spin-triplet pairing amplitude is given by
\begin{equation}
F_{t}^{j}(z,\bm{k})
= 2\Delta\,M_{\bm{k}}^{j}
\begin{cases}
\dfrac{z}{D^{j}(z)}, & j~\text{even},\\[12pt]
-\dfrac{\xi_{\bm{k}}}{D^{j}(z)}, & j~\text{odd},
\end{cases}
\label{eq_ft}
\end{equation}
where the denominator
\begin{equation}
D^{j}(z) =
\begin{cases}
\left( z^{2}+M_{\bm{k}}^{j2}-\xi _{\bm{k}}^{2}-\Delta ^{2}\right)
^{2}-\left( 2zM_{\bm{k}}^{j}\right) ^{2} & j\text{ even,} \\[6pt]
\left( z^{2}-M_{\bm{k}}^{j2}-\xi _{\bm{k}}^{2}-\Delta ^{2}\right)
^{2}-\left( 2M_{\bm{k}}^{j}\xi _{\bm{k}}\right) ^{2} & j\text{ odd,}%
\end{cases}
\label{eq_Dj}
\end{equation}
is an even function of the complex frequency $z$, regardless of the parity $j$.
Therefore, the frequency symmetry (odd or even in $z$) of the pairing amplitude is determined entirely by the numerators in Eq.\,(\ref{eq_fs}) and Eq.\,(\ref{eq_ft}). 
The spin-singlet amplitude $F_{s}^{j}(z,\bm{k})$ is even-frequency and even-momentum, 
thereby satisfying the fermionic antisymmetry condition for Cooper pairs \cite{FukayaCayaoReviewUMs,RevModPhys.91.045005}. 
Interestingly, the parity effect appears explicitly in the spin-triplet amplitude $F_{t}^{j}(z,\bm{k})$, 
which is proportional to the magnetic term $M_{\bm{k}}^{j}$: 
For even-parity unconventional magnets ($d$-, $g$-, and $i$-wave), 
$F_{t}^{j}(z,\bm{k})$ is \textit{odd} in frequency and \textit{even-parity} in momentum. 
In contrast, for odd-parity unconventional magnets ($p$- and $f$-wave), 
$F_{t}^{j}(z,\bm{k})$ is \textit{even} in frequency and \textit{odd-parity} in momentum. 
Thus, the momentum-space symmetry of the spin-triplet amplitude is inherited directly from that of the unconventional magnet \cite{FukayaCayaoReviewUMs}. 
For instance, in the static state, the spin-triplet amplitude of an $f$-wave magnet with spin-singlet $s$-wave superconductivity exhibits a characteristic $f$-wave pattern [Fig.\,\ref{figureR2}(b–i)]: 
the lobes along $+k_{x}$ (red arrow) possess the same magnitude but opposite sign as those along $\theta_{k} = \pi/3$ and $\pi$ (blue arrows), consistent with the threefold symmetry of the $f$-wave order. 
Similar parity-dependent symmetries are observed for the static $p$-, $d$-, $g$-, and $i$-wave cases with $A_0=0$ in Fig.\,\ref{figureR2}, which are indicated by the equal-thickness red and blue arrows.

The spin-triplet pair amplitudes given by Eq.\,(\ref{eq_ft}) in static unconventional magnets with higher angular momentum are, therefore, consistent with earlier theoretical results reporting that the momentum symmetry of induced triplet correlations is inherited from the underlying unconventional magnetic feature \cite{FukayaCayaoReviewUMs}.
This suggests that even more exotic Cooper pairs are very likely to appear when light is applied to unconventional magnets with higher parities, as we address below.

\begin{figure}
    \centering
    \includegraphics[width=0.5\linewidth]{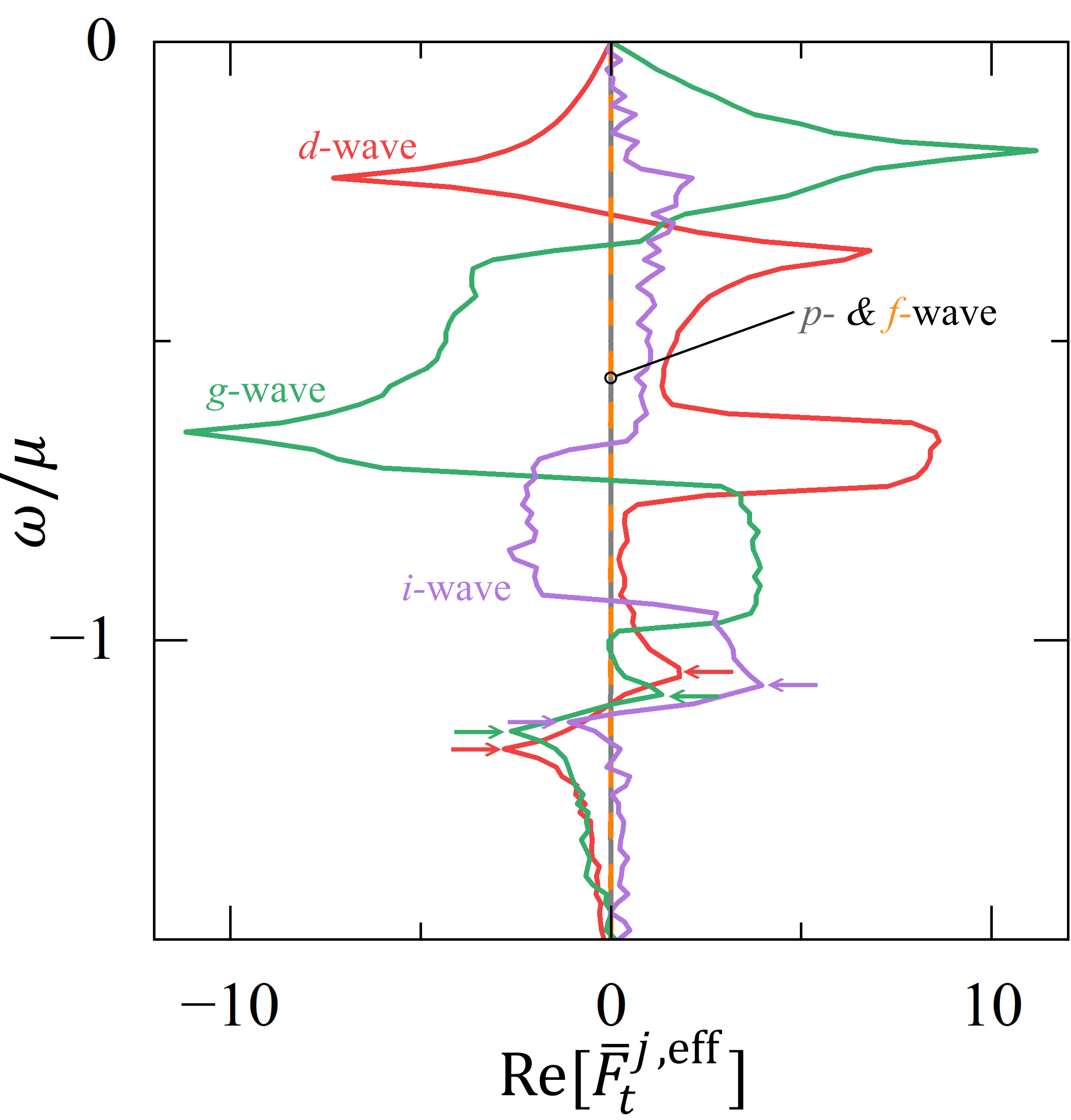}
    \caption{Energy dependence of the real part of spin-triplet amplitude [Eq.\,(\ref{eq_fteff})] for various driven unconventional magnets with conventional $s$-wave superconductivity. 
    The colored arrows indicate the peak-dip structures probing the strength and orientation of the altermagnetism, Eq.\,(\ref{Eq_Mdomega})-(\ref{Eq_Miomega}). Parameters: same as Fig.\,\ref{figureR2} with finite driving fields.
    }
    \label{figureR3}
\end{figure}

\subsection{High-frequency effective Hamiltonian in the driven regime: BdG spectrum and spin-triplet pair amplitude} \label{SecIIB}
Here, we follow the Floquet approach to obtain the effective Hamiltonian in the high-frequency regime.
For this, we use the approach given in Sec.\,\ref{SecI}, especially Eq.\,(\ref{eq_hcpleff})--(\ref{eq_BjkO}). Thus, the effective Hamiltonian for circularly/linearly polarized light can be directly obtained from Eq.\,(\ref{eq_hcpleff}).
%The conclusions obtained for the driven normal (non-superconducting) states can be naturally extended to their superconducting counterparts: 
Under circularly polarized driving, the high-frequency effective Hamiltonian remains identical to the static one [see Eq.\,(\ref{eq_hcpleff})], 
whereas the effective Hamiltonian for an unconventional magnet with $s$-wave spin-singlet superconductivity under linearly polarized light is given by
\begin{equation}
\mathcal{H}_{\text{eff}}^{j,s}(\bm{k})
= \xi_{\bm{k}}\tau_{z}\sigma_{0}
- \Delta\,\tau_{y}\sigma_{y}
+ 
\begin{cases}
\big(M_{\bm{k}}^{j} + M_{\Omega}^{j} + B_{\bm{k}\Omega}^{j}\big)\tau_{z}\sigma_{z}, & j~\text{even}, \\[4pt]
\big(M_{\bm{k}}^{j} + B_{\bm{k}\Omega}^{j}\big)\tau_{0}\sigma_{z}, & j~\text{odd}.
\end{cases}
\label{eq_hjseff}
\end{equation}
Here, $M_{\bm{k}}^{j}$ is the unconventional magnetic field given by Eq.\,(\ref{eq_mjk}), while $M_{\Omega}^{j}$ and $B_{\bm{k}\Omega}^{j}$ are given by Eq.\,(\ref{eq_mjO}) and Eq.\,(\ref{eq_BjkO}) and denote the light-induced Zeeman and unconventional magnetic fields, respectively.
%(i) a light-induced Zeeman field $M_{\Omega}^{j}$ in even-parity unconventional magnets, see {\color{magenta}Eq.\,(\ref{eq_mjO})}, 
%and (ii) a lower-order unconventional magnetic field $B_{\bm{k}\Omega}^{j}$ originating from higher-order ones, see Eq.\,(\ref{eq_BjkO}). 
%The linearly polarized case is therefore of particular interest, yielding the effective Hamiltonian as
%\begin{equation}
%\mathcal{H}_{\text{eff}}^{j,s}(\bm{k})
%= \xi_{\bm{k}}\tau_{z}\sigma_{0}
%- \Delta\,\tau_{y}\sigma_{y}
%+ 
%\begin{cases}
%\big(M_{\bm{k}}^{j} + M_{\Omega}^{j} + B_{\bm{k}%\Omega}^{j}\big)\tau_{z}\sigma_{z}, & j~\text{even}, \\[4pt]
%\big(M_{\bm{k}}^{j} + B_{\bm{k}\Omega}^{j}\big)\tau_{0}\sigma_{z}, & j~\text{odd},
%\end{cases}
%\label{eq_hjseff}
%\end{equation}
%where {\color{magenta}$M_{\Omega}^{j}$ and $B_{\bm{k}\Omega}^{j}$ are given by Eq.\,(\ref{eq_mjO}) and Eq.\,(\ref{eq_BjkO})} and denote the light-induced Zeeman and unconventional magnetic fields, respectively.
At this point, it is worth noting that the light-induced Zeeman field, $M_{\Omega}^{j}$, appears in even-parity unconventional magnets, but is absent in their odd-parity counterpart, as demonstrated in the $d$- and $p$-wave magnets in the main text. 
Interestingly, the unconventional magnetic field $B_{\bm{k}\Omega}^{j}$ was absent in our previous results for $d$- and $p$-wave magnets because it appears only for unconventional magnets with $j \geq 3$, see more details in Sec.\,\ref{SecI}.

To further analyze the effect of the parity in a driven unconventional magnet with $s$-wave spin-singlet superconductivity, we inspect the BdG spectrum.
It reads
\begin{equation}
E_{\nu}^{j, s, \text{eff}} =
\begin{cases}
\nu\big(M_{\bm{k}}^{j} + M_{\Omega}^{j} + B_{\bm{k}\Omega}^{j}\big)
\pm \sqrt{\xi_{\bm{k}}^{2} + \Delta^{2}}, & j~\text{even}, \\[4pt]
\pm \sqrt{\big[\xi_{\bm{k}} + \nu\big(M_{\bm{k}}^{j} + B_{\bm{k}\Omega}^{j}\big)\big]^{2} + \Delta^{2}}, & j~\text{odd},
\end{cases}
\label{eq_ebdgeff}
\end{equation}
which clearly includes both the light-induced Zeeman term $M_{\Omega}^{j}$ and the unconventional magnetic contribution $B_{\bm{k}\Omega}^{j}$, given by Eq.\,(\ref{eq_mjO}) and Eq.\,(\ref{eq_BjkO}), respectively. 
To visualize the energy spectra, in Fig.\,\ref{figureR2}(a-i, b-i, c-i, d-i, e-i), we plot $E_{\nu}^{j, s, \text{eff}}$ as a function of momentum for different unconventional magnets. 
For comparison, in the same figures, we also show the energy spectra in the static case (dashed lines).
The overall spectral structure in the driven case closely resembles that of the static state, 
retaining similar gap profiles and spin splitting at finite momentum. 
Notably, in even-parity altermagnets, an additional spin splitting appears at $k=0$ due to $M_{\Omega}^{j}$, as depicted by the solid lines in Fig.\,\ref{figureR2}(c-i, d-i, e-i). 

We now analyze the emergent superconducting correlations. In the main text, we showed that the spin splitting due to $M_{\Omega}^{j}$ leads to a finite spin-triplet pair amplitude in $j=d$.
As we show below,  $M_{\Omega}^{j}$ turns out to be the key for generating finite spin-triplet correlations in all \textit{all} even-parity unconventional magnets.
Following the same procedure used to derive Eq.\,(\ref{eq_ft}) from Eq.\,(\ref{eq_hjl}), 
the spin-triplet amplitude obtained from the effective Hamiltonian [Eq.\,(\ref{eq_hjseff})] reads
\begin{equation}
F_{t}^{j,s,\text{eff}}(z,\bm{k})
= 2\Delta
\!\times\!
\begin{cases}
\displaystyle
\frac{M_{\bm{k}}^{j} + M_{\Omega}^{j} + B_{\bm{k}\Omega}^{j} }
{D^{j,s,\text{eff}}(z)} z , & j~\text{even},\\[10pt]
\displaystyle
-\frac{M_{\bm{k}}^{j} + B_{\bm{k}\Omega}^{j}}
{D^{j,s,\text{eff}}(z)}\,\xi_{\bm{k}}, & j~\text{odd},
\end{cases}
\label{eq_fteff}
\end{equation}
where $D^{j,s, \text{eff}}(z)$ is obtained from Eq.\,(\ref{eq_Dj}) by replacing 
$M_{\bm{k}}^{j}$ with $M_{\bm{k}}^{j} + M_{\Omega}^{j} + B_{\bm{k}\Omega}^{j}$ for $j$ even and $M_{\bm{k}}^{j}$ with $M_{\bm{k}}^{j} + B_{\bm{k}\Omega}^{j}$ for $j$ odd.
In both cases, $D^{j,s, \text{eff}}(z)$ remains an even function of $z$ and $\bm k$. 
The pair amplitude in the high-frequency driven regime given by Eq.\,(\ref{eq_fteff}) has the same form as the one in the static regime given by Eq.\,(\ref{eq_ft}). This also shows that the frequency dependence is tied to the parity of unconventional magnets: it is odd-frequency (even-frequency) for even-parity (odd-parity) magnets. 
However, despite the similarities, there exist some important differences that can be seen by comparing the numerators of Eq.\,(\ref{eq_fteff}) and (\ref{eq_ft}). In fact, the spin-triplet pair amplitude given by Eq.\,(\ref{eq_fteff}) contains three additive contributions in the numerator: i) from the static unconventional magnetic field described by $M_{\bm{k}}^{j}$, ii) from the light-induced Zeeman field $M_{\Omega}^j$, and iii) from $B_{\bm{k}\Omega}^{j}$ that characterizes a light-induced unconventional magnetic field; see Eq.\,(\ref{eq_mjO}) and Eq.\,(\ref{eq_BjkO}) for the details on the orgin of $M_{\Omega}^j$ and $B_{\bm{k}\Omega}^{j}$, respectively. As a result, the light-induced fields $M_{\Omega}^j$ and $B_{\bm{k}\Omega}^{j}$ enrich the momentum parity symmetries of the emergent spin-triplet Cooper pairs.
Importantly, the components of Eq.\,(\ref{eq_fteff}) propotional to the light-induced unconventional magnetic field, $B_{\bm{k}\Omega}^{j}$, appears only in unconventional magnets with parity order $j \geq 3$, see Eq.\,(\ref{eq_BjkO}).
This contribution enables the coexistence of Cooper pairs having distinct parities as a result of higher-order light-matter interactions in $j\geq3$.
We note that this contribution is absent for $d$- and $p$-wave magnets that we discussed in main text. 

The coexistence of Cooper pairs with different parities can be seen by taking explicit values of $j\geq3$.
First for a $p$-wave unconventional magnet ($j=1$), the spin-triplet pair amplitude in Eq.\,(\ref{eq_fteff}) does not develop any contribution from the light-induced field $B_{\bm{k}\Omega}^{j}$. To show this, in Fig.\,\ref{figureR2}(a-ii, a-iii) we plot the real part of the spin-triplet pair amplitude given by Eq.\,(\ref{eq_fteff}) for $j=1$ as a function of momenta in the static and driven regimes: the static and the driven spin-triplet pair amplitudes are the same. reflecting that light does not induce any contribution. The situation becomes more interesting when the parity is higher.
For example, in the driven $f$-wave unconventional magnets ($j=3$), an effective $p$-wave component emerges. This is shown in Fig.\,\ref{figureR2}(b-ii, b-iii): where we present the real part of the spin-triplet pair amplitude given by Eq.\,(\ref{eq_fteff}) as a function of momenta. We see that the $f$-wave symmetry leads to a lobe at $\theta_{k}=0$ (thick red arrow) with different magnitudes from the one at $\theta_{k}=\pi/3$ (thin blue arrow), but opposite to the one at $\theta_{k}=\pi$ (thick blue arrow); this reflects a mixed $(f+p)$-wave parity. 
Note, however, that, because of the odd momentum dependence, the integrated spin-triplet pair amplitude vanishes:
\begin{equation}
\bar{F}_{t}^{f,s, \text{eff}}(z)
= \frac{1}{(2\pi)^2}\int d\bm{k}\,F_{t}^{f,s, \text{eff}}(z,\bm{k}) = 0,
\end{equation}
which also applies to the $p$-wave case, see Fig.\,\ref{figureR2}(a).
%, consistent with the vanishing spin density in driven odd-parity magnets. 
The integrated spin-triplet pair amplitude is shown in Fig.\,\ref{figureR3}, for odd- and even-parity unconventional magnets.

In the case of even-parity unconventional magnets, the emergent spin-triplet pair amplitude acquires the contribution of both light-induced fields, $M_{\Omega}^{j} $ and $ B_{\bm{k}\Omega}^{j}$, see Eq.\,(\ref{eq_fteff}). On top of this, there is a contribution from the static unconventional magnet $M_{\bm k}^{j} $. Since  $M_{\Omega}^{j} $ and $ B_{\bm{k}\Omega}^{j}$ have different momentum dependence, the spin-triplet pair amplitude acquires mixed parities under driving. For instance, in the driven $d$-wave case, discussed in the main text, a coexistence of $(d+s)$-wave parity appears;
The $s$-wave parity term comes from the light-induced Zeeman field $M_{\Omega}^{j} $, as we showed in the main text. Also, the $s$-wave term changes relative lobe sizes, as shown in Fig.\,\ref{figureR2}(c–iii) and indicated by the thick blue and thin red arrows.
Moreover, the spin-triplet pair amplitude in a $g$-wave magnet ($j=4$) is shown in Figs.\,\ref{figureR2}(d–ii) and (d–iii) for static and driven cases, respectively. 
In the static state, the spin-triplet pair amplitude given be Eq.\,(\ref{eq_fteff}) with $A_0=0$ displays $g$-wave symmetry, with equal values at $\theta_{k}=0$ and $\pi/2$ (thin blue arrows) and an opposite sign at $\theta_{k}=\pi/4$ (thin red arrow). 
Under driving, a $d$-wave component appears, enhancing (reducing) the amplitude along $\theta_{k}=0$ ($\theta_{k}=\pi/2$) and thereby breaking the original $g$-wave symmetry (thick dashed blue arrow). 
Additionally, the light-induced Zeeman term introduces an $s$-wave component, see Eq.\,(\ref{eq_mjO}), leading to unequal magnitudes between $\theta_{k}=0$ and $\theta_{k}=\pi/4$. 
Consequently, the driven $g$-wave magnet exhibits a mixed $(g+d+s)$-wave parity: the $g$-wave term comes from symmetry of the unconvention magnet $M_{\bm k}^{g} $, the $d$-wave term comes from the light-induced unconvention magnetic field $B_{\bm k \Omega}^{g}$ and the $s$-wave parity comes from the light-induced Zeeman field $M_{\Omega}^{g}$.
A similar argument applies to the driven $i$-wave case [Fig.\,\ref{figureR2}(e)], 
which displays a hybrid $(i+g+d+s)$-wave symmetry; the $i$-wave term comes from the unconvention magnet $M_{\bm k}^{i} $, the $s$-wave parity comes from the light-induced Zeeman field $M_{\Omega}^{i}$ and $g+d$-wave parity comes from the light-induced unconvention magnetic field $B_{\bm k \Omega}^{i}$.
This, therefore, demonstrates that even-parity magnets with conventional $s$-wave spin-singlet superconductivity host Cooper pairs with multiple parities with a number of parity symmetries that is higher than their odd-parity counterparts.

The additional parity terms emerging as a result of the linearly polarized light drive have further consequences.
As demonstrated in the main text, the induced $s$-wave component breaks the original compensation in the static $d$-wave state and results in a finite integrated triplet amplitude
\begin{equation}
\bar{F}_{t}^{j,s, \text{eff}}(z)
= \frac{1}{(2\pi)^2}\int d\bm{k}\,F_{t}^{j,s, \text{eff}}(z,\bm{k}) \neq 0,
\quad j~\text{even},
\end{equation}
and similar situations occur for the $g$- and $i$-wave magnets, see Fig.\,\ref{figureR3}.

A crucial consequence of the applied light is that the light-induced Zeeman field $M_{\Omega}^j$ induces a spin splitting in the quasiparticle spectrum and therefore give raise a characteristic peak–dip structure in the integrated spin-triplet pair amplitude around $\omega/\mu = -1$, see the colored arrows in Fig.\,\ref{figureR3}. 
Analogous to the behavior of the spin density in normal states as discussed in the main text, these peak–dip features enable the identification of both the strength and orientation of the underlying altermagnetic order, following the relations
\begin{eqnarray}
M\cos\!\left(2\Theta_{d}\right) &=& \left(\frac{\hbar k_{F}}{eA_{0}}\right)^{2}\delta\omega_{1}, \label{Eq_Mdomega} \\[4pt]
M\cos\!\left(4\Theta_{g}\right) &=& \frac{4}{3}\left(\frac{\hbar k_{F}}{eA_{0}}\right)^{4}\delta\omega_{1}, \label{Eq_Mgomega} \\[4pt]
M\cos\!\left(6\Theta_{i}\right) &=& \frac{8}{5}\left(\frac{\hbar k_{F}}{eA_{0}}\right)^{6}\delta\omega_{1}, \label{Eq_Miomega}
\end{eqnarray}
where $\Theta_{j}=\theta_j-\phi$ and $\delta\omega_{1}$ denotes the Zeeman splitting between the peak and dip energies, similar to the normal static shown in Fig. 2 in the main text for $d$-wave magnets and Fig.\,\ref{figureR1} for $g$- and $i$-wave magnets.
The relation between the Zeeman splitting $\delta\omega_{1}$ and the even-parity altermagnetism,  Eq.\,(\ref{Eq_Mdomega})--(\ref{Eq_Miomega}), can be obtained likewise from Eq.\,(\ref{eq_mgomega})--(\ref{eq_miomega}).

The relations between $M$ and $\Theta_{j}$ and $\omega_{1}$ in Eqs.\,(\ref{Eq_Mdomega})-(\ref{Eq_Miomega}) are consistent with the discussion in the spin-densities in the driven normal states, where the peak-dip structures also provide the information of the altermagnetism.
This is because the spin-density in the normal states is related to the spin-triplet amplitude in the superconducting states, as demonstrated in Eq.\,(18) in the main text. 
Therefore, the conclusion of the main text remains valid: 
The strength and orientation of even-parity unconventional magnets (the $d$-, $g$-, and $i$-wave altermagnets) can be directly extracted from the peak–dip structures of the light-induced spin-triplet amplitude.

\subsection{Low-frequency driven Cooper pair correlations}  \label{SecIIC}

As discussed above and in the main text, spin-triplet odd-frequency pairing only appears in driven $d$-wave UMs in the high-frequency regime, which provides a measurable finite spin-triplet amplitude absent in the static regime.
In this part, we explore the low-frequency driving field and demonstrate the emergence for unusual Cooper pairs that are absent in the static and high-frequency regimes.

In the low-frequency regime, the superconducting pair amplitudes are encoded in the anomalous Floquet Green’s function, associated with the Floquet BdG Hamiltonian derived from Eq. (3) in the main text\,\cite{PhysRevB.103.104505, PhysRevB.109.134517}.
This Green’s function satisfies  the antisymmetry condition: $F^{\sigma_1,\sigma_2}(\bm{k}_1,\bm{k}_2;t_1,t_2) = -F^{\sigma_2,\sigma_1}(\bm{k}_2,\bm{k}_1;t_2,t_1)$. 
When applied to the two-time Floquet Green’s function\,\cite{PhysRevB.103.104505}, the antisymmetry condition leads to the constraint
\begin{equation}
F_{n,m}^{\sigma_1,\sigma_2}(\bm{k}_1,\bm{k}_2;z) = -F_{-m,-n}^{\sigma_2,\sigma_1}(\bm{k}_2,\bm{k}_1;-z),
\label{eqFnm}
\end{equation}
under the exchange of Floquet indices, spin, momentum, and frequency.
Here, Floquet anomalous Green's functions $F_{n,m}^{\sigma_{1}\sigma_{2}}(\bm{k}_1, \bm{k}_2; z)$,  represent   electron pairing   between  the Floquet bands ($n,m$) via the    absorption/emission of $|n-m|$ photons.
Since this additional Floquet index is absent in the static and high-frequency regimes,
Eq.\,(\ref{eqFnm}), thus, expands the set of symmetry-allowed superconducting pairings in driven systems. 
In particular, up to eight distinct classes of Floquet Cooper pairs can arise, classified by their parities under exchange of spin, Floquet indices, frequency, and momentum \cite{PhysRevB.103.104505}.
To characterize these pairings, we define the even and odd combinations under Floquet index exchange:
\begin{equation}
F_{n,m}^{\sigma_1,\sigma_2, \pm}(z, \bm{k}) = \tfrac{1}{2} \big[ F_{n,m}^{\sigma_1,\sigma_2}(z, \bm{k}) \pm F_{-m,-n}^{\sigma_1,\sigma_2}(z, \bm{k}) \big],
\label{eq_Foe}
\end{equation}
where the transformation $(n,m) \leftrightarrow (-m,-n)$ captures the Floquet exchange symmetry \cite{PhysRevB.103.104505}. The spin-singlet and spin-triplet components are further extracted as\,\cite{maeda2025pair}:
\begin{align}
F_{n,m}^{s,\pm}(z,\bm{k}) &= \tfrac{1}{2} \big[ F_{n,m}^{\uparrow\downarrow,\pm}(z,\bm{k}) - F_{n,m}^{\downarrow\uparrow,\pm}(z,\bm{k}) \big], \\
F_{n,m}^{t,\pm}(z,\bm{k}) &= \tfrac{1}{2} \big[ F_{n,m}^{\uparrow\downarrow,\pm}(z,\bm{k}) + F_{n,m}^{\downarrow\uparrow,\pm}(z,\bm{k}) \big].
\label{eq_Soe}
\end{align}
Here, $F_{n,m}^{s,\pm}$ ($F_{n,m}^{t,\pm}$) denote spin-singlet (spin-triplet) even/odd-Floquet pairings. 
This decomposition provides a systematic framework for identifying all symmetry-allowed Cooper pair channels in time-periodically driven superconductors. 
The full classification, including frequency and momentum parities, can be verified numerically, allowing the identification of all possible Floquet pairing symmetries in the system.

\begin{figure}[t]
\centering
\includegraphics[width=0.5\linewidth]{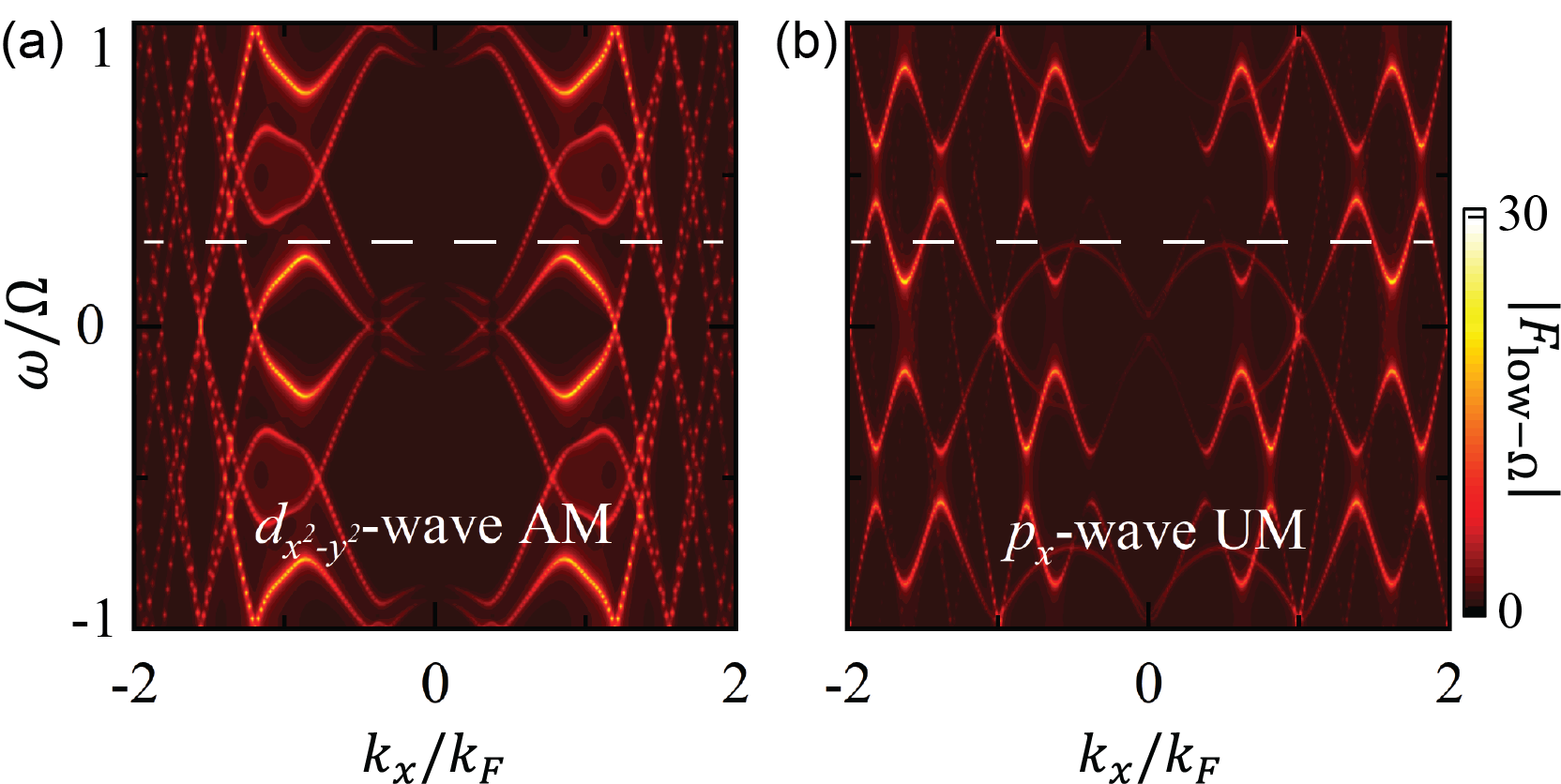}
\caption{
(a) Absolute value of an emergent spin-triplet odd-Floquet odd-frequency odd-parity pair amplitude $F_{\rm low-\Omega}$ as a function of $\omega$ and $k_{x}$ at $k_{y}=0$ and $M/\mu=0.5$  in a $d_{x^{2}-y^{2}}$-wave AM ($\theta_d=0$) with $s$-wave superconductivity under low-frequency LPL. 
(b) Same as in (a) but for an emergent spin-triplet odd-Floquet even-frequency even-parity pair amplitude in a $p$-wave UM ($\theta_p=0$). 
Parameters:  $\phi=0$, $\Delta=0.5$, while the rest same as in Fig. 3 in the main text. 
The white dashed line indicates $\omega=\Omega/4$ related to Fig. 4 in the main text.
}
\label{FigureS1}
\end{figure}

Fig.\,\ref{FigureS1} exhibits the absolute value of an emergent spin-triplet odd-Floquet odd-frequency (even-frequency) odd-parity (even-parity) pair amplitude $F_{\rm low-\Omega}=\sum_{n,m}F^{\rm t, -}_{n,n+2m+1}$ for driven $d_{x^{2}-y^{2}}$-wave ($p$-wave) magnet as a function of $\omega$ and $k_{x}$, which originate from the pairing between electrons different Floquet sideband through $|2m+1|$-photon processes. 
These pairing amplitudes are also shown in Fig.\,4 in the main text, demonstrating the joint effect of unconventional magnetism and time-periodic driving in conventional superconductors.

%\begin{figure}[!t]
%\centering
%\includegraphics[width=0.5\linewidth]{../maintext/Figure4.png}
%\caption{(a) Spin-triplet odd-Floquet odd-frequency odd-parity pair amplitude pair amplitude $\bar{F}_{\rm low- \Omega}$ integrated in $\bm{k}$ as a function of $A_{0}$ for distinct $M$   in a $d_{x^{2}-y^{2}}$-wave AM  with conventional superconductivity under low-$\Omega$ LPL. 
%(b) Same as in (a) but for an emergent spin-triplet odd-Floquet even-frequency even-parity pairing in a $p_{x}$-wave UM. The insets show the momentum dependence of $F_{\rm low- \Omega}$.  Parameters: $\phi=0$, $\Delta=0.5$, and $\omega = \Omega/2$, while the rest same as in Fig. 3 in the main text. 
%}
%\label{Figure4}
%\end{figure}

\subsection{Summary}  \label{SecIID}
In summary, our extended analysis establishes that the key phenomena reported in the main text for $d$-wave altermagnets are universal across all even-parity unconventional magnets. 
Linearly polarized driving fields generate a light-induced Zeeman term producing characteristic peak–dip structures in spin-triplet amplitudes. 
These signatures encode the magnitude and orientation of the intrinsic altermagnetic field, 
providing a robust and experimentally accessible probe for detecting altermagnetism in both normal and superconducting states. 
In contrast, for odd-parity magnets, such as $p$- and $f$-wave systems, the absence of the light-induced Zeeman field leads to vanishing spin density and triplet amplitude, preserving vanishing magnetization even under external driving.
Furthermore, pairing between different Floquet sidebands, assisted by a low-frequency driving field, enables Cooper pairs that are absent in the static and high-frequency regimes, and thus offer a broader ground for engineering dynamical Cooper pairs in UMs.

\section{Assessing  conditions for  experimental realization: homogenous drive, screening effect, benefits of conventional superconductivity}
The results presented in the previous sections demonstrate the universality of the proposed light-induced spin density and spin-triplet pair amplitude in unconventional magnets without and with conventional $s$-wave spin-singlet superconductivity.
This Floquet engineering technique aims to provide an experimentally accessible proposal to detect the magnitude and orientation of the intrinsic of even-parity unconventional magnetic field.
In this section, we provide some of important aspects about the experimental conditions in favor of the realization of our findings.
In particular, we discuss the validity of the spatially uniform driving field assumption made in this work in Sec.\,\ref{SecIIIA} and estimate the realistic parameters of the driving field, including the frequency $\Omega$ and driving amplitude $A_0$ in Sec.\,\ref{SecIIIB}, that are required to generate measurable spin density and spin-triplet pair amplitude in the proposal. 
The advantage of using $s$-wave superconductivity proximity effect is discussed in Sec.\,\ref{SecIIIC}.

\subsection{Validity of the spatially uniform driving field} \label{SecIIIA}
Throughout this work, the driving field is considered to be spatially uniform: $\bm{A}(\bm{r},t)=\bm{A}(t)$.
This is justified by comparing several characteristic length scales to satisfy the condition: $d \ll \delta \ll \lambda$ and $L \ll \lambda$.
Here $d$ and $L$ denote the sample thickness and lateral size, respectively, $\lambda$ is the wavelength of the driving light, and $\delta$ is the optical penetration depth, which depends on both the material properties and the drive frequency. 
As discussed below, these conditions are well satisfied for the characteristic lengths relevant to our work.

(i) Sample dimensions.
In Floquet-engineering experiments on antiferromagnets such as Mn$_3$Sn \cite{Matsuda2020Room}, Cr$_2$O$_3$ \cite{Zhang2023Light}, and MnPS$_3$ \cite{Shan2021Giant}, the typical lateral size and thickness of exfoliated or epitaxial flakes are $L \sim 1$–$10~\mu$m and $d \lesssim 50$ nm, respectively. 
Among these, Mn$_3$Sn is a representative even-parity unconventional (altermagnetic) magnets \cite{Jungwirth2025Symmetry}. 
Recent experiments have also synthesized thin flakes of NiI$_2$ \cite{Song2025Electrical} and Gd$_3$Ru$_4$Al$_{12}$ \cite{Yamada2025Metallic}, exhibiting unconventional $p$-wave magnetism and geometries consistent with the dimensions assumed in our theoretical model.

(ii) Wavelength of the driving field.
The spin-triplet states proposed here are driven by mid-infrared or terahertz radiation with photon energies $\hbar\Omega = 0.05$–$0.2$~eV \cite{Dehghani2021Light,Kennes2019Light}, corresponding to wavelengths $\lambda = 2\pi c/\Omega \simeq 10$–$100~\mu$m. 
The wavelength, therefore, exceeds both the lateral size $L$ and thickness $d$ of the sample by at least three orders of magnitude, i.e., $\lambda \gg L \gg d$.

(iii) Optical penetration depth.
For a good conductor, the penetration depth is given by
\begin{equation}
\delta \simeq \sqrt{\frac{2}{\mu_{0}\Omega\sigma}},
\label{eq_Opd}
\end{equation}
where $\mu_{0}$ is the vacuum permeability, $\Omega$ is the angular frequency of the drive, and $\sigma$ is the electrical conductivity \cite{Ashcroft1976Solid,Jackson1999Classical}. 
For metallic conductivities $\sigma \sim 10^{6}$–$10^{7}$~S\,m$^{-1}$ and mid-infrared or terahertz frequencies $\hbar\Omega = 0.05$–$0.2$~eV ($\Omega \sim 7\times10^{13}$–$3\times10^{14}$~s$^{-1}$), 
one obtains $\delta \simeq 50$–$200$~nm, consistent with experimental determinations of the skin depth in metallic films at similar frequencies~\cite{Luo2021LightInduced,McIver2020Light}.
This penetration depth is much smaller than the wavelength of the driving ($\lambda \gg \delta$) but typically larger than the thickness of the films ($\delta\gg d$).

Therefore, based on the above discussion, the driving field can be regarded as spatially uniform, 
since the wavelength of the driving light greatly exceeds both the lateral size $L$ and the thickness $d$ of the sample, 
as well as the penetration depth. 
These parameters satisfy the condition $d \ll \delta \ll \lambda$ and $L \ll \lambda$.

At this point, we note that, even though the wavelength and frequency of the applied field are related to its spatial uniformity, which we justified in the previous paragraph, the relevant length scales are not directly related to the detection of the proposed light-induced spin-triplet states. 
This is because the found light-induced phases originate from a light-induced Zeeman field
\begin{equation}
M^{d}_{\Omega} = \frac{1}{2} \left(\frac{eA_{0}}{\hbar k_{F}} \right)^2 M, \label{eq_R2}
\end{equation}
where $M$ is the intrinsic strength of the $d$-wave altermagnetism; This light-induced Zeeman field is the second term in the first expression of Eqs. (4) in the main text. 
As we clearly see in Eq.\,(\ref{eq_R2}), the light-induced Zeeman field is determined by the amplitude of the drive $A_{0}$.
%which is determined by the \textbf{amplitude} of the drive. 
The light-induced Zeeman field in Eq.\,(\ref{eq_R2}) is also responsible for giving rise to the peak-dip structures shown in Figs. 2-4  in the main text.

Using representative values for altermagnets \cite{Jungwirth2025Symmetry}, $M \simeq 0.1$–$0.3$ eV. Then, taking  
$\frac{eA_{0}}{\hbar k_{F}} \simeq 0.5$ as adopted in this work, 
the corresponding light-induced Zeeman field is estimated to be 
$M^{d}_{\Omega} \simeq 12.5$–$37.5$ meV. 
This energy scale lies well within the resolution of current pump–probe and tr-ARPES techniques:  
Light-induced band splittings around the order of $10$--$50$ meV have already been resolved in graphene \cite{McIver2020Light,Merboldt2025Observation}, 
ZrTe$_5$ \cite{Luo2021LightInduced}, and Bi$_2$Te$_3$ \cite{aeschlimann2021survival}. 
These experimental advances confirm the feasibility of detecting the predicted Floquet-induced spin-triplet states.

Therefore, since $\lambda$ greatly exceeds the sample dimensions and $\delta > d$ in this regime, 
the conditions $d \ll \delta \ll \lambda$ and $L \ll \lambda$ are well satisfied for the parameter range considered here. 
The electromagnetic field can thus be regarded as homogeneous both through the film thickness and across the in-plane directions, fully validating the uniform-field approximation employed in our calculations. 
Under such a spatially uniform driving field, the proposed light-induced spin-triplet states should be experimentally detectable under current conditions. 
As a final remark, we note that a spatially nonuniform driving field could, in principle, generate a pseudo-magnetic field leading to light-induced Landau quantization. 
Such effects may modify the detailed quasiparticle spectrum but would not suppress the primary light-induced Zeeman splitting responsible for the spin-triplet states discussed in this work.

\subsection{Realistic parameter estimation for driving field} \label{SecIIIB}
Our calculations are parameterized by the dimensionless coupling $eA_0/(\hbar k_F)=0.1$–$1$ in the mid-infrared/terahertz range $\hbar\Omega=0.05$–$0.2$~eV. 
Using $E_0=A_0\Omega$ and a representative Fermi wave vector $k_F\!\sim\!10^{10}$~m$^{-1}$, we obtain
\begin{equation}
A_0\approx(6.6\times10^{-6}\text{–}6.6\times10^{-5})~\mathrm{V\,s/m}, \qquad
\Omega\approx(7.6\times10^{13}\text{–}3.0\times10^{14})~\mathrm{s^{-1}},
\end{equation}
and therefore
\begin{equation}
E_0=A_0\Omega \approx 0.2\text{–}3~\text{MV/cm}.
\end{equation}
Because screening reduces the internal electric field relative to the incident vacuum field, the effective field acting on the electrons is typically smaller by a factor of about two. 
In Floquet-engineering experiments on graphene, for instance, the applied driving field is roughly twice the effective field inside the material~\cite{Merboldt2025Observation}. 
Applying this correction, the required applied field to generate detectable spin-triplet states is estimated as $E_{\text{app}}\!\approx\!0.4$–$6~\text{MV/cm}$. 
This range is consistent with current experimental capabilities, including Floquet studies in graphene~\cite{McIver2020Light,Merboldt2025Observation}, topological insulators~\cite{aeschlimann2021survival}, and semimetals~\cite{Luo2021LightInduced}, where applied fields of $0.3$–$2~\text{MV/cm}$ have been routinely achieved without sample damage. 
Therefore, the field amplitudes corresponding to our chosen $eA_0/(\hbar k_F)$ values are quantitatively realistic for the frequency range of interest and are compatible with present experimental conditions in conductive and (anti)ferromagnetic thin films.

\subsection{Screening effect of the driving field} \label{SecIIIC}

The effect of light illuminated to a metallic system might be reduced by the screening effect \cite{Ashcroft1976Solid,Jackson1999Classical}.
While a comprehensive study of screening effects in driven unconventional magnets warrants a dedicated investigation, which we plan to pursue in a forthcoming work extending a previous study \cite{Mizushima2025Detecting}, there are solid physical arguments indicating that the main results presented here remain robust, as elaborated below.

Recent Floquet experiments have directly addressed this issue in both conductors \cite{Esin2020Floquet} and semimetallic materials \cite{Merboldt2025Observation,McIver2020Light,Luo2021LightInduced}, showing that coherent light–matter coupling remains effective despite the presence of free carriers. 
In particular, Floquet–Bloch sidebands and a light-induced anomalous Hall effect have been observed in monolayer graphene, demonstrating that Floquet states survive even in the presence of metallic screening \cite{McIver2020Light,Merboldt2025Observation}. 
Quantitatively, the screening effect mainly renormalizes the amplitude of the electric field acting inside the material, which can be phenomenologically described as 
 $E_{\text{eff}}=f_{F}E_{\text{app}}$ \cite{Merboldt2025Observation}, where $E_{\text{eff}}$ is electric field effective interacting with the material, $E_{\text{app}}$ is applied field in experiments, and $f_{F}$ is a factor from Fresnel equations.
Thus, the materials interact with an effective electric field that is reduced by the dielectric screening due to the refraction between the vacuum and the conducting layer.
However, this screening-induced electric field reduction does not destroy observation of Floquet states \cite{McIver2020Light,Luo2021LightInduced}, and our results are therefore expected to remain under screening effects.

Based on the analysis above and the estimation of the driving amplitude expected in the current work, the effective field is about $E_{\text{eff}}\!\sim\!3$~MV/cm to produce measurable light-induced spin-triplet states 
This corresponds to an applied field of $E_{\text{app}}\!\sim\!6$~MV/cm, which is within the range of current experimental capabilities \cite{Merboldt2025Observation,McIver2020Light,Luo2021LightInduced}. 
For comparison, pump–probe and trARPES measurements have already achieved peak fields of $1$–$5$~MV/cm without sample damage or degradation of coherence. 
Moreover, Floquet-engineered states have also been observed in antiferromagnetic insulators such as Mn$_3$Sn \cite{Matsuda2020Room}, Cr$_2$O$_3$ \cite{Zhang2023Light}, and MnPS$_3$ \cite{Shan2021Giant}, where Mn$_3$Sn is a candidate altermagnet with noncollinear even-parity magnetic order \cite{Jungwirth2025Symmetry}. 
These experimental and material advances demonstrate that realizing Floquet-engineered spin-triplet states in unconventional magnets is experimentally feasible under present conditions. 
Therefore, although screening is unavoidable in metallic or semimetallic systems, its quantitative effect is moderate and does not preclude the observation of the proposed Floquet-induced phenomena.

\subsection{Advantages of using $s$-wave superconductivity} \label{SecIIID}
Although the unconventional magnets coupled with conventional $s$-wave superconductors is discussed here, it is not strictly necessary to restrict the parent superconductor to conventional $s$-wave pairing \cite{FukayaCayaoReviewUMs}. 
Nevertheless, choosing an $s$-wave superconductor offers advantages both experimentally and theoretically for elucidating the essential effects arising from the interplay between light and unconventional magnetism. 

(i) Experimental accessibility.
Although other pairing channels are possible, conventional $s$-wave superconductivity is by far the most thoroughly understood, both theoretically and experimentally. 
Its synthesis, control, and proximity effect are well established in materials such as Al, Nb, and NbO$_2$, which serve as standard building blocks for hybrid superconducting devices, including ferromagnet–superconductor heterostructures \cite{Golubov2025Physics,Vagov2023Intertype}, 
antiferromagnet–superconductor hybrids \cite{Machida1980Theory}, 
and topological systems \cite{Prada2020From}. 
In contrast, for unconventional spin-singlet pairing channels such as $d$-wave superconductivity, the pairing mechanism and interface behavior remain less explored \cite{Zareapour2012Proximity,Lofwander2004Proximity,kashiwaya2000}. 
Although numerous theoretical studies and high-$T_c$ cuprate materials demonstrate intrinsic $d$-wave pairing, their integration with unconventional magnets under light irradiation is still largely unexplored and experimentally untested.

(ii) Theoretical transparency.
Unconventional magnets are characterized by direction-dependent spin-split Fermi surfaces. 
When a periodic driving field is applied, additional symmetries emerge in the Fermi surface through the light-induced Zeeman and unconventional magnetic fields [see Fig.\,\ref{figureR2}]. 
When such a magnet is coupled to an \emph{isotropic} $s$-wave superconductor, the resulting spin-triplet correlations acquire momentum-space symmetries that directly reflect the static unconventional magnetic field and its light-induced modifications. 
This makes it possible to clearly attribute the observed parity and orientation of the triplet amplitude to the underlying magnetic structure \cite{FukayaCayaoReviewUMs}. 
In contrast, coupling to a $d$-wave superconductor would introduce an additional intrinsic anisotropy due to the $d$-wave gap function, making the origin of the induced triplet symmetry far less transparent, as it would arise from a synergic effect of the $d$-wave pairing, unconventional magnetism, and the drive. 
Moreover, the microscopic interaction between unconventional (e.g., $d$-wave) superconductivity and time-periodic driving remains an open question that warrants a separate and detailed investigation \cite{Dehghani2021Light,Kennes2019Light,Kitamura2022Floquet,yokoyama2025FloquetSC,PhysRevB.109.134517}, which lies beyond the scope of the present work.

Therefore, employing an $s$-wave superconductor allows a clear identification of the key physical mechanism and offers an experimentally feasible route to realize the light-induced spin-triplet correlations.

\bibliography{biblio}